\documentclass[
    journal,
    twocolumn,
    ]{IEEEtran}
\usepackage{graphicx}
\usepackage{amsmath}
\usepackage{amssymb}
\usepackage{epsfig}
\usepackage{epstopdf}
\usepackage{amsthm}
\usepackage{afterpage}
\usepackage{amsmath,amssymb,amsfonts}
\usepackage{verbatim}
\usepackage{psfrag}
\usepackage{color}
\usepackage{cite}
\usepackage{setspace}
\usepackage{adjustbox}
\usepackage{epsfig}
\usepackage{epstopdf}
\usepackage{footmisc}
\usepackage{fmtcount}
\usepackage{stackrel}
\usepackage{float}
\usepackage{breqn}
\usepackage{orcidlink}
\usepackage{hyperref}
\usepackage{enumerate}
\usepackage{graphicx}
\usepackage{subcaption}
\usepackage{flushend}
\usepackage{soul}
\usepackage{xcolor}
\usepackage[T1]{fontenc}
\usepackage{graphicx}

\usepackage{tabularx,booktabs}

\begin{document}
%\doublespace

\title{\title{Semantic Communication for the Internet of Underwater Things: Architectures, Applications, Challenges, and Future Directions}}
\author{Ruhul Amin Khalil, \IEEEmembership{Member, IEEE}, Asiya Jehangir, \IEEEmembership{Student Member, IEEE}, Hanane Lamaazi, Saddaf Rubab, \IEEEmembership{Senior Member, IEEE}, and Nasir Saeed, \IEEEmembership{Senior Member, IEEE}
\thanks{This work was supported by the Office of the Associate Provost for Research at the United Arab Emirates University (UAEU), UAE. \textit{(Corresponding author: Ruhul Amin Khalil e-mail: ruhulamin@uaeu.ac.ae)}}
\thanks{R. A. Khalil and N. Saeed are with the Engineering Requirement Unit (ERU) and Department of Electrical and Communication Engineering, College of Engineering, United Arab Emirates University, Al-Ain 15551, UAE.}
\thanks{A. Jehangir is with the Department of Electrical Engineering, University of Engineering and Technology, Peshawar 25120, Pakistan.}
\thanks{H. Lamaazi is with the College of Information Technology, United Arab Emirates University, Al-Ain 15551, UAE.}
\thanks{S. Rubab is with the Department of Computer Engineering, College of Computing and Informatics, University of Sharjah, Sharjah 27272, UAE.}
\thanks{Manuscript received~X, X~X; revised ~X, X~X.}}
\markboth{}%
{Shell \MakeLowercase{\textit{et al.}}: Bare Demo of IEEEtran.cls
for Journals}

\maketitle

\begin{abstract}
The Internet of Underwater Things (IoUT) supports marine sensing, environmental monitoring, subsea inspection, and autonomous underwater operations. However, IoUT communication is constrained by limited bandwidth, long propagation delay, time-varying underwater channels, intermittent connectivity, and strict energy budgets. Semantic Communication (SC) offers a promising alternative by transmitting task-relevant meaning rather than raw data, thereby improving communication efficiency in resource-constrained underwater networks. This paper presents a critical and feasibility-aware survey of SC for IoUT, focusing on opportunities, challenges, limitations, and future research directions. We first review the fundamentals of SC-enabled IoUT systems, including semantic representations, layered architectures, semantic channel modeling, and task-oriented evaluation metrics. We then examine learning-driven approaches based on machine learning (ML), knowledge graphs (KGs), vision-language models (VLMs), generative models, and federated learning (FL), with emphasis on their feasibility under underwater edge constraints. Representative applications, including environmental monitoring, marine ecology, subsea infrastructure inspection, disaster response, and autonomous underwater vehicle (AUV) coordination, are analyzed from an SC perspective. Finally, we identify key research directions involving standardized semantic models, reproducible testbeds, compute--communication trade-offs, trustworthy reconstruction, hybrid underwater links, energy-aware edge intelligence, semantic security, digital twins (DTs), and cross-domain interoperability. This survey provides a structured foundation for developing reliable, efficient, and meaning-driven IoUT communication systems.
\end{abstract}


\section{Introduction}
\label{sec:introduction}

The Internet of Underwater Things (IoUT) extends the Internet of Things (IoT) paradigm to underwater and subsea environments by interconnecting underwater sensors, autonomous underwater vehicles (AUVs), remotely operated vehicles (ROVs), seabed observatories, surface buoys, underwater gateways, and shore- and cloud-based infrastructures. The IoUT concept was introduced to support persistent underwater sensing, communication, and decision-making in marine environments where terrestrial networking models cannot be directly applied \cite{Domingo2012IoUT}. IoUT systems are relevant to environmental monitoring, disaster prevention, offshore industry, military surveillance, underwater archaeology, aquaculture, and ocean exploration, as discussed in comprehensive studies of underwater networking applications \cite{Kao2017IoUT,Felemban2015USNApplications,khalil2020toward}. Recent smart-ocean architectures further emphasize that practical IoUT deployments require the integration of sensing, communication, networking, data fusion, and application layers \cite{Qiu2020SmartOcean}. However, underwater environments impose severe constraints on communication, localization, synchronization, computation, and energy management. These constraints arise from water-dependent propagation, node mobility, long deployment periods, limited physical access, and the difficulty of underwater maintenance \cite{Heidemann2012UnderwaterSensors}. Consequently, IoUT networks remain fundamentally different from terrestrial IoT systems and require communication methods that are efficient, adaptive, and aware of the application-level value of information \cite{Akyildiz2005UASN}.

\begin{figure}[ht!]
\centering
\includegraphics[width=1\columnwidth]{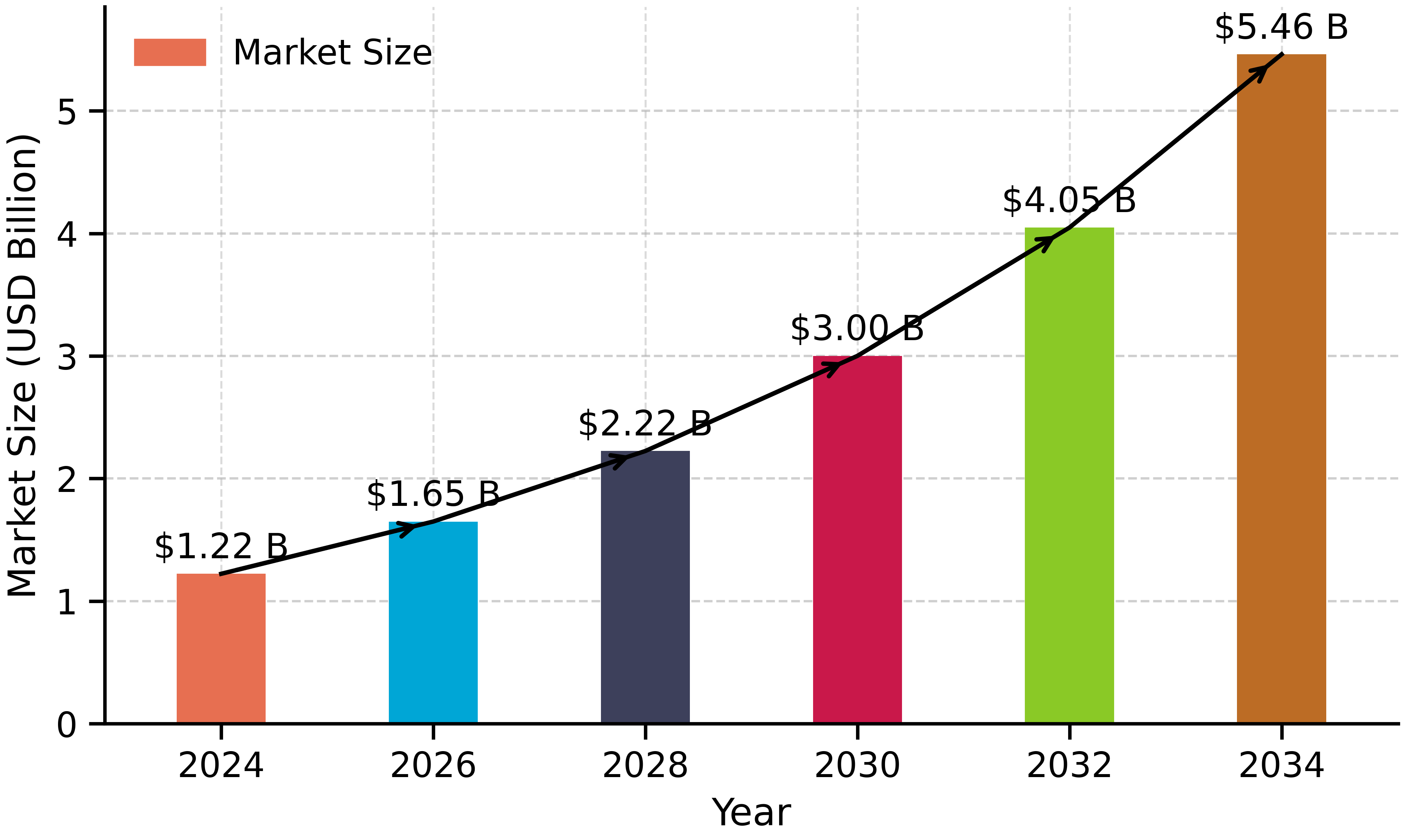}
\caption{Projected growth trend of underwater/marine IoT markets. Data adapted from \cite{MRFUnderwaterMarineIoT2026}; intermediate values are interpolated using the reported CAGR of 16.15\%.}
\label{markettrends}
\end{figure}

The increasing demand for marine sensing, subsea automation, offshore infrastructure inspection, and oceanographic intelligence has also strengthened the commercial relevance of underwater and marine IoT technologies. Fig.~\ref{markettrends} presents an indicative market-growth trend for underwater/marine IoT systems based on the cited market report \cite{MRFUnderwaterMarineIoT2026}. Since commercial market estimates may vary across providers, this figure is used only to motivate the growing demand for underwater connectivity and should not be interpreted as technical evidence for the performance of any specific IoUT architecture. To maintain research integrity, the source of the plotted data is explicitly acknowledged in the caption.

Underwater communication is commonly achieved through acoustic, optical, radio-frequency (RF), magnetic induction (MI), or hybrid links. A broad survey of underwater wireless communication technologies shows that acoustic, optical, and RF links differ significantly in range, bandwidth, propagation delay, hardware complexity, and environmental sensitivity \cite{Gussen2016UWC}. Acoustic communication remains the dominant option for long-range underwater connectivity because sound waves propagate much farther than optical or RF signals in water \cite{Stojanovic2009UAC,khalil2021bayesian}. Nevertheless, acoustic links suffer from limited bandwidth, slow propagation speed, multipath propagation, Doppler spread, and strong dependence on temperature, depth, salinity, and mobility \cite{Chitre2008UnderwaterAcoustic}. A recent review of underwater acoustic communication further highlights open challenges in channel modeling, adaptive signal processing, medium access control, and network protocol design \cite{Li2025UnderwaterAcoustic}. Underwater optical wireless communication can provide much higher data rates over short distances, but its performance is sensitive to absorption, scattering, turbidity, alignment, and line-of-sight requirements \cite{Kaushal2016UOWC}. A detailed survey of underwater optical wireless communication shows that optical links are attractive for high-throughput local exchange, yet they cannot replace long-range acoustic connectivity \cite{Zeng2017UOWCSurvey}. Therefore, underwater optical networking and localization methods are commonly considered complementary components of heterogeneous IoUT systems \cite{Saeed2019UOWC}. RF and MI links can support specific short-range, near-surface, or seabed scenarios, but their deployment flexibility is restricted by conductivity, antenna or coil design, alignment, and attenuation effects \cite{Akyildiz2015MI}. Recent reviews confirm that practical underwater networks are increasingly moving toward hybrid and adaptive communication architectures \cite{Theocharidis2025UnderwaterComm}.

Beyond propagation impairments, IoUT nodes are usually constrained by limited battery capacity, restricted memory, modest processing power, sparse deployment, and difficult physical access. Energy-efficient medium access remains a central issue because underwater acoustic networks require careful control of channel access, retransmissions, collisions, and idle listening \cite{Hasan2025EnergyMAC}. A review of underwater wireless sensor network (UWSN) issues identifies packet-size selection, localization, routing, channel utilization, and power control as persistent challenges \cite{Awan2019UWSNReview}. Similarly, a taxonomy-oriented survey of UWSNs reports that underwater sensing platforms must address dynamic topology, low throughput, limited bandwidth, and energy imbalance \cite{Fattah2020UWSNSurvey}. At the data level, IoUT deployments can generate large volumes of marine data from cameras, sonars, hydrophones, chemical sensors, and mobile vehicles, making raw-data forwarding expensive in terms of bandwidth and energy \cite{Jahanbakht2021IoUTBigData}. IoUT communication architectures therefore require not only improved physical links but also intelligent data reduction, context-aware transmission, and application-aware forwarding \cite{Bello2022IoUTCommunication}. Security and privacy concerns further complicate deployment because underwater devices may be exposed to jamming, routing attacks, Sybil attacks, blackhole attacks, physical capture, and privacy leakage \cite{Raj2021UIoTSecurity}. Routing is also difficult because underwater links are intermittent, topology changes are frequent, and forwarding energy can be high \cite{Luo2021RoutingUWSN}. These constraints motivate communication paradigms that transmit fewer but more useful messages rather than attempting to deliver every raw observation.

SC has emerged as a promising paradigm in which the goal is not only to reproduce transmitted bits but also to preserve the meaning, relevance, or task utility of information. This idea is consistent with the classical view that communication can be studied at technical, semantic, and effectiveness levels \cite{ShannonWeaver1949}. In modern wireless systems, SC has been revisited as a means to move beyond purely bit-oriented communication toward meaning-aware and goal-oriented information exchange \cite{Strinati2021SemanticGoal}. Deep learning (DL) and artificial intelligence (AI) have accelerated this direction by enabling trainable semantic encoders and decoders that extract task-relevant features from text, image, speech, and multimodal data \cite{Xie2021DeepSC}. Several surveys classify SC into semantic-oriented, goal-oriented, and semantic-aware communication models \cite{Luo2022SemanticOverview}. Task-oriented communication further emphasizes that the quality of communication should be evaluated by the usefulness of the information received for a downstream decision or action \cite{Gunduz2023BeyondBits,Uysal2022DataSignificance}. In the broader future-Internet context, SC has been identified as a key enabler of intelligent, context-aware, and application-based networking \cite{Yang2023SemanticFutureInternet}. A tutorial-style survey on semantics-empowered communication also shows that SC requires careful treatment of semantic representation, semantic metrics, datasets, reasoning mechanisms, and implementation constraints \cite{Lu2024SemanticsEmpowered}. Recent survey work further highlights that SC design must consider not only semantic extraction and transmission, but also semantic noise, model training, security, privacy, and resource management \cite{Liu2024SemanticSurvey}.

In an SC-enabled IoUT system, the transmitter may avoid forwarding the complete raw observation $x$ and instead extract a compact semantic representation,
\begin{equation}
    s = f_{\theta}(x,c,g),
    \label{eq:intro_semantic_representation}
\end{equation}
where $c$ denotes environmental or network context, $g$ denotes the application goal, and $f_{\theta}(\cdot)$ is a semantic encoder. This encoder may be implemented using rules, ontologies, knowledge graphs, lightweight machine learning models, deep neural networks, vision-language models, or hybrid mechanisms. The representation $s$ may take the form of event labels, object attributes, region-of-interest descriptors, semantic segmentation maps, compressed feature embeddings, knowledge-graph triples, anomaly alerts, or task-specific decisions. Recent theoretical work on SC has formalized semantic encoding, semantic decoding, and semantic distortion-cost trade-offs, which are highly relevant when communication efficiency must be balanced against information loss \cite{Shao2024TheorySemantic}. For IoUT, this formulation is important because the most useful message is often not the raw signal itself, but the task-relevant interpretation of the signal.

\begin{figure*}[t]
\centering
\includegraphics[width=0.95\textwidth]{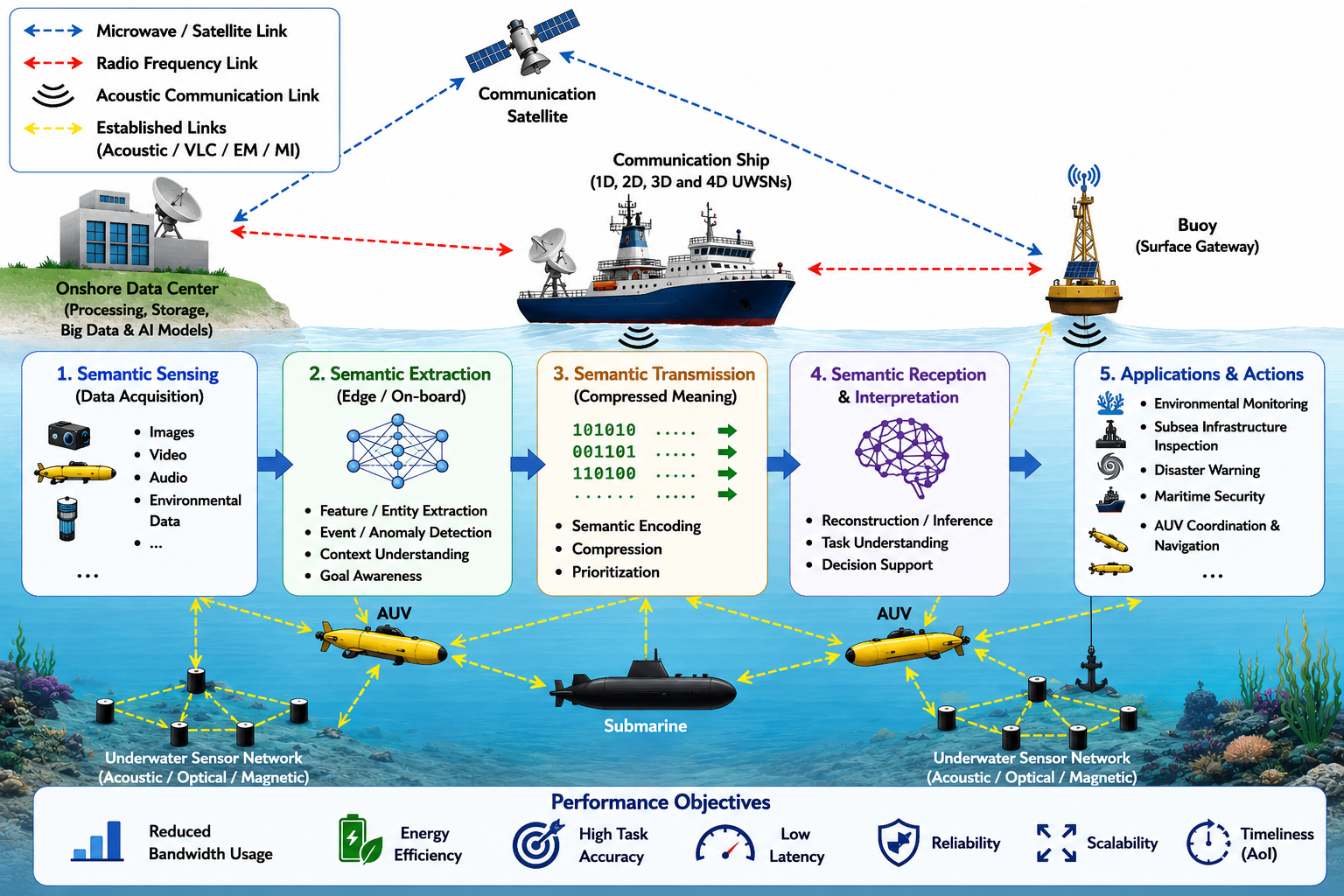}
\caption{Generic flow of SC in IoUT.}
\label{fig:sc_iout_flow}
\end{figure*}

The relevance of SC is particularly clear in underwater applications. For example, an environmental monitoring node may not need to transmit every temperature, salinity, dissolved oxygen, turbidity, or pollutant sample if the application only requires detection of abnormal ecological changes. AUV-based inspection may similarly prioritize semantic descriptors such as defect type, crack location, corrosion severity, leakage probability, confidence score, and timestamp rather than continuously transmitting raw images or sonar frames. Early work on SC for underwater communication discusses the potential advantages of semantic transmission in low signal-to-noise ratio (SNR) and narrowband acoustic channels \cite{Zhang2023UnderwaterSemComSurvey}. The Semantic-Oriented Architecture with Generative AI (SAGE) framework investigated an image-to-text-to-image approach in which semantic descriptions are transmitted instead of full underwater images \cite{Loureiro2024SAGE}. 

Underwater wireless optical SC has also been experimentally demonstrated using deep-learning-based semantic feature extraction for image transmission \cite{Xu2024UWOSC}. More recently, underwater acoustic SC has been studied for image transmission through compact textual semantics and entity-level representations \cite{Zhang2025UASC}. Environment semantics have further been introduced into underwater optical SC to support channel-state-aware adaptation \cite{Xu2025ESA_UWOSC}. Semantic-oriented federated learning (FL) has also been proposed for hybrid ground--aqua computing systems, indicating that distributed learning may reduce raw-data uploading while supporting privacy-aware semantic model training \cite{Picano2024SemanticFL}. A recent resilience analysis of underwater semantic wireless communication shows that semantic reconstruction can remain meaningful under some text-level channel errors, while also highlighting the need for realistic impairment models and evaluation metrics \cite{Loureiro2025Resilience}. Large language model (LLM)-assisted underwater image transmission has also been explored, but such approaches should be interpreted as emerging and computationally demanding rather than immediately deployable on all underwater edge devices \cite{Chen2024LLMUnderwater}.

Fig.~\ref{fig:sc_iout_flow} illustrates a generic flow of SC in an IoUT environment. The figure is an author-created conceptual illustration that builds on common IoUT architectural components, including underwater sensing nodes, AUVs, surface gateways, satellite or terrestrial backhaul, and shore/cloud processing units. Similar architectural components are discussed in recent UWSN modernization studies \cite{Muhammad2025UWSNModernization}. In this flow, raw underwater observations are first processed at underwater nodes, AUVs, or edge gateways to extract task-dependent semantic information. The semantic representation is then transmitted through a suitable underwater link, such as acoustic, optical, RF, MI, or a hybrid link. At the receiver side, semantic interpretation, task inference, approximate visualization, or decision support is performed using shared context, learned models, or domain knowledge.

Despite its potential, SC should not be treated as a universal solution for underwater communication. Semantic extraction can reduce the number of transmitted bits, but it introduces additional computation, memory usage, model-update overhead, and inference latency at underwater or edge devices. This creates a communication--computation trade-off that is especially important in battery-limited IoUT systems. Moreover, semantic compression may discard information that later becomes important for a different task, while generative reconstruction may produce visually plausible but incorrect details. Security and privacy surveys on SC show that semantic representations may leak sensitive meaning even when raw data are not directly transmitted \cite{Guo2024SemComNet}. Resource management, security, and privacy must therefore be considered jointly when SC is deployed in practical networks \cite{Won2025SemComSecurityPrivacy}. Secure SC surveys further emphasize that adversarial perturbations, semantic leakage, model inversion, and semantic manipulation attacks create risks beyond conventional bit-level communication security \cite{Meng2025SecureSemCom}. Hence, a rigorous SC-enabled IoUT design must jointly consider bandwidth savings, edge-computation cost, model size, semantic fidelity, task accuracy, latency, energy consumption, reliability, trustworthiness, and security.

\subsection{Related Surveys}
\label{subsec:related_surveys}

Several surveys have investigated IoUT, UWSNs, underwater communication technologies, big marine data analytics, routing, security, and general SC. Foundational IoUT surveys introduced underwater sensing architectures, application domains, channel models, and deployment challenges \cite{Domingo2012IoUT,Kao2017IoUT}, while more recent IoUT reviews have emphasized smart-ocean integration, marine data analytics, and heterogeneous underwater communication infrastructures \cite{Qiu2020SmartOcean,Jahanbakht2021IoUTBigData}. Similarly, UWSN surveys have provided useful taxonomies of underwater sensing requirements, routing constraints, energy limitations, and network-layer challenges \cite{Fattah2020UWSNSurvey,Awan2019UWSNReview}. However, these studies mainly focus on conventional underwater networking and do not systematically examine semantic representations, task-oriented communication, or semantic reliability.

In parallel, several surveys have examined SC for terrestrial, vehicular, aerial, edge-intelligent, and sixth-generation (6G) communication systems. General SC surveys discuss semantic information theory, semantic coding, task-oriented communication, and semantic-aware networking \cite{Luo2022SemanticOverview,Yang2023SemanticFutureInternet}. Other studies focus on context-aware communication, semantics-empowered networks, and the role of AI in extracting task-relevant information from multimodal data \cite{Gunduz2023BeyondBits,Lu2024SemanticsEmpowered}. More recent works also analyze resource management, privacy, security, and trustworthy SC networks \cite{Guo2024SemComNet,Won2025SemComSecurityPrivacy,Meng2025SecureSemCom}. Nevertheless, most of these studies are not centered on underwater communication, where severe bandwidth limitations, long propagation delay, acoustic/optical/RF heterogeneity, edge-computation constraints, and mission-critical uncertainty substantially change the design requirements.

Therefore, the intersection of SC and IoUT still lacks a focused and feasibility-aware synthesis. In particular, the joint interaction among semantic representation, underwater channel constraints, edge-computation cost, generative reconstruction risk, task-oriented metrics, and practical IoUT deployment remains insufficiently analyzed. Table~\ref{tab:intro_related_surveys} positions this paper relative to representative surveys and closely related studies.

\begin{table*}[ht!]
\centering
\footnotesize
\caption{Positioning of this survey relative to representative surveys and related studies.}
\label{tab:intro_related_surveys}
\renewcommand{\arraystretch}{1.22}
\setlength{\tabcolsep}{3.8pt}
\begin{tabular}{p{3.2cm} p{4.2cm} p{4.4cm} p{4.4cm}}
\toprule
\textbf{Reference} 
& \textbf{Primary Focus} 
& \textbf{Gap with Respect to SC-enabled IoUT} 
& \textbf{How This Paper Differs} \\
\midrule

Domingo \cite{Domingo2012IoUT}; Kao \textit{et al.} \cite{Kao2017IoUT}; Mohsan \textit{et al.} \cite{Mohsan2022IoUTSurvey,Mohsan2023IoUTReview} 
& Foundational IoUT concepts, applications, challenges, architectures, and channel models 
& Provide broad IoUT motivation and application coverage, but do not study SC, semantic representation, or task-oriented semantic delivery 
& Focuses on SC as a meaning-driven communication paradigm for IoUT systems and analyzes its feasibility under underwater constraints \\
\midrule

Qiu \textit{et al.} \cite{Qiu2020SmartOcean} 
& Smart-ocean architecture and open issues for underwater IoT 
& Emphasizes sensing, networking, data fusion, and system architecture, but does not focus on semantic extraction or semantic channel modeling 
& Explains how semantic sensing, semantic transmission, and task-oriented reasoning can be inserted into IoUT layers \\
\midrule

Jahanbakht \textit{et al.} \cite{Jahanbakht2021IoUTBigData} 
& IoUT and big marine data analytics 
& Discusses marine data growth, AI, and analytics, but not SC-specific encoding, reconstruction reliability, or semantic metrics 
& Links marine data-volume challenges with semantic compression, task relevance, edge intelligence, and compute--communication trade-offs \\
\midrule

Bello and Zeadally \cite{Bello2022IoUTCommunication} 
& IoUT communication architecture, technologies, challenges, and opportunities 
& Covers underwater communication media and network architectures, but not semantic reliability, semantic noise, or generative reconstruction risk 
& Connects underwater communication modalities with SC design choices, semantic utility, and deployment feasibility \\
\midrule

Fattah \textit{et al.} \cite{Fattah2020UWSNSurvey}; Awan \textit{et al.} \cite{Awan2019UWSNReview} 
& UWSN requirements, taxonomy, recent advances, and open challenges 
& Focus on underwater communication and networking issues without task-level semantic representation or semantic-efficiency analysis 
& Relates UWSN constraints to semantic representation, semantic compression, semantic routing, and task-oriented evaluation \\
\midrule

Luo \textit{et al.} \cite{Luo2021RoutingUWSN}; Khan \textit{et al.} \cite{Khan2018RoutingUWSN} 
& Routing protocols for UWSNs 
& Focus on forwarding, energy, topology, and reliability, but not meaning-aware routing or semantic prioritization 
& Identifies semantic-aware routing and task-level information forwarding as key directions for SC-enabled IoUT \\
\midrule

Saeed \textit{et al.} \cite{Saeed2019UOWC}; Zeng \textit{et al.} \cite{Zeng2017UOWCSurvey} 
& Underwater optical wireless communication, networking, and localization 
& Focus mainly on optical communication and localization rather than semantic extraction, semantic utility, or task-level information delivery 
& Considers optical links as part of hybrid SC-enabled IoUT architectures involving acoustic, optical, RF, MI, and edge processing \\
\midrule

Theocharidis and Kavallieratou \cite{Theocharidis2025UnderwaterComm}; Li \textit{et al.} \cite{Li2025UnderwaterAcoustic} 
& Underwater communication technologies and acoustic communication foundations 
& Review physical/link-level underwater communication, but not SC-specific semantic modeling, semantic outage, or task-oriented evaluation 
& Uses underwater communication constraints as the basis for analyzing when and how SC can provide application-level benefits \\
\midrule

Luo \textit{et al.} \cite{Luo2022SemanticOverview}; Yang \textit{et al.} \cite{Yang2023SemanticFutureInternet} 
& General SC foundations, technologies, applications, and challenges 
& Focus primarily on future Internet and generic wireless scenarios rather than underwater systems 
& Maps SC concepts to underwater sensing, constrained links, edge intelligence, and marine use cases \\
\midrule

G{\"u}nd{\"u}z \textit{et al.} \cite{Gunduz2023BeyondBits}; Lu \textit{et al.} \cite{Lu2024SemanticsEmpowered} 
& Task-oriented, context-aware, and semantics-empowered communication 
& Provide broad theoretical and architectural insights, but do not specialize in IoUT channel, energy, and deployment constraints 
& Extends task-oriented SC concepts to underwater communication, semantic edge processing, and IoUT applications \\
\midrule

Liu \textit{et al.} \cite{Liu2024SemanticSurvey}; Getu \textit{et al.} \cite{Getu2023SemanticGoalSurvey} 
& SC technologies, solutions, applications, and semantic/goal-oriented research landscape 
& Offer broad SC coverage, but limited analysis of underwater communication channels, marine sensing, and IoUT deployment realities 
& Provides a focused synthesis of semantic representations, metrics, applications, and feasibility barriers in IoUT \\
\midrule

Zhang \textit{et al.} \cite{Zhang2023UnderwaterSemComSurvey}; Loureiro \textit{et al.} \cite{Loureiro2024SAGE} 
& Early underwater SC studies and generative AI-based semantic image transmission 
& Individual studies with specific assumptions, datasets, and reconstruction objectives 
& Provides a broader comparison across semantic representations, models, links, metrics, and deployment conditions \\
\midrule

Xu \textit{et al.} \cite{Xu2024UWOSC}; Zhang \textit{et al.} \cite{Zhang2025UASC} 
& Underwater optical/acoustic SC for image transmission 
& Focus on specific transmission modalities and image-oriented scenarios 
& Places these early systems within a wider IoUT taxonomy and feasibility-aware research agenda \\
\midrule

Guo \textit{et al.} \cite{Guo2024SemComNet}; Won \textit{et al.} \cite{Won2025SemComSecurityPrivacy}; Meng \textit{et al.} \cite{Meng2025SecureSemCom} 
& SC networks, resource management, security, and privacy 
& Do not address underwater-specific threats, edge constraints, or marine operational risks in detail 
& Highlights semantic security and privacy as core requirements for trustworthy SC-enabled IoUT \\
\midrule

\textbf{This paper} 
& SC for IoUT 
& --- 
& Provides a focused survey of opportunities, challenges, semantic representations, architectures, metrics, feasibility trade-offs, applications, and future research directions for SC-enabled IoUT \\
\bottomrule
\end{tabular}
\end{table*}

\subsection{Contributions}
\label{subsec:contributions}

The main contributions of this paper are summarized as follows:
\begin{itemize}
    \item Firstly, we present a focused survey of SC for IoUT networks, emphasizing how underwater channel constraints, limited energy resources, intermittent connectivity, heterogeneous communication media, and marine application requirements affect the design of meaning-driven underwater communication systems.

    \item Secondly, we develop a structured taxonomy of SC-enabled IoUT systems based on semantic representation type, communication modality, network layer, learning paradigm, and application objective. This taxonomy distinguishes symbolic semantics, feature-level semantics, knowledge-driven semantics, generative semantics, and task-oriented semantic decisions.

    \item Then, we analyze learning-driven SC techniques for IoUT, including ontologies, knowledge graphs, lightweight DL, transformers, vision-language models, generative models, and federated learning, with particular attention to feasibility under underwater edge constraints.

    \item Furthermore, we examine the communication--computation trade-off in SC-enabled IoUT systems by discussing the joint impact of sensing cost, semantic inference cost, transmission cost, model-update overhead, reconstruction complexity, latency, energy consumption, and reliability.

    \item Finally, we identify representative applications and future research directions for practical SC-enabled IoUT systems, including environmental monitoring, subsea infrastructure inspection, marine ecology, disaster response, AUV coordination, standardized semantic models, realistic testbeds, trustworthy generative reconstruction, semantic security, and cross-domain interoperability.
\end{itemize}

\subsection{Organization of the Paper}
\label{subsec:organization}

\begin{figure*}[ht!]
\centering
\includegraphics[width=18.25cm, height=11cm]{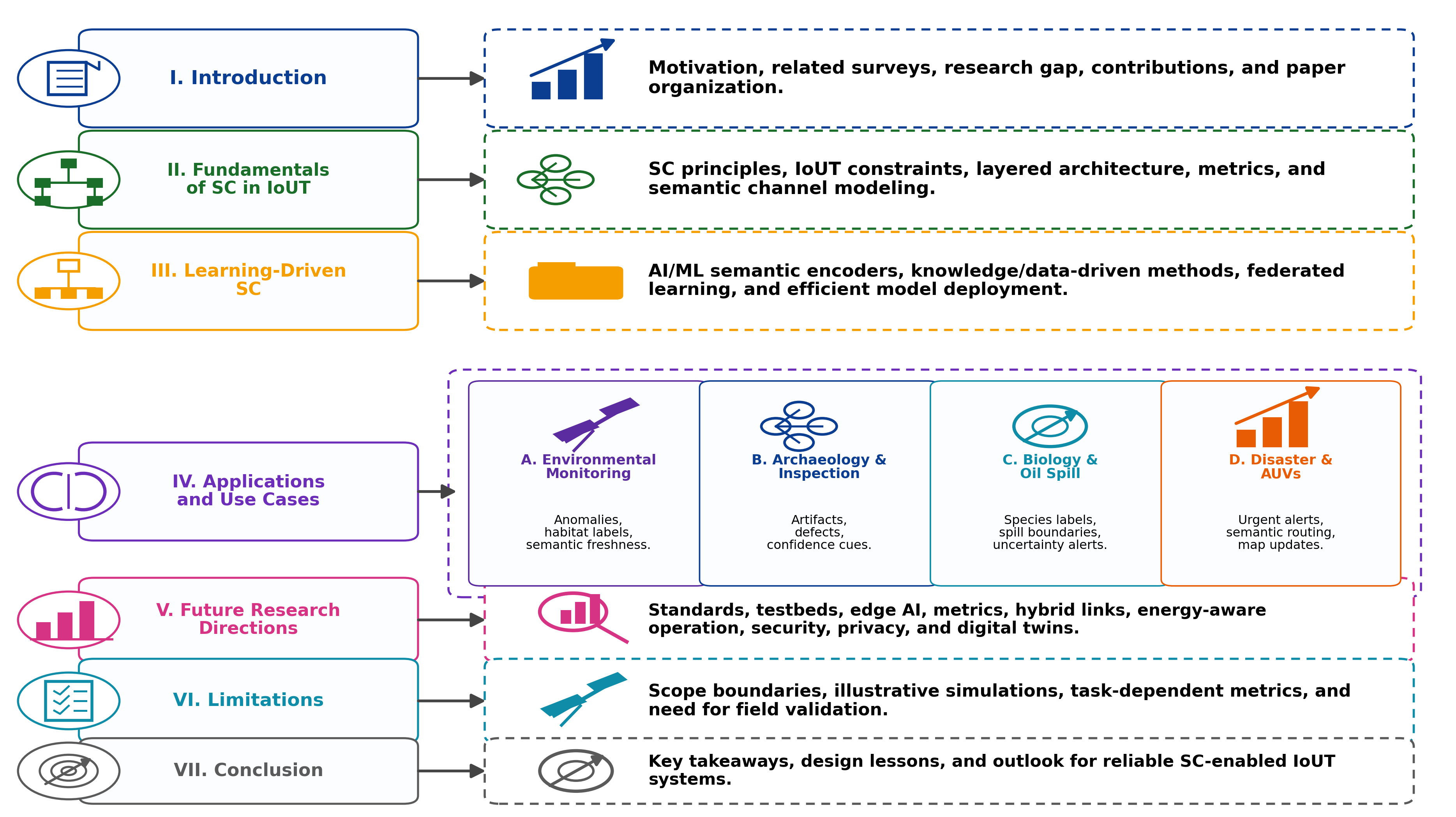}
\caption{Systematic organization of this paper.}
\label{organization}
\end{figure*}

The remainder of this paper is organized as follows. Section~\ref{sec02} introduces the fundamentals of SC in IoUT, including underwater communication constraints, semantic representations, layered system architecture, semantic channel modeling, and evaluation metrics. Section~\ref{sec03} reviews learning-driven SC approaches, including knowledge-driven, data-driven, generative, and federated methods. Section~\ref{sec04} discusses representative IoUT applications and use cases, including environmental monitoring, marine ecology, underwater archaeology, subsea infrastructure inspection, disaster response, and AUV coordination. Section~\ref{sec05} presents future research directions for practical SC-enabled IoUT deployments, covering standardization, reproducible validation, lightweight edge intelligence, semantic metrics, hybrid underwater links, energy-aware operation, security, privacy, and cross-domain orchestration. Section~\ref{sec06} discusses the scope and limitations of this survey. Finally, Section~\ref{sec07} concludes the paper. For clarity, Fig.~\ref{organization} summarizes the organization of this paper, while Table~\ref{tab:abbreviations} lists the main acronyms used throughout the manuscript.

\begin{table}[h!]
\centering
\footnotesize
\caption{List of acronyms used in this paper.}
\label{tab:abbreviations}
\renewcommand{\arraystretch}{1.12}
\begin{tabular}{p{2.1cm}p{5.7cm}}
\toprule
\textbf{Acronym} & \textbf{Description} \\
\midrule
AI & Artificial Intelligence \\
AoI & Age of Information \\
AUV & Autonomous Underwater Vehicle \\
BER & Bit Error Rate \\
BLEU & Bilingual Evaluation Understudy \\
BLIP & Bootstrapping Language-Image Pre-training \\
CLIP & Contrastive Language--Image Pre-training \\
CNN & Convolutional Neural Network \\
DTN & Delay-Tolerant Networking \\
FL & Federated Learning \\
IoT & Internet of Things \\
IoUT & Internet of Underwater Things \\
LLM & Large Language Model \\
MAC & Medium Access Control \\
MI & Magnetic Induction \\
ML & Machine Learning \\
PSNR & Peak Signal-to-Noise Ratio \\
RF & Radio Frequency \\
RIS & Reconfigurable Intelligent Surface \\
ROV & Remotely Operated Vehicle \\
SAGE & Semantic-Oriented Architecture with Generative AI \\
SAR & Synthetic Aperture Radar \\
SC & Semantic Communication \\
SNR & Signal-to-Noise Ratio \\
SSIM & Structural Similarity Index Measure \\
UWSN & Underwater Wireless Sensor Network \\
VLM & Vision-Language Model \\
\bottomrule
\end{tabular}
\end{table}

\section{Fundamentals of SC in IoUT}
\label{sec02}
SC in IoUT aims to shift underwater networking from raw-data delivery to task-relevant, context-aware, and meaning-preserving information exchange. This shift is motivated by the fact that underwater networks differ substantially from terrestrial radio networks in deployment cost, propagation behavior, node mobility, and energy availability, as highlighted in practical underwater networking studies \cite{Partan2006PracticalUnderwater,Sozer2000UnderwaterAcousticNetworks}. In addition, UWSNs must address constraints across the physical, MAC, and routing layers, including low bandwidth, long delay, and unreliable connectivity \cite{Climent2014UnderwaterAcousticWSN}. Therefore, the fundamentals of SC-enabled IoUT require a joint understanding of semantic representation, underwater propagation, system architecture, semantic metrics, and semantic channel modeling. Rather than treating SC as a simple compression mechanism, this section presents it as a cross-layer design principle to improve the usefulness of information under constrained underwater resources. Table~\ref{tab:fundamentals_sc_iout} summarizes the main concepts, models, metrics, and design implications discussed in this section.

\begin{table*}[ht!]
\centering
\footnotesize
\caption{Fundamentals of SC in IoUT: Concepts, Models, Metrics, and Design Implications.}
\label{tab:fundamentals_sc_iout}
\renewcommand{\arraystretch}{1.20}
\setlength{\tabcolsep}{3.5pt}
\begin{tabular}{p{2.8cm} p{4.1cm} p{4.0cm} p{3.2cm} p{2.6cm}}
\toprule
\textbf{Fundamental Aspect} 
& \textbf{Role in SC-enabled IoUT} 
& \textbf{Representative Model / Expression} 
& \textbf{Key Metrics} 
& \textbf{Representative References} \\
\midrule

Semantic representation 
& Converts raw underwater observations such as images, sonar, acoustic signals, and water-quality data into task-relevant meanings, labels, embeddings, events, or descriptors 
& $s=f_{\theta}(x,c,g)$, where $x$ is raw data, $c$ is context, and $g$ is the task goal
& Semantic similarity, task relevance, compression ratio, confidence score 
& \cite{Bao2011TheorySemantic,Tishby1999InformationBottleneck,Farsad2018TextJSCC} \\
\midrule

Semantic decoding and inference 
& Interprets received semantic information to produce a task output, alert, decision, semantic map, or approximate visualization 
& $\hat{y}=h_{\phi}(\hat{s},c,g,\mathcal{K})$, where $\mathcal{K}$ denotes shared knowledge or ontology 
& Task accuracy, false alarm rate, missed detection rate, uncertainty 
& \cite{Weng2021SpeechSC,ZhangSTBL2023TaskUnawareSC,Shi2021SemanticAwareNetworking} \\
\midrule

Underwater channel constraints 
& Determines how semantic packets are affected by acoustic, optical, RF, or MI links under bandwidth, delay, fading, and Doppler limitations 
& $\tau(d)=d/v_s$, with $v_s \approx 1500~\mathrm{m/s}$; 
$R_m=B_m\log_2(1+\gamma_m)$
& SNR, BER, packet delivery ratio (PDR), latency, channel availability 
& \cite{Partan2006PracticalUnderwater,Climent2014UnderwaterAcousticWSN,Sendra2015AcousticModems} \\
\midrule

Semantic compression 
& Reduces payload size by transmitting task-relevant descriptors instead of raw observations, while controlling semantic information loss
& $\rho_s=b_s/b_x$, where $b_s$ and $b_x$ are semantic and raw payload sizes
& Compression ratio, semantic distortion, payload size, reconstruction utility 
& \cite{Bourtsoulatze2019DeepJSCC,Loureiro2024SAGE,Zhang2025UASC} \\
\midrule

Compute--communication trade-off 
& Determines whether semantic extraction is worthwhile by comparing inference cost with transmission savings
& $E_{\mathrm{SC}}=E_{\mathrm{enc}}+b_s e_{\mathrm{tx}}+E_{\mathrm{dec}}$; SC is beneficial if $E_{\mathrm{SC}}<E_{\mathrm{raw}}$
& Inference energy, transmission energy, model size, node lifetime 
& \cite{Satyanarayanan2017EdgeComputing,Sherlock2022MiniModems,Kairouz2021FL} \\
\midrule

Semantic packet design 
& Adds task, context, confidence, location, time, priority, and provenance metadata to support meaning-aware forwarding
& $p_s=\{s,g,\ell,t,q,u,\kappa\}$, where $q$ is confidence and $u$ is urgency
& Priority level, confidence, provenance, semantic delivery probability 
& \cite{Pompili2006RoutingUWSN,Xie2006VBF,Yan2008DBR} \\
\midrule

Semantic-aware routing and link selection 
& Selects paths and communication modalities according to semantic value, urgency, energy, latency, and channel quality 
& $m^{\star}=\arg\max_{m\in\mathcal{M}}\mathcal{U}_m$, where $\mathcal{U}_m$ is semantic link utility
& Semantic utility, delay, energy cost, link reliability, semantic outage 
& \cite{Pompili2006RoutingUWSN,Zhou2022AdaptiveHARQ,Uysal2022DataSignificance} \\
\midrule

Semantic fusion and reasoning 
& Combines semantic information from multiple underwater nodes, AUVs, buoys, and heterogeneous sensors to infer a higher-level system state
& $z=\mathcal{F}(s_1,s_2,\ldots,s_N,c,\mathcal{K})$
& Multimodal consistency, inference accuracy, knowledge consistency 
& \cite{Hogan2021KnowledgeGraphs,McMahan2017FedAvg,Xiao2023ImplicitSemantic} \\
\midrule

Semantic efficiency 
& Evaluates how much task utility is obtained per unit bandwidth, time, or energy rather than only measuring transmitted bits
& $\eta_B=\mathbb{E}[U_g(\hat{y},y)]/(BT)$; 
$\eta_E=\mathbb{E}[U_g(\hat{y},y)]/E_{\mathrm{tot}}$ 
& Semantic bits/Joule, utility/Hz, task success rate, useful throughput 
& \cite{Papineni2002BLEU,Wang2004SSIM,Radford2021CLIP} \\
\midrule

Freshness and timeliness 
& Captures whether received semantic information is still useful for time-sensitive IoUT tasks such as AUV control or disaster warning
& $\Delta(t)=t-u(t)$; 
$\bar{\Delta}=\lim_{T\rightarrow\infty}\frac{1}{T}\int_0^T \Delta(t)dt$
& AoI, semantic AoI, deadline miss ratio, event response time 
& \cite{Yates2021AoISurvey,Felemban2015USNApplications} \\
\midrule

Semantic reliability and outage 
& Measures whether the received meaning remains acceptable for the task despite physical noise, model mismatch, or context mismatch
& $P_{\mathrm{s\text{-}out}}(\epsilon_g)=\Pr[d_g(s,\hat{s})>\epsilon_g]$; 
$R_s=1-P_{\mathrm{s\text{-}out}}$
& Semantic outage, task failure probability, robustness, confidence 
& \cite{Lu2021Rethinking,Zhou2022AdaptiveHARQ,Bao2011TheorySemantic} \\
\midrule

Trust, security, and provenance 
& Protects semantic information from manipulation, leakage, false interpretation, and untrusted model updates in sensitive marine applications
& $\kappa$ in $p_s$ may encode authentication, provenance, access rights, or trust metadata
& Trust score, attack detection rate, privacy leakage, semantic integrity 
& \cite{Bonawitz2017SecureAggregation,Kairouz2021FL} \\
\bottomrule
\end{tabular}
\end{table*}

\subsection{Key principles of SC}
Traditional communication systems are primarily optimized for the accurate transfer of symbols or bits through a noisy channel. SC changes this objective by focusing on preserving meaning, relevance, and task utility. Earlier studies on semantic information considered how semantic content can be compressed and transmitted reliably beyond purely syntactic information \cite{Bao2011TheorySemantic}. In modern AI-enabled communication systems, this idea is often implemented through learned representations, in which the transmitter extracts the information most useful for the receiver-side task. The information bottleneck principle provides one useful perspective: a representation should retain information relevant to a target variable while discarding irrelevant details from the original observation \cite{Tishby1999InformationBottleneck}.

Let $x \in \mathcal{X}$ denote a raw underwater observation, such as an image, sonar return, acoustic sample, or water-quality measurement. In an SC-enabled IoUT system, the transmitter does not necessarily send $x$ directly. Instead, it generates a semantic representation
\begin{equation}
    s = f_{\theta}(x,c,g),
    \label{eq:semantic_encoder}
\end{equation}
where $c$ denotes environmental or network context, $g$ denotes the application goal, and $f_{\theta}(\cdot)$ is a semantic encoder. Depending on the application, $s$ may be a symbolic label, a feature vector, a region-of-interest descriptor, a semantic segmentation map, a knowledge-graph triple, an anomaly score, or a task-level decision. Deep joint source-channel coding (Deep JSCC) studies have shown that neural encoders can learn representations that are robust to channel impairments without strictly separating source and channel coding \cite{Farsad2018TextJSCC}. Similar ideas have been applied to image transmission, where learned encoders map visual content directly into channel symbols \cite{Bourtsoulatze2019DeepJSCC}.

At the receiver, the objective is not always to reconstruct the exact original source. Instead, the receiver may infer a task result, update a semantic map, trigger an alarm, or generate an approximate visualization:
\begin{equation}
    \hat{y} = h_{\phi}(\hat{s},c,g,\mathcal{K}),
    \label{eq:semantic_decoder}
\end{equation}
where $\hat{s}$ is the received semantic representation, $h_{\phi}(\cdot)$ is the semantic decoder or inference model, and $\mathcal{K}$ denotes shared knowledge such as an ontology, codebook, semantic memory, or domain-specific knowledge graph. For speech transmission, SC systems have demonstrated that learned encoders can prioritize essential content rather than only minimizing waveform-level distortion \cite{Weng2021SpeechSC}. For dynamic data and task-unaware transmitters, receiver-guided semantic extraction has also been investigated to reduce mismatch between transmitter observations and receiver tasks \cite{ZhangSTBL2023TaskUnawareSC}.

A generic SC optimization problem for IoUT can be expressed as
\begin{equation}
\begin{aligned}
    \min_{\theta,\phi,\pi} \quad
    & \mathbb{E}\!\left[
    \lambda_s d_s(s,\hat{s})
    + \lambda_g d_g(y,\hat{y})
    + \lambda_E E_{\mathrm{tot}}
    + \lambda_T T_{\mathrm{tot}}
    \right] \\
    \mathrm{s.t.} \quad
    & B \leq B_{\max}, \quad
      E_{\mathrm{tot}} \leq E_{\max}, \quad
      T_{\mathrm{tot}} \leq T_{\max},
\end{aligned}
\label{eq:sc_objective}
\end{equation}
where $d_s(\cdot)$ is semantic distortion, $d_g(\cdot)$ is task-level loss, $E_{\mathrm{tot}}$ is total energy consumption, $T_{\mathrm{tot}}$ is end-to-end latency, and $B$ is occupied bandwidth. The policy $\pi$ may include decisions on sampling, scheduling, routing, link selection, and retransmission. Task-oriented communication research emphasizes that such decisions should be optimized according to task-level utility rather than only throughput or BER \cite{Wen2023TaskOriented}. In underwater networks, this means that an urgent semantic alert may deserve priority even if its payload is small, while a large raw-data stream may be delayed or suppressed if it contributes little to the mission objective.

Several principles follow from this formulation. First, SC is \emph{task-aware}: the representation should depend on the intended application. Second, SC is \emph{context-aware}: turbidity, node mobility, residual energy, link quality, and mission urgency should influence what is transmitted. Third, SC is \emph{uncertainty-aware}: semantic inference may be wrong, incomplete, or overconfident. Fourth, SC is \emph{resource-aware}: semantic extraction is beneficial only if communication savings exceed the additional computation cost. These principles are central to IoUT because underwater nodes often operate for extended periods with limited opportunities for maintenance.

\subsection{Characteristics of IoUT Environments}
IoUT environments impose physical and operational constraints that directly affect SC design. Underwater acoustic modems are widely used for long-range communication, but practical modem studies show that their performance depends strongly on frequency, range, transmit power, environmental variability, and hardware design \cite{Sendra2015AcousticModems}. The propagation delay over a distance $d$ can be approximated as
\begin{equation}
    \tau(d) = \frac{d}{v_s},
    \label{eq:propagation_delay}
\end{equation}
where $v_s \approx 1500~\mathrm{m/s}$ is the speed of sound in water. This delay is much larger than radio propagation delay in air, which complicates feedback, retransmission, time synchronization, and closed-loop control.

The achievable physical-layer rate of a link using communication modality $m$ can be idealized as
\begin{equation}
    R_m = B_m \log_2(1+\gamma_m),
    \label{eq:link_rate}
\end{equation}
where $B_m$ is available bandwidth and $\gamma_m$ is received SNR. In practice, both $B_m$ and $\gamma_m$ are highly variable in underwater environments. The relationship between capacity, distance, and acoustic frequency further explains why long-range underwater links are bandwidth-constrained \cite{Stojanovic2007CapacityDistance}. Recent miniature acoustic modem prototypes indicate that low-cost, low-power designs are feasible, but they still entail significant trade-offs among range, data rate, power consumption, and robustness \cite{Sherlock2022MiniModems}. Hence, SC must be evaluated not only by how much it compresses data, but also by whether the semantic payload remains useful under realistic link and hardware constraints.

Let $b_x$ denote the number of bits required for transmitting a raw observation and $b_s$ denote the number of bits required for the corresponding semantic representation. The semantic compression ratio is
\begin{equation}
    \rho_s = \frac{b_s}{b_x}, \quad 0 < \rho_s \leq 1.
    \label{eq:compression_ratio}
\end{equation}
A smaller $\rho_s$ indicates greater compression, but aggressive compression may remove information that later becomes important for another task. This issue is critical in underwater monitoring, where sensor networks may support pollution monitoring, seismic observation, equipment inspection, and habitat analysis simultaneously \cite{Felemban2015USNApplications}. Therefore, SC should not be designed as a one-size-fits-all compression method; it should select semantic granularity based on the task, channel conditions, energy budget, and acceptable information loss. Table~\ref{tab:iout-characteristics} summarizes major IoUT characteristics and their implications for SC design.

\begin{table*}[ht!]
\centering
\footnotesize
\caption{Characteristics of IoUT environments and their implications for semantic communication design.}
\label{tab:iout-characteristics}
\renewcommand{\arraystretch}{1.20}
\setlength{\tabcolsep}{3.5pt}
\begin{tabular}{p{2.5cm} p{4.5cm} p{5.0cm} p{3.25cm} p{1.5cm}}
\toprule
\textbf{IoUT Characteristic} 
& \textbf{Impact on Underwater Communication} 
& \textbf{Implication for SC Design} 
& \textbf{Representative Metrics} 
& \textbf{Refs.} \\
\midrule

Limited bandwidth 
& Acoustic links provide constrained data rates, making continuous transmission of raw images, sonar maps, video, and dense sensor streams inefficient
& Extract and transmit compact semantic units such as events, object labels, region-of-interest descriptors, compressed embeddings, or anomaly summaries
& Payload size, semantic compression ratio, useful throughput, task accuracy 
& \cite{Partan2006PracticalUnderwater,Climent2014UnderwaterAcousticWSN} \\
\midrule

Long propagation delay 
& The low speed of sound in water causes large delays, limiting feedback, retransmission, synchronization, and real-time control
& Use delay-tolerant semantic packets, event-triggered reporting, priority-aware forwarding, and semantic retransmission only when task utility is insufficient 
& End-to-end latency, AoI, semantic AoI, deadline miss ratio 
& \cite{Sendra2015AcousticModems,Yates2021AoISurvey,Fall2003DTN} \\
\midrule

Multipath and Doppler effects 
& Reflections from the surface/seabed and node mobility introduce fading, packet corruption, and time-varying channels, especially for AUV-based links
& Apply robust semantic encoding, task-aware redundancy, confidence-aware decoding, and semantic error tolerance instead of relying only on bit-level recovery
& Semantic outage, BER/PDR, semantic reliability, task success rate 
& \cite{Climent2014UnderwaterAcousticWSN,Zhou2022AdaptiveHARQ} \\
\midrule

Energy constraints 
& Underwater nodes are commonly battery-powered, while battery replacement, recharging, or recovery is costly and sometimes infeasible
& Perform local semantic extraction only when saved transmission energy exceeds computation and decoding cost; use lightweight models and event-based sensing
& Energy per useful message, inference energy, node lifetime, semantic bits/Joule 
& \cite{Sherlock2022MiniModems,Kairouz2021FL} \\
\midrule

Dynamic topology 
& Ocean currents, AUV mobility, intermittent links, and sparse deployment cause frequent changes in connectivity
& Use semantic-aware routing, opportunistic forwarding, store-carry-forward mechanisms, and link selection based on semantic priority
& Delivery probability, routing cost, link availability, semantic delivery ratio 
& \cite{Pompili2006RoutingUWSN,Xie2006VBF,Yan2008DBR} \\
\midrule

Heterogeneous sensing 
& IoUT systems collect multimodal data, including images, sonar, hydrophone signals, chemical readings, pressure, temperature, salinity, and navigation data
& Use multimodal semantic fusion, ontology-supported metadata, task-specific representations, and confidence-aware interpretation
& Multimodal consistency, semantic similarity, fusion accuracy, uncertainty 
& \cite{Felemban2015USNApplications,Hogan2021KnowledgeGraphs} \\
\midrule

Intermittent connectivity 
& Links may be unavailable due to mobility, channel blockage, turbidity, modem range, or gateway unavailability
& Prioritize mission-critical semantics, cache low-priority summaries, and exploit delay-tolerant or AUV-assisted forwarding
& Contact probability, buffering delay, semantic packet loss, delivery latency 
& \cite{Pompili2006RoutingUWSN,Uysal2022DataSignificance,Fall2003DTN} \\
\midrule

Environmental variability 
& Turbidity, salinity, temperature, pressure, ambient noise, and biological activity affect sensing quality and communication reliability
& Adapt semantic granularity, sensing rate, link modality, and confidence thresholds according to environmental context
& Context-aware accuracy, channel availability, semantic distortion, confidence score 
& \cite{Sendra2015AcousticModems,Shi2021SemanticAwareNetworking} \\
\midrule

Safety-critical decisions 
& Incorrect semantic interpretation may affect AUV navigation, subsea inspection, disaster response, or maritime security
& Attach confidence, provenance, uncertainty, and task-risk metadata to semantic packets; avoid treating generative outputs as exact evidence
& False alarm rate, missed detection rate, trust score, semantic reliability 
& \cite{Wen2023TaskOriented,Xiao2023ImplicitSemantic} \\
\midrule

Security and privacy exposure 
& Semantic packets may reveal sensitive mission meaning, asset condition, vehicle location, or environmental-event information 
& Incorporate semantic authentication, provenance tracking, secure aggregation, privacy-preserving learning, and access control
& Privacy leakage, attack detection rate, semantic integrity, trustworthiness 
& \cite{Bonawitz2017SecureAggregation,Kairouz2021FL} \\
\bottomrule
\end{tabular}
\end{table*}

SC can mitigate several IoUT constraints by prioritizing data relevance over data volume. For example, a water-quality node may suppress normal measurements and transmit only abnormal chemical changes. Similarly, an imaging node may send object labels, bounding boxes, or compressed feature embeddings instead of full-resolution images. However, the value of SC depends on reliable semantic encoders, appropriate task metrics, and feasible local processing. Thus, SC should be interpreted as a resource-allocation and decision-making framework, not merely as a data-compression method.

\subsection{SC Architecture for IoUT}
\begin{figure*}[t]
\centering
\includegraphics[width=0.98\textwidth]{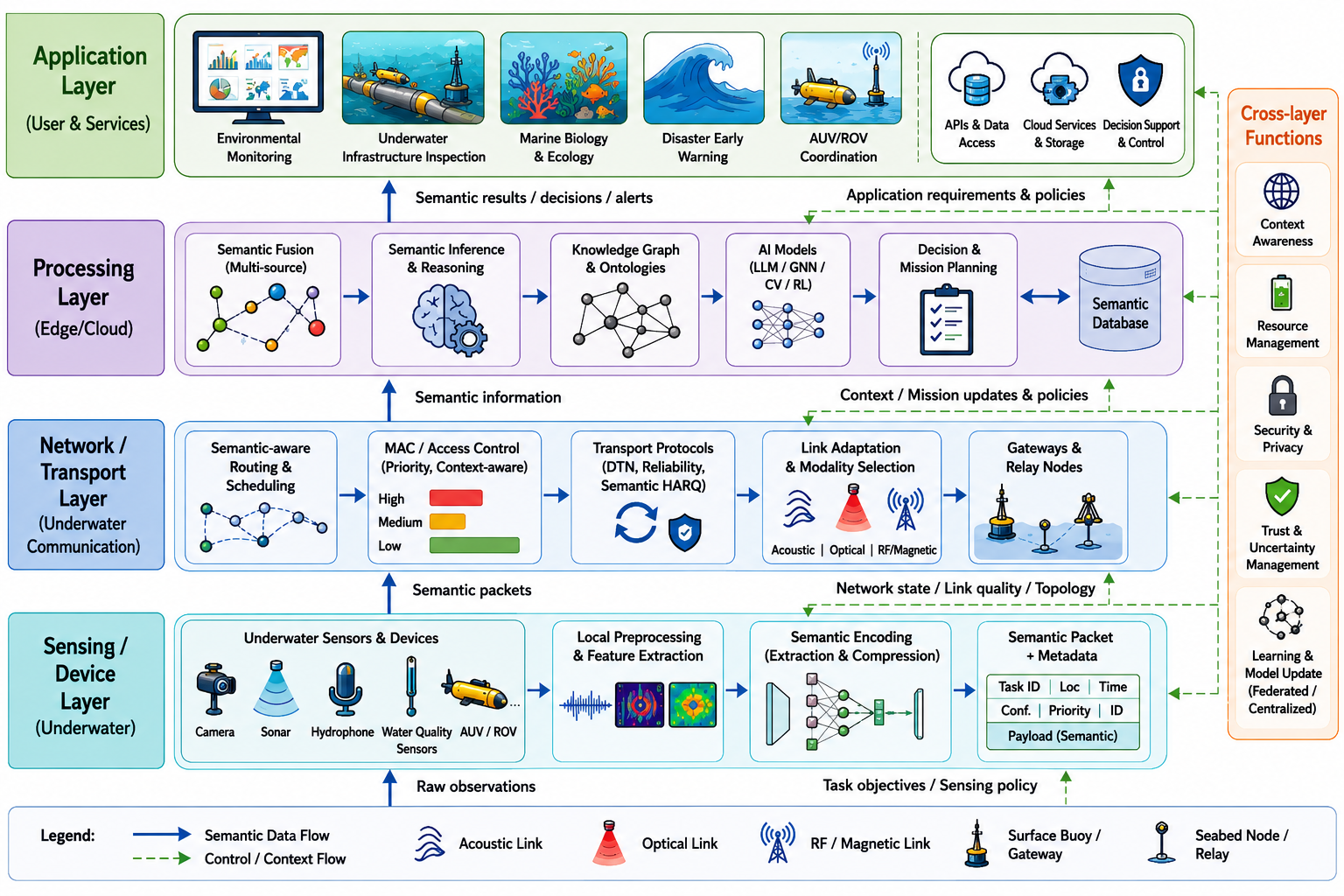}
\caption{Layered SC architecture for IoUT systems.}
\label{fig:layered_sc_iout_architecture}
\end{figure*}

A practical SC architecture for IoUT should support sensing, semantic extraction, semantic transmission, semantic interpretation, and application-level decision-making. The architecture can be organized into four interacting layers: the Sensing/Device Layer, the Network/Transport Layer, the Processing Layer, and the Application Layer. This layered view clarifies where semantic information is generated, transmitted, fused, and ultimately used by applications. Semantic-aware networking studies also emphasize that semantic extraction, knowledge modeling, and reasoning may be distributed across devices, edge nodes, and cloud infrastructures \cite{Shi2021SemanticAwareNetworking}. The overall layered architecture is illustrated in Fig.~\ref{fig:layered_sc_iout_architecture}.

\subsubsection{Sensing/Device Layer}
The Sensing/Device Layer includes underwater cameras, hydrophones, sonars, chemical sensors, pressure sensors, salinity sensors, seabed nodes, AUVs, and ROVs. These devices observe physical, chemical, biological, acoustic, and visual properties of the underwater environment. In conventional IoUT systems, raw observations are often forwarded to a gateway or surface station. In SC-enabled IoUT, the sensing layer performs local preprocessing, feature extraction, event detection, or semantic summarization before transmission.

Let $x_t$ denote the observation collected at time $t$. A semantic sensing device may decide whether to sleep, sense, process, or transmit according to
\begin{equation}
    a_t = \pi_s(x_t,c_t,e_t,g),
    \label{eq:sensing_policy}
\end{equation}
where $a_t$ is the sensing/transmission action, $c_t$ is environmental context, $e_t$ is residual energy, and $g$ is the mission goal. For example, a habitat-monitoring camera may transmit a semantic alert only when the probability of an ecological anomaly exceeds a threshold. The information bottleneck view suggests that the local encoder should preserve task-relevant content while discarding nuisance factors such as background texture or illumination variation \cite{Tishby1999InformationBottleneck}. Because this layer is constrained by computation, memory, and energy, rule-based filters, lightweight convolutional neural networks, quantized models, and event-triggered sensing are often more realistic than large transformers or diffusion models. Local semantic packets should also retain metadata, including timestamp, location, confidence, and sensor identity, because these elements are essential for interpretation and trust evaluation.

\subsubsection{Network/Transport Layer}
The Network/Transport Layer forwards semantic messages through underwater links and gateways. It includes acoustic modems, optical links, RF/MI links, relay nodes, surface buoys, AUV-assisted forwarding, and hybrid communication mechanisms. Routing studies show that underwater forwarding decisions must consider application delay, link quality, energy, and topology changes rather than only shortest paths \cite{Pompili2006RoutingUWSN}. Vector-based and depth-based routing further illustrate how geometry and depth information help handle sparse and dynamic underwater deployments \cite{Xie2006VBF,Yan2008DBR}.

A semantic packet can be represented as
\begin{equation}
    p_s = \{s, g, \ell, t, q, u, \kappa\},
    \label{eq:semantic_packet}
\end{equation}
where $s$ is the semantic payload, $g$ is the task identifier, $\ell$ is location, $t$ is timestamp, $q$ is confidence or quality score, $u$ is urgency, and $\kappa$ contains security or provenance information. This structure allows the network to prioritize high-value semantic content. For example, a disaster-warning packet should be forwarded before routine environmental summaries, while low-confidence semantic predictions may require additional verification.

For hybrid underwater links, the communication modality can be selected through a semantic-aware cost function:
\begin{equation}
    m^{\star} =
    \arg\min_{m \in \mathcal{M}}
    \left[
    \lambda_E E_m
    + \lambda_T T_m
    + \lambda_D \mathbb{E}\{d_s^{(m)}\}
    - \lambda_U U_g^{(m)}
    \right],
    \label{eq:link_selection}
\end{equation}
where $\mathcal{M}$ is the set of available modalities, $E_m$ is expected energy cost, $T_m$ is latency, $d_s^{(m)}$ is expected semantic distortion, and $U_g^{(m)}$ is task utility. In SC-enabled IoUT, the routing decision should account for semantic priority, acceptable distortion, and retransmission value. Adaptive semantic retransmission is relevant because incremental-knowledge hybrid automatic repeat request (HARQ) can improve semantic interpretation rather than merely recover corrupted bits \cite{Zhou2022AdaptiveHARQ}. DTN principles are also useful because underwater links may be partitioned or intermittently available \cite{Fall2003DTN}.

\subsubsection{Processing Layer}
The Processing Layer performs semantic fusion, context reasoning, model inference, knowledge-base updating, and decision support. It may be implemented on AUVs, surface buoys, edge gateways, shore stations, or cloud servers. Edge computing research shows that placing computation near the data source can reduce bandwidth usage and latency, but it also introduces resource-management challenges \cite{Satyanarayanan2017EdgeComputing}. In IoUT, this trade-off is more pronounced because underwater nodes and gateways may have limited power and intermittent backhaul links.

Semantic fusion can be represented as
\begin{equation}
    z = \mathcal{F}(s_1,s_2,\ldots,s_N,c,\mathcal{K}),
    \label{eq:semantic_fusion}
\end{equation}
where $s_i$ is the semantic representation received from node $i$, $c$ is context, $\mathcal{K}$ is prior knowledge, and $z$ is the fused semantic state. Knowledge graphs provide a natural structure for representing entities, relations, constraints, and contextual knowledge across heterogeneous data sources \cite{Hogan2021KnowledgeGraphs}. For example, a marine ecology system may connect semantic observations such as ``high turbidity,'' ``coral bleaching,'' and ``temperature anomaly'' to infer ecological risk.

The Processing Layer may also support distributed learning. Federated averaging enables distributed nodes to train a shared model without transmitting raw data \cite{McMahan2017FedAvg}. However, FL in practical networks must address non-independent and identically distributed (non-IID) data, unreliable participation, privacy risks, and communication overhead, as discussed in broader FL surveys \cite{Kairouz2021FL}. In IoUT, these challenges are amplified by link intermittency and energy imbalance. Hence, FL-based semantic learning should be designed with adaptive participation, lightweight model updates, and energy-aware aggregation. Implicit semantic-aware communication further suggests that meaning may include hidden relations and reasoning paths, not only explicit labels or features \cite{Xiao2023ImplicitSemantic}; therefore, this layer should manage context, uncertainty, and task-dependent interpretation.

\subsubsection{Application Layer}
The Application Layer converts semantic information into user-facing services, alerts, control actions, and decision support. Representative applications include environmental monitoring, marine ecology, underwater archaeology, subsea infrastructure inspection, disaster response, AUV coordination, aquaculture, and maritime security. In task-oriented communication, the usefulness of information is measured according to how much it improves the downstream decision or action \cite{Wen2023TaskOriented}. Therefore, this layer should define the semantic units, acceptable information loss, latency requirements, and risk tolerance for each mission.

For an application goal $g$, the receiver may choose an action
\begin{equation}
    a^{\star} = \arg\max_{a \in \mathcal{A}}
    \mathbb{E}\left[ U_g(a,z) \mid \hat{s},c,\mathcal{K} \right],
    \label{eq:application_action}
\end{equation}
where $\mathcal{A}$ is the action set and $U_g(a,z)$ is task utility. Examples include triggering an environmental alarm, requesting a higher-resolution retransmission, updating a semantic map, scheduling an AUV inspection, or activating emergency response. Data-significance-oriented communication similarly emphasizes that useful information should be delivered to the right computation or actuation point at the right time \cite{Uysal2022DataSignificance}. This principle is particularly relevant underwater, where late or misinterpreted information may be useless even if it is bit-wise correct.

\subsection{Metrics for Semantic Efficiency}
Traditional metrics such as BER, PDR, throughput, and delay remain important, but they are insufficient for SC-enabled IoUT because they do not directly measure whether the received content is meaningful or useful. Semantic-efficiency metrics should capture the relationship among transmitted meaning, task outcome, communication cost, computation cost, and timeliness.

For text-oriented SC, the BLEU score can measure overlap between generated and reference text, although it does not fully capture meaning \cite{Papineni2002BLEU}. For image-oriented IoUT applications, SSIM is widely used to evaluate perceived structural quality \cite{Wang2004SSIM}. VLMs such as CLIP can support semantic alignment between images and text, making them useful for evaluating whether a reconstructed underwater image is semantically consistent with its description \cite{Radford2021CLIP}. However, such metrics should not be treated as universal measures of safety-critical correctness.

If $e(s)$ and $e(\hat{s})$ denote embeddings of the transmitted and received semantic representations, cosine semantic similarity can be written as
\begin{equation}
    \mathrm{Sim}_s(s,\hat{s})
    =
    \frac{e(s)^{\mathrm{T}} e(\hat{s})}
    {\|e(s)\|_2 \|e(\hat{s})\|_2}.
    \label{eq:semantic_similarity}
\end{equation}
The corresponding semantic distortion is
\begin{equation}
    d_s(s,\hat{s}) = 1 - \mathrm{Sim}_s(s,\hat{s}).
    \label{eq:semantic_distortion}
\end{equation}

\begin{figure*}[h!]
\centering
\begin{subfigure}{0.48\textwidth}
    \centering
    \includegraphics[width=\linewidth]{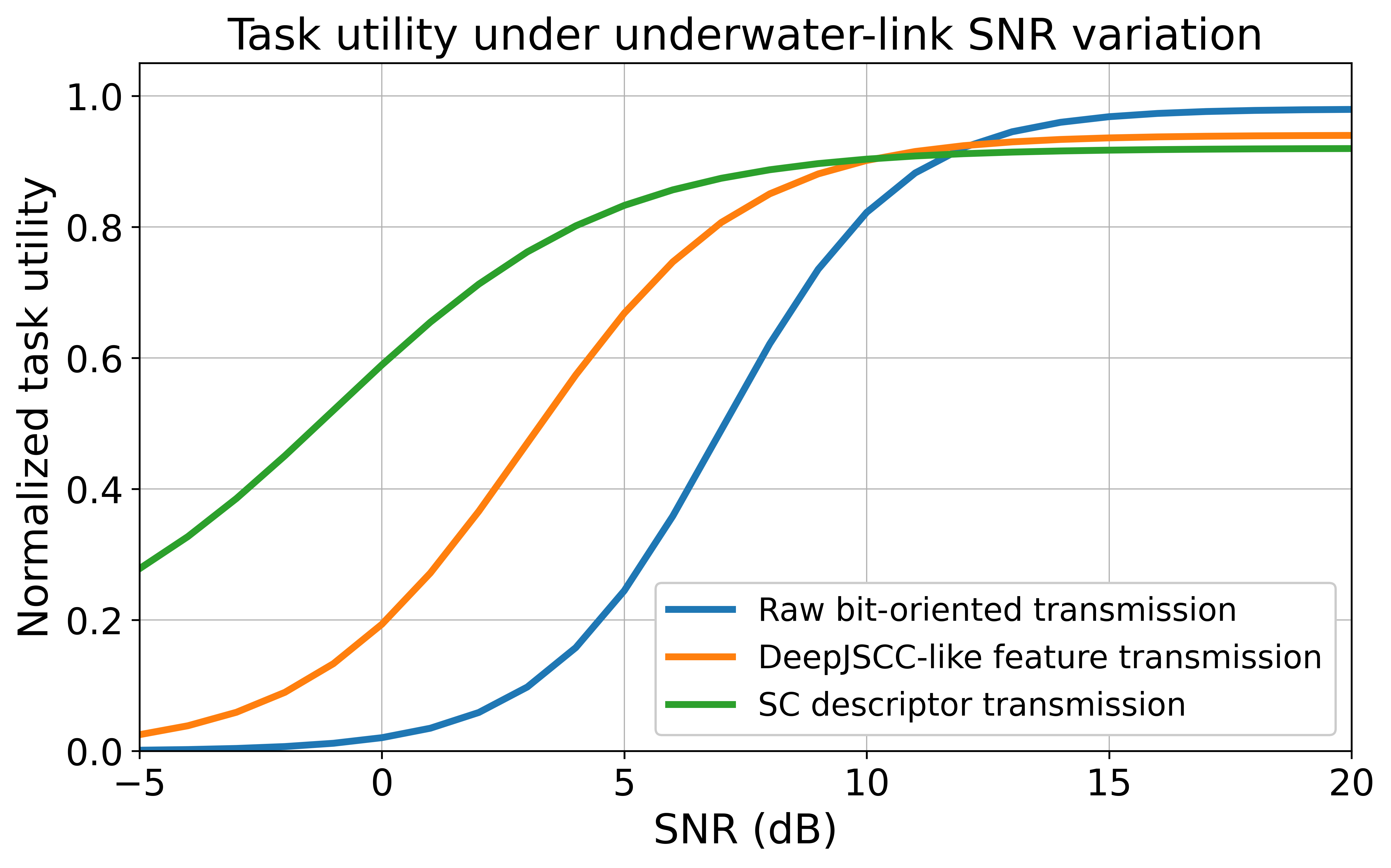}
    \caption{Normalized task utility versus SNR.}
    \label{fig:task_utility_snr}
\end{subfigure}
\hfill
\begin{subfigure}{0.48\textwidth}
    \centering
    \includegraphics[width=\linewidth]{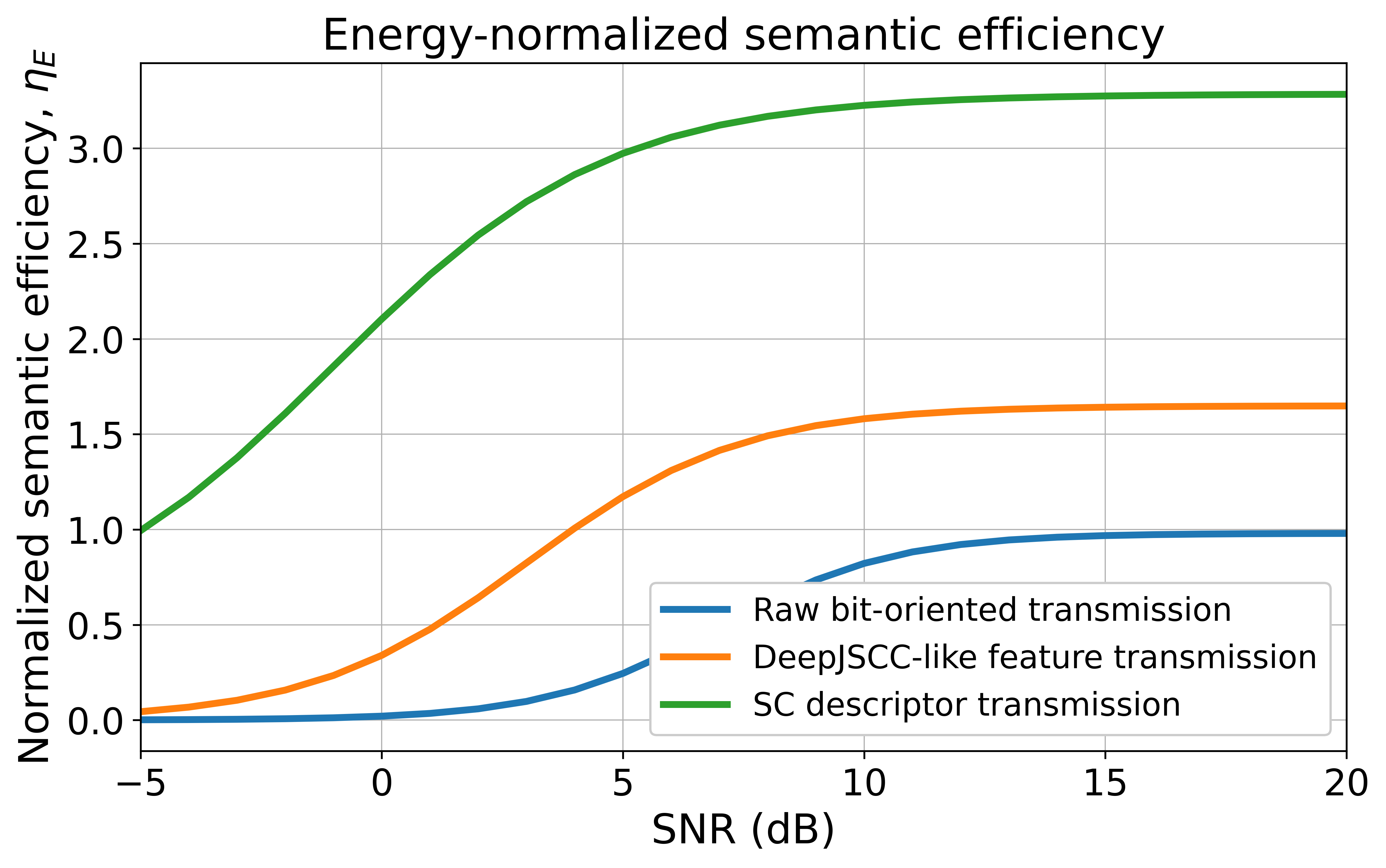}
    \caption{Energy-normalized semantic efficiency versus SNR.}
    \label{fig:semantic_efficiency_snr}
\end{subfigure}
\caption{Illustrative normalized comparison of raw bit-oriented transmission, DeepJSCC-like feature transmission, and SC descriptor transmission for IoUT. The plots are generated from the metric models in Eqs.~\eqref{eq:sim_task_utility} and \eqref{eq:sim_energy_efficiency} using the parameters in Table~\ref{tab:illustrative_simulation_parameters}; they illustrate metric behavior rather than measured modem or field-test results.}
\label{fig:semantic_efficiency_simulation}
\end{figure*}

For classification or detection tasks, semantic similarity may be less important than false-alarm rate, missed-detection probability, F1-score, or task success rate. Thus, the metric must be selected according to the application objective.

Semantic efficiency can be defined from a bandwidth perspective as
\begin{equation}
    \eta_B = \frac{\mathbb{E}[U_g(\hat{y},y)]}{B T},
    \label{eq:bandwidth_semantic_efficiency}
\end{equation}
where $U_g(\hat{y},y)$ is task utility, $B$ is bandwidth, and $T$ is transmission time. An energy-aware version is
\begin{equation}
    \eta_E = \frac{\mathbb{E}[U_g(\hat{y},y)]}{E_{\mathrm{tot}}},
    \label{eq:energy_semantic_efficiency}
\end{equation}
where $E_{\mathrm{tot}}$ includes sensing, semantic inference, transmission, reception, and semantic decoding energy. Deep JSCC studies show that learned transmission systems can behave differently from conventional, separately designed source-channel systems at low SNR, reinforcing the need for end-to-end, task-aware metrics \cite{Bourtsoulatze2019DeepJSCC}.

\paragraph{Illustrative metric comparison}
To illustrate why SC-enabled IoUT systems require task-aware and energy-aware metrics, we consider a normalized simulation comparing raw bit-oriented transmission, DeepJSCC-like feature transmission, and SC descriptor transmission. This simulation is not intended to benchmark a specific modem or underwater dataset; rather, it illustrates how semantic-efficiency metrics behave under different payload and computation assumptions. The normalized task utility of scheme $i$ is modeled as
\begin{equation}
    U_i(\Gamma_{\mathrm{dB}})
    =
    U_i^{\min}
    +
    \frac{U_i^{\max}-U_i^{\min}}
    {1+\exp[-a_i(\Gamma_{\mathrm{dB}}-\Gamma_i)]},
    \label{eq:sim_task_utility}
\end{equation}
where $\Gamma_{\mathrm{dB}}$ is the received SNR in dB, $a_i$ controls the transition slope, and $\Gamma_i$ denotes the operating point at which task utility begins to improve. The raw scheme is modeled as requiring a larger payload and a higher SNR, while the DeepJSCC-like scheme exhibits the graceful degradation behavior observed in learned joint source-channel coding \cite{Bourtsoulatze2019DeepJSCC}. The SC descriptor scheme uses a smaller semantic payload but includes additional semantic encoding and decoding cost.

For each scheme, the energy-normalized semantic efficiency is computed as
\begin{equation}
    \eta_E^{(i)}
    =
    \frac{U_i(\Gamma_{\mathrm{dB}})}
    {E_{\mathrm{enc}}^{(i)} + b_i e_{\mathrm{tx}} + E_{\mathrm{dec}}^{(i)}},
    \label{eq:sim_energy_efficiency}
\end{equation}
where $b_i$ is the normalized payload size, $e_{\mathrm{tx}}$ is the normalized transmission energy per bit, and $E_{\mathrm{enc}}^{(i)}$ and $E_{\mathrm{dec}}^{(i)}$ denote semantic encoding and decoding energy. In the simulation, $b_i=\{1.0,0.45,0.10\}$ for raw, DeepJSCC-like, and SC descriptor transmission, respectively, while the corresponding computation costs are $\{0,0.12,0.18\}$. Fig.~\ref{fig:semantic_efficiency_simulation} shows that SC descriptor transmission can provide higher energy-normalized semantic efficiency when reduced communication load compensates for inference cost. However, if semantic encoding or decoding becomes too expensive, the advantage of SC may diminish.

The compute--communication trade-off is central to SC in IoUT. Let $e_{\mathrm{tx}}(d,m)$ denote the energy required to transmit one bit over distance $d$ using modality $m$. A simplified raw-transmission energy model is
\begin{equation}
    E_{\mathrm{raw}} = b_x e_{\mathrm{tx}}(d,m),
    \label{eq:raw_energy}
\end{equation}
where $b_x$ is the raw payload size. For SC, the total energy can be approximated as
\begin{equation}
    E_{\mathrm{SC}}
    =
    E_{\mathrm{enc}}
    + b_s e_{\mathrm{tx}}(d,m)
    + E_{\mathrm{dec}},
    \label{eq:sc_energy}
\end{equation}
where $E_{\mathrm{enc}}$ is semantic extraction energy, $b_s$ is semantic payload size, and $E_{\mathrm{dec}}$ is receiver-side semantic decoding or inference energy. SC is energy-beneficial only if
\begin{equation}
    E_{\mathrm{SC}} < E_{\mathrm{raw}},
    \quad \Rightarrow \quad
    (b_x-b_s)e_{\mathrm{tx}}(d,m)
    >
    E_{\mathrm{enc}}+E_{\mathrm{dec}}.
    \label{eq:sc_energy_condition}
\end{equation}
This condition is important because large AI models may reduce transmitted bits but consume non-negligible inference energy.

\begin{table*}[ht!]
\centering
\footnotesize
\caption{Evaluation metrics for SC-enabled IoUT systems: purpose, use cases, and limitations.}
\label{tab:semantic_metrics_iout}
\renewcommand{\arraystretch}{1.20}
\setlength{\tabcolsep}{3.5pt}
\begin{tabular}{p{2.6cm} p{4.2cm} p{3.5cm} p{4.5cm} p{1.5cm}}
\toprule
\textbf{Metric Category} 
& \textbf{Representative Metrics / Expressions} 
& \textbf{Suitable IoUT Use Cases} 
& \textbf{Interpretation and Main Caveat} 
& \textbf{Refs.} \\
\midrule

Physical-layer reliability 
& SNR, BER, block error rate, PDR, throughput, link availability
& Modem evaluation, link adaptation, acoustic/optical/RF comparison
& Measures whether symbols or packets are delivered reliably, but not whether content is semantically useful
& \cite{Climent2014UnderwaterAcousticWSN,Sendra2015AcousticModems} \\
\midrule

Semantic fidelity 
& Embedding similarity, ontology consistency, concept accuracy, semantic distortion $d_s(s,\hat{s})$
& Text descriptors, semantic labels, semantic maps, multimodal metadata, knowledge-graph messages
& Measures meaning preservation, but may not directly reflect downstream task success
& \cite{Bao2011TheorySemantic,Hogan2021KnowledgeGraphs} \\
\midrule

Task-level utility 
& Accuracy, precision, recall, F1-score, mean intersection-over-union (mIoU), missed-detection rate, false-alarm rate
& Object recognition, anomaly detection, subsea inspection, ecological event detection, AUV mission execution
& Most relevant for SC, but requires task-specific ground truth and decision thresholds
& \cite{Islam2020SUIM,Wen2023TaskOriented} \\
\midrule

Perceptual and visual quality 
& PSNR, SSIM, CLIP-based image--text alignment, operator assessment
& Image reconstruction, semantic visualization, remote inspection, operator situational awareness
& Visually plausible outputs may still be incorrect; not evidence of exact recovery
& \cite{Wang2004SSIM,Radford2021CLIP} \\
\midrule

Freshness and timeliness 
& End-to-end latency, AoI, semantic AoI, deadline miss ratio, event response time
& AUV control, disaster warning, adaptive monitoring, time-sensitive environmental alerts
& Captures usefulness over time; semantic freshness is task-dependent
& \cite{Yates2021AoISurvey,Uysal2022DataSignificance} \\
\midrule

Resource efficiency 
& Energy per useful semantic message, semantic utility/Joule, inference energy, model memory, model-update overhead
& Battery-limited sensors, AUVs, edge gateways, FL
& Must include sensing, inference, transmission, reception, decoding, and model-update cost
& \cite{Sherlock2022MiniModems,Kairouz2021FL} \\
\midrule

Robustness and generalization 
& Performance under SNR variation, turbidity, packet loss, domain shift, mobility, sensor drift, and missing semantic packets
& Long-term monitoring, mobile AUV missions, multi-site ecological sensing, infrastructure inspection
& Results from one dataset or channel model may not generalize to another environment
& \cite{Li2020WaterNet,Islam2020FUnIEGAN} \\
\midrule

Semantic reliability and outage 
& Semantic outage probability $P_{\mathrm{s\text{-}out}}=\Pr[d_g(s,\hat{s})>\epsilon_g]$, semantic delivery ratio, task failure probability
& Mission-critical alerts, pipeline inspection, disaster response, AUV navigation
& More meaningful than packet reliability, but requires a task-dependent distortion threshold
& \cite{Lu2021Rethinking,Zhou2022AdaptiveHARQ} \\
\midrule

Trustworthiness, security, and provenance 
& Confidence calibration, uncertainty score, provenance metadata, semantic integrity, privacy leakage, adversarial robustness, trust score
& Multi-stakeholder IoUT, defense, infrastructure monitoring, environmental compliance, privacy-sensitive sensing 
& Requires explicit threat models, calibrated uncertainty, and provenance-aware semantic packet design
& \cite{Bonawitz2017SecureAggregation,Kairouz2021FL} \\
\bottomrule
\end{tabular}
\end{table*}

Freshness is another important metric. The AoI at time $t$ is
\begin{equation}
    \Delta(t) = t - u(t),
    \label{eq:aoi}
\end{equation}
where $u(t)$ is the generation time of the most recently received useful update. The time-average AoI is
\begin{equation}
    \bar{\Delta}
    =
    \lim_{T \rightarrow \infty}
    \frac{1}{T}
    \int_{0}^{T} \Delta(t)\,dt.
    \label{eq:average_aoi}
\end{equation}
AoI has become a standard freshness metric for real-time status updating \cite{Yates2021AoISurvey}. In SC-enabled IoUT, however, freshness should be interpreted semantically: a habitat map may remain useful for hours, while an AUV collision warning may become useless within seconds. To provide a structured view of these evaluation dimensions, Table~\ref{tab:semantic_metrics_iout} summarizes the major metrics for SC-enabled IoUT systems, including physical-layer reliability, semantic fidelity, task-level utility, perceptual quality, freshness, resource efficiency, robustness, semantic reliability, and trustworthiness.

\subsection{Semantic Channel Modeling in IoUT}
Semantic channel modeling extends conventional channel modeling by including semantic encoding, task relevance, context mismatch, model uncertainty, and semantic distortion. In a conventional link, the physical channel maps transmitted symbols $u$ to received symbols $r$ according to $P(r|u,c)$. In an SC system, the end-to-end semantic channel can be modeled as
\begin{equation}
    P(\hat{s}|s,c,g)
    =
    \sum_{u,r}
    P_{\phi}(\hat{s}|r,c,g,\mathcal{K})
    P(r|u,c)
    P_{\psi}(u|s,c,g),
    \label{eq:semantic_channel}
\end{equation}
where $P_{\psi}(u|s,c,g)$ represents semantic-to-symbol encoding, $P(r|u,c)$ represents the underwater physical channel, and $P_{\phi}(\hat{s}|r,c,g,\mathcal{K})$ represents semantic decoding. This model makes clear that semantic errors may arise not only from physical noise but also from poor semantic encoding, model mismatch, outdated context, or inconsistent shared knowledge. Rethinking modern communication from semantic coding to semantic communication further highlights that semantic processing and channel noise should be considered jointly rather than independently \cite{Lu2021Rethinking}.

A semantic outage event can be defined as the case where task-level semantic distortion exceeds an application-dependent threshold:
\begin{equation}
    P_{\mathrm{s\text{-}out}}(\epsilon_g)
    =
    \Pr\left[d_g(s,\hat{s}) > \epsilon_g \right],
    \label{eq:semantic_outage}
\end{equation}
where $\epsilon_g$ is the maximum tolerable semantic distortion for task $g$. The corresponding semantic reliability is
\begin{equation}
    R_s(\epsilon_g) = 1 - P_{\mathrm{s\text{-}out}}(\epsilon_g).
    \label{eq:semantic_reliability}
\end{equation}
This reliability measure differs from packet-level reliability. A packet may contain bit errors but still convey the correct semantic alert, while a perfectly delivered packet may be semantically insufficient if it omits task-critical context.

\begin{figure}[h!]
\centering
\includegraphics[width=0.98\columnwidth]{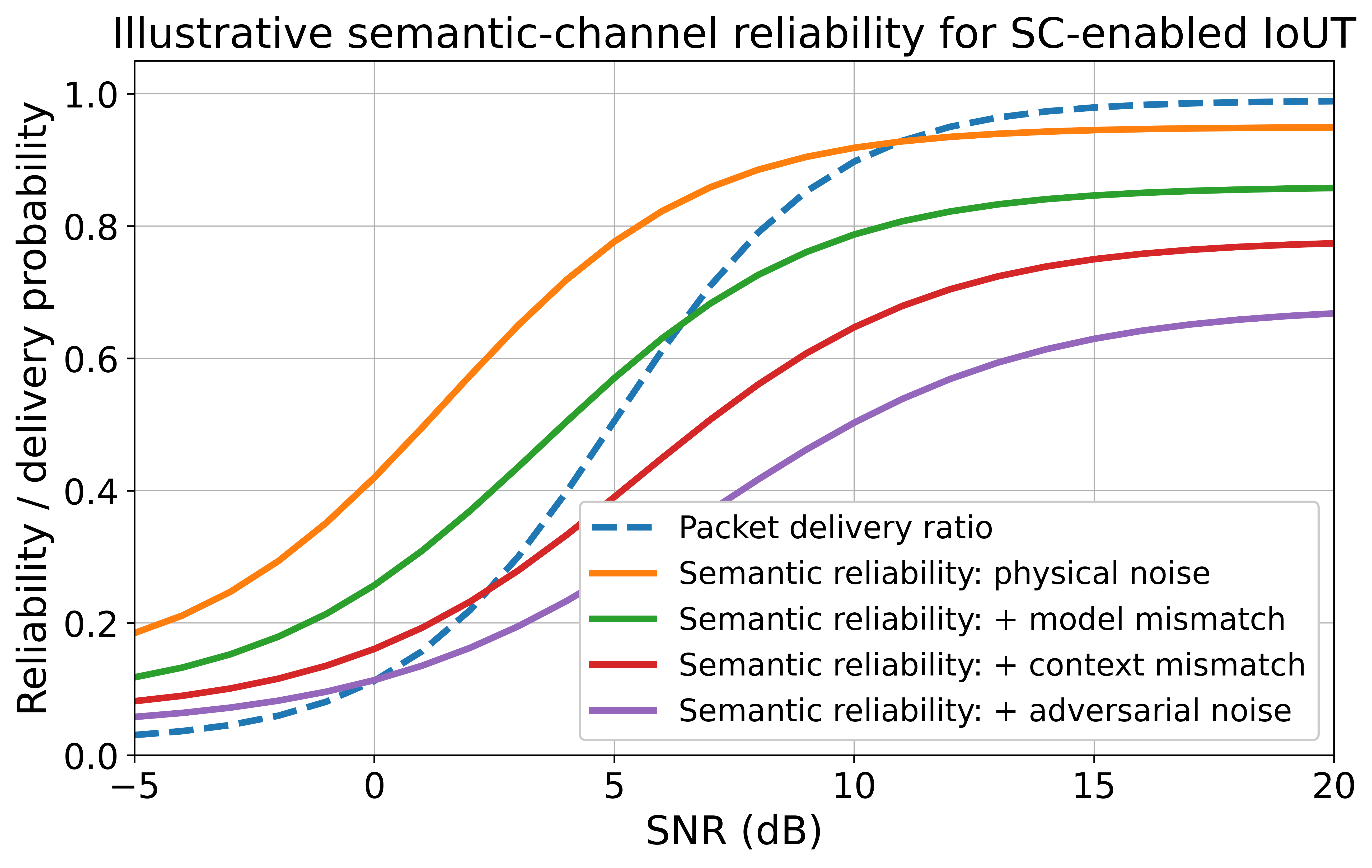}
\caption{Illustrative semantic-channel reliability for SC-enabled IoUT. The curves are generated using the normalized reliability models and mismatch penalties listed in Table~\ref{tab:illustrative_simulation_parameters}.}
\label{fig:semantic_channel_reliability}
\end{figure}

Fig.~\ref{fig:semantic_channel_reliability} provides an illustrative comparison showing that packet-level reliability and semantic reliability may evolve differently as SNR changes, especially when model mismatch, context mismatch, or adversarial semantic noise is present.

Semantic noise in IoUT can be categorized into four types. First, physical semantic noise occurs when underwater channel impairments corrupt the transmitted representation. Second, model semantic noise occurs when the encoder or decoder misinterprets the source because of domain shift or insufficient training. Third, context semantic noise occurs when the transmitter and receiver use different environmental assumptions, ontologies, or task goals. Fourth, adversarial semantic noise occurs when an attacker manipulates semantic labels, confidence values, or model updates. The early theory-oriented view of SC already recognized that reliable semantic transfer requires both data-level and meaning-level consistency \cite{Bao2011TheorySemantic}.

For hybrid IoUT systems, semantic channel modeling should also support modality selection. Acoustic links may be preferred for long-range, low-rate alerts; optical links for short-range visual descriptors; RF links for near-surface connectivity; and MI links for localized seabed or near-field communication. The semantic utility of modality $m$ can be written as
\begin{equation}
    \mathcal{U}_m
    =
    \alpha_g U_g^{(m)}
    - \alpha_E E_m
    - \alpha_T T_m
    - \alpha_D \mathbb{E}[d_s^{(m)}],
    \label{eq:semantic_link_utility}
\end{equation}
where $U_g^{(m)}$ is task utility, $E_m$ is energy cost, $T_m$ is latency, and $d_s^{(m)}$ is semantic distortion. The system can then select
\begin{equation}
    m^{\star} = \arg\max_{m \in \mathcal{M}} \mathcal{U}_m.
    \label{eq:utility_link_selection}
\end{equation}
This model highlights why SC-enabled IoUT should not be optimized solely by physical throughput. The best link is the one that provides the most useful semantic outcome under the current mission, channel, and resource constraints.
\section{Learning-Driven Semantic Communication for IoUT}
\label{sec03}

Learning-driven SC uses trainable encoders, decoders, reasoning modules, and control policies to transform raw underwater observations into task-relevant semantic representations. Instead of treating communication as the delivery of all source symbols, learning-driven SC identifies the portion of information that is useful for inference, decision-making, or actuation. DeepSC demonstrated this principle for text transmission by jointly learning semantic and channel coding modules \cite{Xie2021DeepSC}, while semantic-aware networking studies later extended this idea from link-level transmission to network-level semantic processing \cite{Shi2021SemanticAwareNetworking}. In IoUT, this shift is especially relevant because underwater nodes often collect high-dimensional images, sonar data, acoustic signals, and environmental measurements, while their communication links and energy resources remain severely limited.

A generic learning-driven SC pipeline for IoUT can be expressed as
\begin{equation}
    s = Q_{\eta}\!\left(f_{\theta}(x,c,g)\right), \quad
    u = c_{\psi}(s), \quad
    \hat{y} = h_{\phi}(r,c,g,\mathcal{K}),
    \label{eq:learning_sc_pipeline}
\end{equation}
where $x$ is the raw underwater observation, $c$ denotes environmental or network context, $g$ is the task objective, $f_{\theta}(\cdot)$ is the semantic encoder, $Q_{\eta}(\cdot)$ is a quantization or tokenization function, $c_{\psi}(\cdot)$ is the channel encoder, $r$ is the received signal, $h_{\phi}(\cdot)$ is the receiver-side inference or reconstruction model, and $\mathcal{K}$ denotes shared domain knowledge. Task-oriented communication research emphasizes that $\hat{y}$ should be evaluated according to its usefulness for the task rather than only according to bit-level fidelity \cite{Wen2023TaskOriented}. For IoUT, this means that communication success should be measured by whether the received semantics support the intended underwater task, such as anomaly detection, AUV coordination, habitat classification, or infrastructure inspection.

The learning objective should therefore include semantic, task, communication, and computation costs. A representative objective is
\begin{equation}
\begin{aligned}
    \min_{\theta,\eta,\psi,\phi} \quad
    \mathbb{E}_{x,c,g,r}
    \big[
    &\mathcal{L}_{g}(y,\hat{y})
    + \lambda_s d_s(s,\hat{s})
    + \lambda_b R(s)  \\
    &+ \lambda_E E_{\mathrm{inf}}(\theta,\phi)
    + \lambda_T T_{\mathrm{inf}}(\theta,\phi)
    \big],
\end{aligned}
\label{eq:learning_sc_loss}
\end{equation}
where $\mathcal{L}_{g}$ is task loss, $d_s(\cdot)$ is semantic distortion, $R(s)$ represents the number of transmitted semantic bits or tokens, $E_{\mathrm{inf}}$ is inference energy, and $T_{\mathrm{inf}}$ is inference latency. This formulation is important for IoUT because it explicitly captures the communication--computation trade-off: semantic extraction can reduce payload size, but the gain is meaningful only when inference cost, latency, and model-update overhead remain acceptable.

\subsection{AI/ML Models for Semantic Encoding}
AI/ML models enable semantic encoding by mapping heterogeneous underwater observations into compact and task-relevant representations. Convolutional neural networks (CNNs) and encoder--decoder networks are widely used for local feature extraction, with U-Net providing a foundational architecture for dense segmentation tasks \cite{Ronneberger2015UNet}. In underwater vision, the SUIM dataset and benchmark provide labeled categories such as fish, reefs, aquatic plants, divers, robots, and seafloor, making it relevant for semantic segmentation in IoUT perception pipelines \cite{Islam2020SUIM}. Underwater image enhancement models such as Water-Net \cite{Li2020WaterNet} and FUnIE-GAN \cite{Islam2020FUnIEGAN} are also useful preprocessing modules because degraded underwater images can reduce the reliability of downstream semantic extraction. More general segmentation foundation models, such as Segment Anything, further demonstrate the potential of promptable segmentation, but domain adaptation for the underwater domain remains necessary before such models can be used reliably in marine environments \cite{Kirillov2023SAM}.

Transformer-based models improve semantic encoding by capturing global dependencies across image patches, text tokens, acoustic streams, or multimodal sequences. The original Transformer architecture introduced self-attention for sequence modeling \cite{Vaswani2017Attention}, while the Vision Transformer (ViT) showed that images can be represented as patch-token sequences \cite{Dosovitskiy2021ViT}. Swin Transformer further introduced hierarchical shifted-window attention, which is useful for dense vision tasks such as detection and segmentation \cite{Liu2021Swin}. For IoUT, these models are attractive because underwater scenes often contain long-range dependencies, such as the spatial relation between a pipeline defect and surrounding infrastructure or the temporal relation between acoustic signatures and AUV movement. However, their computational and memory requirements may exceed the capacity of small underwater nodes, so deployment should be conditioned on hardware capabilities, latency budgets, and mission criticality.

Vision-language models (VLMs) offer another important direction for semantic encoding by linking visual observations to textual descriptions. CLIP learns transferable image--text representations from natural-language supervision \cite{Radford2021CLIP}, and BLIP supports image understanding and captioning \cite{Li2022BLIP}. BLIP-2 further reduces training cost by bootstrapping VLMs with frozen image encoders and LLMs, making it relevant to semantic description generation under limited adaptation data \cite{Li2023BLIP2}. In IoUT, such models can transform an underwater image into compact semantic descriptors, for example, ``coral bleaching region,'' ``pipeline crack,'' or ``AUV obstacle ahead.'' Recent underwater-specific systems such as AquaVLM explore mobile VLMs for context-aware underwater messaging \cite{Tian2025AquaVLM}, while LLM-assisted underwater image transmission has been studied as an emerging SC direction \cite{Chen2024LLMUnderwater}. These approaches are promising, but they should be treated as early-stage and computationally demanding rather than immediately deployable on all underwater sensor nodes.

Generative models can be used at the receiver to produce approximate visualizations or semantic reconstructions from compact representations. Denoising diffusion probabilistic models provide a foundation for iterative generative reconstruction \cite{Ho2020DDPM}, while latent diffusion models reduce generation cost by operating in a compressed latent space \cite{Rombach2022LDM}. ControlNet enables conditional control over generated images using structural or semantic guidance \cite{Zhang2023ControlNet}. In underwater communication, SAGE-type systems and underwater acoustic SC approaches use image-to-text or entity-list representations before receiver-side reconstruction \cite{Loureiro2024SAGE,Zhang2025UASC}. However, generative outputs may be plausible but incorrect. Therefore, in safety-critical IoUT tasks, such as pipeline inspection, AUV navigation, or disaster response, generative models should be used for task-oriented visualization or decision support, not as proof of exact signal recovery.

Quantization, pruning, distillation, and parameter-efficient adaptation are essential for deploying learning-driven SC in resource-constrained underwater systems. Knowledge distillation transfers information from a large teacher model to a smaller student model \cite{Hinton2015Distillation}, while deep compression combines pruning, quantization, and coding to reduce model size \cite{Han2016DeepCompression}. Integer quantization enables efficient inference on embedded processors \cite{Jacob2018Quantization}, and post-training quantization methods, such as GPTQ, further reduce the cost of large-model inference \cite{Frantar2023GPTQ}. Low-rank adaptation (LoRA) provides parameter-efficient fine-tuning with fewer trainable parameters \cite{Hu2021LoRA}, while QLoRA extends this idea to quantized LLMs \cite{Dettmers2023QLoRA}. More recently, selective state-space models such as Mamba have attracted attention for their linear-in-sequence-length scaling \cite{Gu2023Mamba}. Vision Mamba investigates efficient visual representation learning using bidirectional state-space models \cite{Zhu2024VisionMamba}, although MambaOut cautions that state-space models are not automatically superior for all vision tasks \cite{Yu2025MambaOut}. This observation is important for IoUT: model choice should be justified by task, hardware, channel constraints, and energy budget rather than by novelty alone.

\begin{figure}[t!]
\centering
\includegraphics[width=0.98\columnwidth]{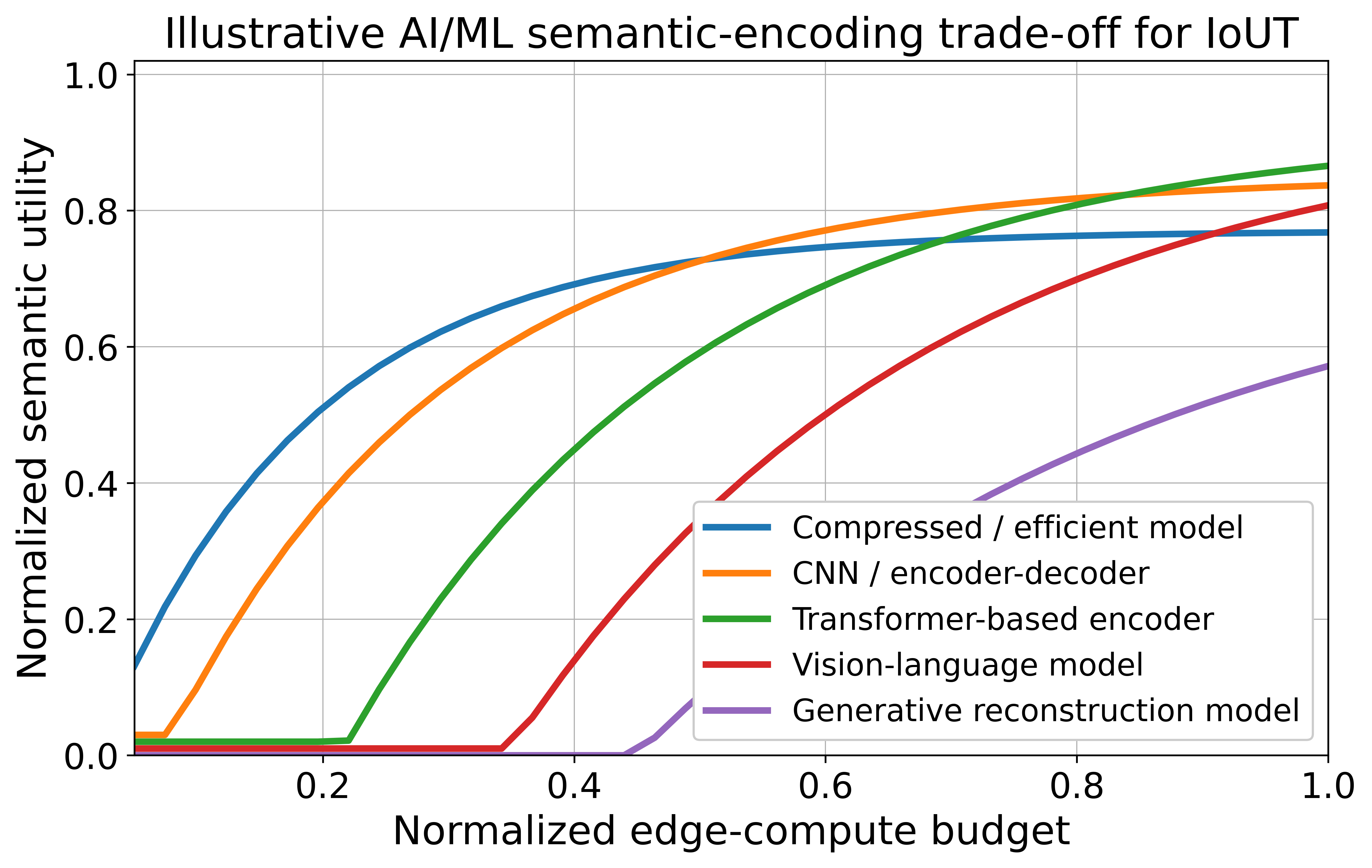}
\caption{Illustrative AI/ML semantic-encoding trade-off for IoUT. The plotted points use normalized compute-budget and semantic-utility scores listed in Table~\ref{tab:illustrative_simulation_parameters}.}
\label{fig:ai_ml_semantic_encoding_tradeoff}
\end{figure}

Fig.~\ref{fig:ai_ml_semantic_encoding_tradeoff} illustrates the deployment trade-off among representative AI/ML model families for SC-enabled IoUT. Lightweight and compressed models are better suited to strict edge-compute constraints, whereas transformers, VLMs, and generative models generally require larger computational budgets to achieve higher semantic abstraction or reconstruction capabilities.

\begin{table*}[ht!]
\centering
\footnotesize
\caption{Illustrative numerical parameters used for Figs.~\ref{fig:semantic_efficiency_simulation}--\ref{fig:ai_ml_semantic_encoding_tradeoff}. All values are normalized and are used only to reproduce the conceptual plots.}
\label{tab:illustrative_simulation_parameters}
\renewcommand{\arraystretch}{1.18}
\setlength{\tabcolsep}{3.5pt}
\begin{tabular}{p{1.5cm} p{2.5cm} p{5.25cm} p{7.5cm}}
\toprule
\textbf{Figure} & \textbf{Quantity} & \textbf{Model / Assumption} & \textbf{Parameter Values} \\
\midrule

Fig.~\ref{fig:semantic_efficiency_simulation} 
& SNR range 
& Received SNR sweep used for normalized task-utility and semantic-efficiency curves 
& $\Gamma_{\mathrm{dB}}\in[-5,20]$ dB with $0.5$ dB step size \\
\midrule

Fig.~\ref{fig:semantic_efficiency_simulation} 
& Task utility 
& Logistic utility model:
$U_i(\Gamma_{\mathrm{dB}})=U_i^{\min}+\frac{U_i^{\max}-U_i^{\min}}{1+\exp[-a_i(\Gamma_{\mathrm{dB}}-\Gamma_i)]}$
& Raw bit-oriented: $(U_i^{\min},U_i^{\max},a_i,\Gamma_i)=(0.05,0.92,0.35,10)$;
DeepJSCC-like: $(0.18,0.92,0.30,4)$;
SC descriptor: $(0.35,0.90,0.22,0)$ \\
\midrule

Fig.~\ref{fig:semantic_efficiency_simulation} 
& Payload and computation cost 
& Energy-normalized semantic efficiency:
$\eta_E^{(i)}=\frac{U_i(\Gamma_{\mathrm{dB}})}
{E_{\mathrm{enc}}^{(i)}+b_i e_{\mathrm{tx}}+E_{\mathrm{dec}}^{(i)}}$
& Normalized payloads:
$b_i=\{1.00,0.45,0.10\}$ for raw, DeepJSCC-like, and SC descriptor transmission.
Normalized $e_{\mathrm{tx}}=1$.
Computation costs:
$E_{\mathrm{enc}}^{(i)}+E_{\mathrm{dec}}^{(i)}=\{0,0.12,0.18\}$, split equally between encoder and decoder when required. \\
\midrule

Fig.~\ref{fig:semantic_channel_reliability} 
& Packet delivery ratio 
& Logistic packet-level reliability model:
$P_{\mathrm{PDR}}(\Gamma_{\mathrm{dB}})=\frac{1}{1+\exp[-0.45(\Gamma_{\mathrm{dB}}-3)]}$
& $\Gamma_{\mathrm{dB}}\in[-5,20]$ dB with $0.5$ dB step size \\
\midrule

Fig.~\ref{fig:semantic_channel_reliability} 
& Semantic reliability 
& Baseline physical-noise semantic reliability:
$R_{\mathrm{phy}}(\Gamma_{\mathrm{dB}})=0.10+\frac{0.88}{1+\exp[-0.35(\Gamma_{\mathrm{dB}}-0.5)]}$.
Mismatch-aware reliability:
$R_s^{(j)}=\mathrm{clip}(R_{\mathrm{phy}}-\delta_j,0,1)$
& Penalty values:
$\delta_j=\{0,0.08,0.16,0.25\}$ for physical-noise-only, model-mismatch, context-mismatch, and adversarial-noise cases, respectively. \\
\midrule

Fig.~\ref{fig:ai_ml_semantic_encoding_tradeoff} 
& Model-family trade-off 
& Normalized deployment score using edge-compute budget on the $x$-axis and semantic utility on the $y$-axis 
& Compressed efficient model: $(0.20,0.38)$;
CNN encoder--decoder: $(0.38,0.55)$;
Transformer-based encoder: $(0.62,0.72)$;
VLM: $(0.80,0.84)$;
Generative reconstruction model: $(0.93,0.90)$ \\
\bottomrule
\end{tabular}
\end{table*}

\subsection{Knowledge-Driven vs. Data-Driven Approaches}
Learning-driven SC can be broadly divided into data-driven, knowledge-driven, and hybrid approaches. Data-driven approaches learn semantic representations directly from datasets. They are effective when large and representative training data are available, as shown by Deep JSCC for image transmission \cite{Bourtsoulatze2019DeepJSCC}. For IoUT, data-driven encoders can learn features for fish detection, coral segmentation, pipeline inspection, underwater image restoration, sonar classification, or acoustic event recognition. However, underwater datasets are often limited, domain-shifted, and affected by turbidity, illumination, color distortion, viewpoint changes, and sensor heterogeneity. Consequently, a model trained in one marine environment may not generalize well to another. Domain-adversarial training provides one way to reduce domain shift by learning representations that are predictive for the task but less dependent on the source domain \cite{Ganin2016DANN}.

Knowledge-driven approaches use structured domain knowledge to define what should be observed, transmitted, and interpreted. The W3C/OGC Semantic Sensor Network ontology provides a vocabulary for sensors, observations, properties, procedures, and features of interest \cite{W3CSSN2017}. Knowledge graphs provide a general framework for representing entities and relations, which is useful when IoUT data must be integrated across marine biology, geology, infrastructure inspection, navigation, and disaster-response systems \cite{Hogan2021KnowledgeGraphs}. In SC-enabled IoUT, a knowledge graph may represent relations such as ``thermal anomaly near pipeline,'' ``coral bleaching associated with temperature rise,'' ``AUV obstacle detected near mission path,'' or ``high turbidity affecting optical-link reliability.'' Such structured knowledge helps receivers interpret compact semantic messages consistently across heterogeneous devices and applications.

A useful hybrid formulation combines data-driven task loss with knowledge-consistency regularization:
\begin{equation}
    \mathcal{L}_{\mathrm{hybrid}}
    =
    \mathcal{L}_{\mathrm{data}}(y,\hat{y})
    +
    \lambda_K
    \sum_{r \in \mathcal{R}}
    \xi_r(s,\mathcal{K}),
    \label{eq:hybrid_learning_loss}
\end{equation}
where $\mathcal{L}_{\mathrm{data}}$ is the supervised or self-supervised learning loss, $\mathcal{R}$ is a set of semantic rules or relations, and $\xi_r(\cdot)$ penalizes violations of domain knowledge. Neuro-symbolic AI studies show that combining neural perception with symbolic reasoning can improve interpretability and reliability \cite{vanHarmelen2021NeuroSymbolic}. A recent neural-symbolic survey further emphasizes that hybrid systems are useful when perception must be combined with logical constraints or explainable reasoning \cite{Yu2023NeuralSymbolic}. For SC-enabled IoUT, this combination is attractive because underwater perception is uncertain, while many mission constraints are rule-based or safety-critical.

For IoUT, hybrid learning is especially useful because many underwater tasks require both perception and reasoning. For example, a purely data-driven detector may identify a bright region as an anomaly, while a knowledge-driven rule may determine whether that anomaly is consistent with coral bleaching, sediment disturbance, lighting variation, or artificial structure reflection. Relational graph neural networks can support reasoning over structured relations in knowledge graphs \cite{Schlichtkrull2018RGCN}, and such mechanisms can be used to align semantic messages across heterogeneous IoUT nodes. The main challenge is that domain knowledge must be standardized, maintained, and shared across devices, missions, and organizations. Table~\ref{tab:key-semantic-tech} summarizes major model families and learning paradigms for SC-enabled IoUT, together with their roles, deployment levels, and design cautions.

%\textbf{Illustrative numerical setup for Figs.~\ref{fig:semantic_efficiency_simulation}--\ref{fig:ai_ml_semantic_encoding_tradeoff}:}
Figs.~\ref{fig:semantic_efficiency_simulation}--\ref{fig:ai_ml_semantic_encoding_tradeoff} are illustrative normalized results intended to clarify the behavior of semantic-efficiency, semantic-reliability, and model-deployment trade-offs. They are not obtained from a specific underwater modem, dataset, or field trial. Instead, the curves and points are generated from the analytical metric models introduced in this paper using the normalized parameters summarized in Table~\ref{tab:illustrative_simulation_parameters}. The SNR-dependent curves in Figs.~\ref{fig:semantic_efficiency_simulation} and \ref{fig:semantic_channel_reliability} use logistic response models to emulate graceful degradation and threshold-like link behavior, while Fig.~\ref{fig:ai_ml_semantic_encoding_tradeoff} uses normalized compute-budget and semantic-utility scores to show relative deployment suitability of representative AI/ML model families. Therefore, these results should be interpreted as reproducible conceptual examples for metric interpretation, not as measured performance claims.

\begin{table*}[ht!]
\centering
\footnotesize
\caption{Learning model families and paradigms for SC-enabled IoUT systems.}
\label{tab:key-semantic-tech}
\renewcommand{\arraystretch}{1.25}
\setlength{\tabcolsep}{4pt}
\small
\begin{tabular}{p{3.0cm} p{4.75cm} p{4.0cm} p{5.25cm}}
\toprule
\textbf{Model Family / Paradigm} 
& \textbf{Role in SC-enabled IoUT} 
& \textbf{Suitable Deployment Level} 
& \textbf{Main Limitations and Design Cautions} \\
\midrule

CNNs and encoder--decoder networks 
& Extract local visual, acoustic, or sensor-level features for semantic segmentation, object detection, anomaly screening, and region-of-interest extraction \cite{Ronneberger2015UNet,Islam2020SUIM}
& Underwater sensor nodes with lightweight models; AUVs/ROVs; edge gateways; surface buoys
& Performance may degrade under turbidity, low illumination, color distortion, and domain shift. Retraining or domain adaptation may be required for different underwater environments \\
\midrule

Transformers and visual transformers 
& Capture long-range spatial, temporal, and multimodal dependencies in images, sonar maps, acoustic sequences, and multi-sensor observations \cite{Vaswani2017Attention,Liu2021Swin}
& AUVs or ROVs with sufficient onboard computing; surface buoys; edge servers; cloud platforms
& High memory, computation, and inference-latency requirements may limit direct deployment on low-power underwater nodes. Model compression is often necessary \\
\midrule

Vision-language models 
& Map underwater visual observations into textual descriptions, object relations, semantic tags, and context-aware messages for compact semantic transmission \cite{Radford2021CLIP,Li2022BLIP} 
& AUVs/ROVs with accelerated processors; surface gateways; cloud or shore stations; selected edge platforms
& Require underwater-domain adaptation and careful validation. Generated captions or semantic descriptions may be incomplete, overconfident, or inaccurate in safety-critical scenarios \\
\midrule

Generative and diffusion models 
& Support approximate semantic reconstruction, visualization, missing-content completion, and receiver-side interpretation from compact semantic prompts or descriptors \cite{Ho2020DDPM,Rombach2022LDM,Zhang2023ControlNet}
& Primarily surface stations, edge/cloud servers, and mission-control centers; limited underwater-edge use only with strong model compression
& Generated outputs are not exact recovery of the original observations. Hallucinated or visually plausible but incorrect details can be harmful in inspection, navigation, or disaster-response tasks \\
\midrule

Knowledge graphs and ontologies 
& Represent domain entities, sensor metadata, mission goals, semantic relations, environmental context, and rule-based constraints for explainable semantic interpretation \cite{W3CSSN2017,Hogan2021KnowledgeGraphs}
& Edge/cloud reasoning layer; surface gateways; shared semantic registry across IoUT stakeholders
& Require standardization, maintenance, and agreement across domains. Inconsistent ontologies may cause semantic mismatch between transmitters and receivers \\
\midrule

Federated and distributed learning 
& Train and update semantic models collaboratively without centralizing raw underwater data, supporting privacy-aware and bandwidth-conscious model adaptation \cite{McMahan2017FedAvg,Picano2024SemanticFL}
& AUV fleets; buoy clusters; coastal edge systems; hybrid ground--aqua computing infrastructures
& Non-independent and non-identically distributed (non-IID) data, intermittent links, straggling nodes, energy imbalance, and model-update overhead can reduce convergence and communication efficiency \\
\midrule

Compressed and efficient models 
& Reduce model memory, inference latency, and adaptation cost through pruning, quantization, distillation, parameter-efficient tuning, or lightweight architectures \cite{Han2016DeepCompression,Hu2021LoRA,Dettmers2023QLoRA}
& Low-power underwater sensor nodes; embedded AUV processors; edge gateways with limited computing resources
& Compression may reduce semantic accuracy, robustness, or calibration quality. Compression should be task-aware rather than based only on parameter reduction \\
\midrule

State-space and sequence-efficient models 
& Process long temporal sequences, acoustic streams, and mission logs with lower sequence-scaling cost than full attention in some settings \cite{Gu2023Mamba,Zhu2024VisionMamba} 
& AUVs, surface buoys, edge processors, and cloud systems handling long-duration sensing streams
& Efficiency gains are task- and architecture-dependent. These models should be benchmarked against CNNs and transformers under realistic underwater datasets and hardware constraints \\
\bottomrule
\end{tabular}
\end{table*}

\subsection{Federated and Distributed Learning in IoUT SC}
The decentralized nature of IoUT makes federated and distributed learning important for SC. In FL, underwater nodes, AUVs, buoys, or coastal edge servers collaboratively train a semantic model while keeping raw data local. FedAvg is the canonical algorithm for communication-efficient learning from decentralized data \cite{McMahan2017FedAvg}. If node $i$ has local dataset $\mathcal{D}_i$, the global objective can be written as
\begin{equation}
    \min_{\theta} F(\theta)
    =
    \sum_{i=1}^{N}
    p_i F_i(\theta),
    \qquad
    p_i =
    \frac{|\mathcal{D}_i|}{\sum_{j=1}^{N}|\mathcal{D}_j|}.
    \label{eq:fl_objective}
\end{equation}
After local training, the server or aggregator updates the global model as
\begin{equation}
    \theta^{(r+1)}
    =
    \sum_{i \in \mathcal{S}_r}
    \frac{|\mathcal{D}_i|}
    {\sum_{j \in \mathcal{S}_r}|\mathcal{D}_j|}
    \theta_i^{(r+1)},
    \label{eq:fedavg_update}
\end{equation}
where $\mathcal{S}_r$ is the set of participating nodes at communication round $r$. In IoUT, this participation set may change due to AUV mobility, acoustic link outages, energy depletion, or scheduling constraints.

Statistical heterogeneity is a major challenge because underwater nodes may observe different species, seafloor types, turbidity levels, or infrastructure conditions. FedProx addresses heterogeneity by adding a proximal term to the local objective \cite{Li2020FedProx}:
\begin{equation}
    \min_{\theta_i}
    F_i(\theta_i)
    +
    \frac{\mu}{2}
    \left\|\theta_i-\theta^{(r)}\right\|_2^2.
    \label{eq:fedprox}
\end{equation}
Client drift is another issue under non-IID data; SCAFFOLD reduces this drift through control variates \cite{Karimireddy2020SCAFFOLD}. Adaptive federated optimization methods further modify server-side updates to improve convergence stability \cite{Reddi2021FedOpt}. FedBN addresses non-IID feature distributions using local batch-normalization statistics, which is relevant when underwater nodes observe different visual or acoustic domains \cite{Li2021FedBN}. These methods are important for IoUT because semantic models trained in one underwater region may drift when updated with data from another.

Distributed learning is also useful when no reliable central aggregator exists. In a decentralized setting, node $i$ can update its model using neighbor models:
\begin{equation}
    \theta_i^{(r+1)}
    =
    \sum_{j \in \mathcal{N}_i}
    W_{ij}\theta_j^{(r)}
    -
    \eta \nabla F_i(\theta_i^{(r)}),
    \label{eq:decentralized_update}
\end{equation}
where $\mathcal{N}_i$ is the neighborhood of node $i$, $W_{ij}$ is a mixing weight, and $\eta$ is the learning rate. Such decentralized learning may be suitable for AUV swarms or buoy-assisted networks where connectivity is intermittent. However, underwater channels can make repeated model exchange expensive, so model-update compression and event-triggered participation are necessary. Deep gradient compression has shown that distributed training traffic can be substantially reduced by transmitting only important gradient updates \cite{Lin2018DeepGradientCompression}, suggesting a useful direction for bandwidth-constrained semantic FL in IoUT.

Privacy and trust must also be considered. FL keeps raw data local, but model updates can still leak information. Secure aggregation protects individual model updates by allowing the server to observe only their aggregate \cite{Bonawitz2017SecureAggregation}. Broader FL surveys show that privacy, robustness, personalization, and systems heterogeneity remain open challenges even in terrestrial settings \cite{Kairouz2021FL}. For IoUT, these issues become more difficult because nodes may be physically exposed, links may be intermittent, and semantic messages may reveal sensitive mission information. Semantic-oriented FL for hybrid ground--aqua systems has already shown that distributed semantic learning is relevant to underwater communication \cite{Picano2024SemanticFL}, but practical SC-enabled IoUT still requires lightweight model updates, robust aggregation, privacy-aware participation, and energy-aware client selection.

\subsection{Key AI Models Enabling SC in IoUT}
The choice of AI model in SC-enabled IoUT should depend on deployment location, task risk, channel modality, and available energy budget. At underwater sensor nodes, lightweight CNNs, rule-based filters, compact vision transformers, and quantized models are often more feasible than large VLMs or diffusion models. Mobile-friendly models such as MobileViT illustrate how transformer-like representations can be adapted for constrained devices \cite{Mehta2022MobileViT}. At AUVs and surface buoys, transformer-based or state-space-based encoders may be practical when hardware acceleration is available. At cloud or shore stations, large VLMs and generative models can support semantic interpretation, visualization, and decision support.

A deployment-aware feasibility condition can be expressed as
\begin{equation}
    M_{\theta} + M_{\mathrm{act}} \leq M_{\mathrm{hw}},
    \qquad
    T_{\mathrm{inf}} + \frac{b_s}{R_m} + \tau(d) \leq T_{\max},
    \label{eq:model_feasibility}
\end{equation}
where $M_{\theta}$ is model memory, $M_{\mathrm{act}}$ is activation memory, $M_{\mathrm{hw}}$ is available hardware memory, $T_{\mathrm{inf}}$ is inference time, $b_s$ is semantic payload size, $R_m$ is the link rate under modality $m$, and $\tau(d)$ is propagation delay. The energy condition is similarly
\begin{equation}
    E_{\mathrm{enc}} + b_s e_{\mathrm{tx}}(d,m)
    <
    b_x e_{\mathrm{tx}}(d,m),
    \label{eq:model_energy_feasibility}
\end{equation}
where $E_{\mathrm{enc}}$ is semantic-encoding energy, $b_x$ is raw-data size, and $e_{\mathrm{tx}}(d,m)$ is per-bit transmission energy. These equations show that an AI model is useful for SC only when its computation cost is justified by communication savings and task improvement.

A practical model hierarchy for IoUT can therefore be organized into three levels. First, sensor-level models perform event-triggered sensing, thresholding, lightweight detection, and feature extraction. Second, AUV- or buoy-level models perform semantic fusion, mission-aware prioritization, and context-aware compression. Third, cloud-level models perform expensive reasoning, long-term learning, generative reconstruction, and cross-domain knowledge integration. This hierarchy avoids the unrealistic assumption that all underwater nodes can run large LLMs or diffusion models. It also supports graceful degradation: if cloud connectivity is lost, local semantic alerts can still be generated by compact edge models.

Recent underwater communication systems illustrate the direction deployment-aware learning is taking. Underwater optical semantic communication has demonstrated image-oriented semantic feature extraction over optical links \cite{Xu2024UWOSC}, while acoustic semantic communication studies show how compact textual semantics can reduce payloads for image transmission \cite{Zhang2025UASC}. AquaScope investigates generative image compression for underwater image transmission on mobile devices \cite{Tian2025AquaScope}, and AquaVLM explores context-aware underwater message generation using mobile VLMs \cite{Tian2025AquaVLM}. These works suggest that learning-driven SC is moving from concept to prototype, but they also reinforce the need for careful reporting of datasets, hardware, channel conditions, model size, inference latency, energy consumption, and task metrics.

Learning-driven SC can make IoUT communication more adaptive and task-aware, but it should be evaluated critically. Large models may improve semantic abstraction but increase memory and energy consumption. Generative models may improve visualization but introduce hallucination risks. FL may preserve local data while increasing model-update traffic. Knowledge-driven models may improve interpretability but require standard ontologies. Therefore, future SC-enabled IoUT systems should combine lightweight perception, domain knowledge, uncertainty estimation, model compression, and distributed training rather than relying on a single AI model family.

\section{Applications and Use Cases}
\label{sec04}
\begin{figure*}[t]
\centering
\includegraphics[width=0.975\textwidth]{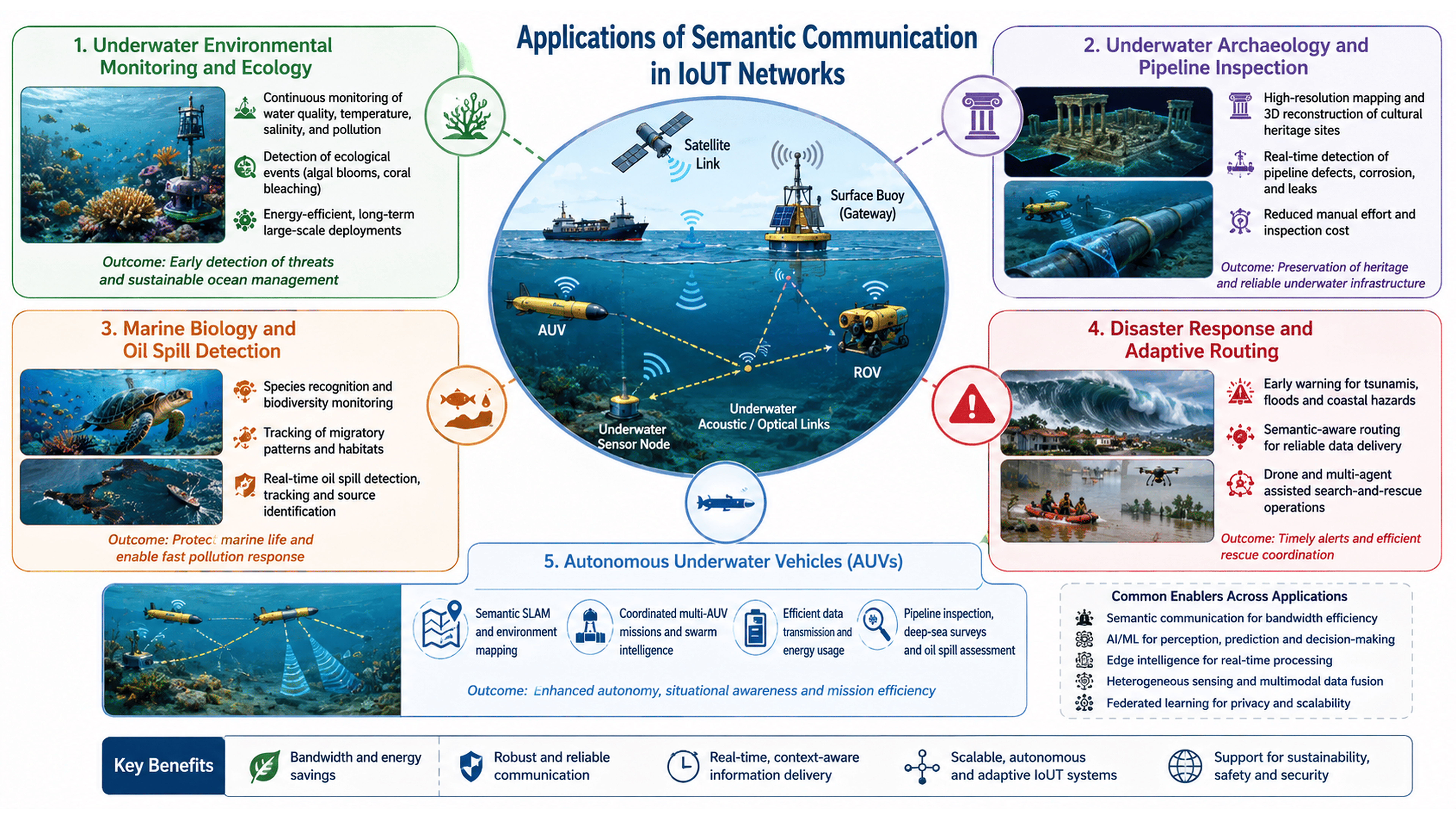}
\caption{Applications of SC in IoUT systems.}
\label{fig:applications}
\end{figure*}

SC can support IoUT applications by transmitting task-relevant meaning instead of continuously forwarding raw underwater observations. However, the role of SC differs across application domains. In environmental monitoring, the semantic output may be an anomaly alert; in marine biology, it may be a species label or a behavioral event; in infrastructure inspection, it may be a defect descriptor; and in AUV operations, it may be a semantic map update or a confidence-aware navigation cue. Therefore, SC-enabled IoUT should be evaluated using application-specific objectives rather than a single universal metric. Fig.~\ref{fig:applications} summarizes representative applications of SC in IoUT networks.

For an application domain $a$, a semantic message can be expressed as
\begin{equation}
    s_a = \{z_a,\ell,t,q,u,\mathcal{P}\},
    \label{eq:application_semantic_message}
\end{equation}
where $z_a$ denotes task-specific semantic content, $\ell$ is location, $t$ is timestamp, $q$ is confidence or uncertainty, $u$ is urgency, and $\mathcal{P}$ denotes provenance, security, or sensor-origin metadata. This representation is useful for IoUT because underwater data are often generated by heterogeneous sensing platforms, including cameras, sonars, hydrophones, chemical sensors, AUVs, ROVs, and seabed nodes \cite{Jahanbakht2021IoUTBigData}. In contrast to raw-data transmission, SC emphasizes whether the received semantic message reliably supports the intended action, such as triggering an alert, updating a map, prioritizing an AUV trajectory, or requesting additional measurements.

Table~\ref{tab:underwater_applications_use_cases} summarizes representative IoUT applications, the semantic units that may be transmitted, the role of SC in each domain, key risks, and suitable evaluation metrics. The table intentionally avoids direct cross-paper numerical performance comparisons because such values are often obtained using different datasets, channel models, hardware platforms, and evaluation protocols.

\begin{table*}[ht!]
\centering
\footnotesize
\caption{Representative applications of SC in IoUT networks: semantic units, design roles, risks, and evaluation metrics.}
\label{tab:underwater_applications_use_cases}
\renewcommand{\arraystretch}{1.25}
\small
\begin{tabular}{p{2.5cm} p{3.65cm} p{3.5cm} p{3.5cm} p{2.65cm}}
\toprule
\textbf{Application} 
& \textbf{Representative Semantic Units} 
& \textbf{Role of SC in IoUT} 
& \textbf{Main Risks / Constraints} 
& \textbf{Metrics and References} \\
\midrule

Environmental monitoring and ecology 
& Temperature/salinity anomalies, dissolved oxygen events, turbidity changes, habitat labels, coral/reef condition, pollutant alerts 
& Event-triggered sensing, selective reporting, semantic summarization of long-term observations, multimodal fusion of environmental and visual data
& Sensor drift, biofouling, sparse deployment, ecological false alarms, loss of long-term raw evidence
& AoI, event-detection accuracy, semantic freshness, energy per useful alert \cite{Felemban2015USNApplications,Jahanbakht2021IoUTBigData,Islam2020SUIM,Dunbabin2012RobotsEnvMonitoring} \\
\midrule

Underwater archaeology and cultural heritage 
& Artifact labels, site-region descriptors, geometric landmarks, photogrammetric features, uncertainty-tagged map updates 
& Transmit survey summaries, candidate artifact locations, and map updates rather than all imagery or sonar scans
& Generated reconstructions may not be valid archaeological evidence; raw data must be retained for archival and scientific verification
& 3D mapping quality, localization uncertainty, semantic-map consistency \cite{Menna2018Underwater3D,JohnsonRoberson2017UnderwaterMapping} \\
\midrule

Subsea infrastructure and pipeline inspection 
& Crack/corrosion labels, leakage cues, pipe centerline, defect location, severity level, inspection confidence, repair priority
& Prioritize defect-related information, transmit region-of-interest descriptors, and support AUV/ROV inspection decisions
& Poor visibility, turbidity, acoustic noise, false defect detection, high consequence of missed defects
& Detection accuracy, missed defect rate, localization error, confidence calibration \cite{Zhang2021PipelineLeak,Bobkov2023PipelineTracking,Li2024UATSSIC} \\
\midrule

Marine biology and biodiversity monitoring 
& Species labels, fish counts, behavior events, migration cues, habitat occupancy, rare-species alerts
& Convert long videos or image streams into compact ecological summaries and confidence-aware species observations
& Domain shift across habitats, class imbalance, low visibility, label uncertainty, limited annotated underwater datasets
& Species-recognition accuracy, F1-score, count error, uncertainty \cite{Saleh2022FishSurvey,Marrable2022FishLabeling,Wang2023VDMOSurvey,Ditria2020FishAbundance} \\
\midrule

Oil spill and pollutant response 
& Spill/no-spill label, slick boundary, source hypothesis, pollutant concentration event, dispersion direction, response priority
& Fuse satellite, surface, and underwater sensing into semantic alerts for coordinated response and monitoring 
& Synthetic Aperture Radar (SAR) look-alikes, weather effects, limited ground truth, cross-domain fusion between underwater and satellite data
& Intersection-over-union (IoU), false alarm rate, detection latency, segmentation reliability \cite{Solberg2012OilSpillRemoteSensing,Shaban2021OilSpillSAR,Dehghani2023OilSpillHybrid,Trujillo2024OilSpillTwoStep} \\
\midrule

Disaster response and adaptive routing 
& Tsunami/seismic alert, pressure anomaly, route priority, link-state summary, survivor/location cue, hazard level
& Prioritize urgent semantic packets, support delay-aware routing, and reduce non-critical traffic during emergency conditions
& Intermittent links, high latency, sparse sensors, uncertain hazard evolution, strict decision deadlines 
& Deadline miss ratio, semantic AoI, alert reliability, delivery probability \cite{Mori2022TsunamiReview,Rim2022GNSSTsunami,Li2024GNSSIRTsunami} \\
\midrule

AUV coordination and mission autonomy 
& Semantic landmarks, map updates, obstacle cues, mission intent, task allocation, local observations, uncertainty estimates
& Share compact semantic maps and mission-relevant observations among AUVs, buoys, and surface stations
& Global Positioning System (GPS)-denied navigation, energy limits, acoustic latency, map inconsistency, multi-agent coordination overhead
& Navigation error, map consistency, task-completion rate, energy per mission \cite{Wynn2014AUVGeoscience,Paull2014AUVNavigation,Zhang2023AUVNavigation,Zhou2023SubseaAUV,Ma2025AUVKeyTech} \\
\bottomrule
\end{tabular}
\end{table*}

\subsection{Underwater Environmental Monitoring and Ecology}
Underwater environmental monitoring requires continuous observation of physical, chemical, biological, and visual parameters, including temperature, salinity, pressure, dissolved oxygen, pH, turbidity, pollutant concentrations, and habitat conditions. UWSN applications have long emphasized environmental monitoring as a major use case because continuous human observation is impractical in remote marine environments \cite{Felemban2015USNApplications}. Robotic systems have also been widely studied for environmental monitoring because autonomous platforms can collect data in regions that are difficult, unsafe, or costly for humans to access \cite{Dunbabin2012RobotsEnvMonitoring}. In modern IoUT systems, the data volume increases further when fixed sensors are combined with underwater cameras, AUV surveys, hydrophones, and surface gateways \cite{Jahanbakht2021IoUTBigData}.

SC can improve such monitoring by converting continuous raw measurements into semantic events. For example, instead of transmitting every salinity or turbidity sample, a node may transmit a message such as ``rapid turbidity increase near reef sector A with confidence $q$'' when the event is relevant to the monitoring objective. Smart-ocean architectures already highlight the need for integrated sensing, communication, and data fusion in marine environments \cite{Qiu2020SmartOcean}; SC adds a decision layer by determining which observations are sufficiently meaningful to transmit under bandwidth and energy constraints. This is important because routine measurements may dominate the channel even when only rare environmental deviations require immediate attention.

Underwater visual monitoring also benefits from semantic extraction. The SUIM dataset provides a benchmark for semantic segmentation of underwater imagery into categories such as fish, reefs, aquatic plants, divers, robots, and seafloor \cite{Islam2020SUIM}. Underwater image enhancement models such as Water-Net \cite{Li2020WaterNet} and FUnIE-GAN \cite{Islam2020FUnIEGAN} can improve visual quality before semantic analysis, although image enhancement should not be confused with verified information recovery. In SC-enabled ecological monitoring, a camera node may therefore transmit a semantic segmentation mask, a habitat label, an event descriptor, or a region-of-interest descriptor instead of a full-resolution frame.

The main limitation is that ecological interpretation often requires long-term, auditable evidence. A semantic alert may be sufficient for real-time response, but scientific analysis may still require raw or lightly compressed data for calibration, trend analysis, and validation. Therefore, SC for environmental monitoring should combine selective semantic reporting with periodic raw-data sampling, calibration metadata, provenance information, and confidence-aware event descriptions. Such a hybrid strategy can reduce routine communication load while preserving enough evidence for later scientific verification.

\subsection{Underwater Archaeology and Pipeline Inspection}
Underwater archaeology requires accurate documentation of submerged cultural heritage while minimizing physical disturbance. Reviews of underwater three-dimensional (3D) recording show that photogrammetry, sonar, laser scanning, and robotic platforms are increasingly used for underwater cultural heritage documentation \cite{Menna2018Underwater3D}. Robotic mapping has also been demonstrated for archaeological sites using AUVs and stereo vision, showing how autonomous or semi-autonomous platforms can support large-area documentation \cite{JohnsonRoberson2017UnderwaterMapping}. In this setting, SC should not replace archival data acquisition; rather, it should support survey coordination by transmitting semantic map updates, candidate artifact locations, coverage gaps, uncertainty estimates, and inspection priorities.

A key point is that generative semantic reconstruction is not equivalent to archaeological evidence. A generated view may help operators understand a site layout, but final documentation should preserve provenance, sensor metadata, calibration parameters, and original measurements. Thus, semantic packets for archaeological applications should include uncertainty, timestamp, sensor type, acquisition conditions, and spatial reference information. This prevents a reconstructed or summarized output from being mistaken for a verified metric record. It also allows mission operators to request additional raw data or higher-resolution scans when a semantic summary is ambiguous.

Subsea pipeline inspection has a different but equally safety-critical objective. Pipeline leaks, corrosion, burial, and mechanical damage can be difficult to detect through periodic manual inspection. AUV-based leak inspection has been studied using multibeam echo sounder and forward-looking sonar data for pipeline following and anomaly detection \cite{Zhang2021PipelineLeak}. Stereo-image-based pipeline recognition and tracking have also been investigated to support AUV-based inspection \cite{Bobkov2023PipelineTracking}. SC can support these systems by transmitting defect-related semantic units, such as ``possible corrosion,'' ``pipe boundary detected,'' ``leakage cue,'' ``inspection confidence,'' and ``repair priority,'' rather than all raw imagery or sonar frames.

For cooperative simultaneous localization and mapping (SLAM) and infrastructure mapping, semantic segmentation-based underwater acoustic image transmission has been proposed to preserve task-relevant sonar information under limited acoustic bandwidth \cite{Li2024UATSSIC}. This direction is relevant to SC because the transmitted content is guided by downstream mapping or detection utility. However, pipeline inspection is a high-risk domain: a missed defect may have environmental, economic, and safety consequences. Therefore, SC should be designed with conservative decision thresholds, confidence reporting, redundancy for safety-critical alerts, and escalation mechanisms that request raw or higher-resolution data when uncertainty is high.

\subsection{Marine Biology and Oil Spill Detection}
Marine biology increasingly relies on large-scale visual and acoustic datasets collected by underwater cameras, baited remote underwater video systems, hydrophones, and mobile robots. DL surveys for fish classification show that underwater video analysis can reduce the burden of manual species identification, but performance is affected by water clarity, lighting, occlusion, class imbalance, and dataset bias \cite{Saleh2022FishSurvey}. Machine-assisted labeling has also been explored to accelerate species recognition from underwater video \cite{Marrable2022FishLabeling}. Object-detection-based methods have further shown that DL can support automated fish abundance estimation from underwater imagery \cite{Ditria2020FishAbundance}. A broader survey of DL-based visual detection of marine organisms emphasizes that underwater visual degradation remains a major challenge for automated ecological analysis \cite{Wang2023VDMOSurvey}.

In SC-enabled marine biology, the semantic message may include species identity, count estimate, behavior event, confidence score, location, and time window. This is more useful for low-bandwidth IoUT links than transmitting all video frames. For instance, an AUV or seabed camera may report ``high-confidence detection of target species near station 3'' or ``unusual aggregation pattern detected'' while storing raw imagery locally for later audit. SC therefore supports rapid ecological awareness while preserving the possibility of detailed offline scientific analysis. This balance is important because ecological conclusions often depend on long-term trends, species interactions, and sampling bias, not only on single-event detections.

Oil spill and pollutant response require broader cross-domain interpretation because observations may come from SAR imagery, aerial platforms, surface vessels, underwater sensors, and coastal monitoring stations. Remote sensing has long been used for ocean oil-spill monitoring because slicks can be observed over wide areas using satellite platforms \cite{Solberg2012OilSpillRemoteSensing}. DL frameworks have been used to detect and segment oil spills from SAR data \cite{Shaban2021OilSpillSAR}, and hybrid CNN--transformer networks have also been studied for oil-spill detection in SAR Earth observations \cite{Dehghani2023OilSpillHybrid}. More recent work evaluates two-step classification and segmentation models using Sentinel-1 SAR imagery \cite{Trujillo2024OilSpillTwoStep}. These studies are not IoUT communication systems by themselves, but they are relevant because SC-enabled IoUT can fuse remote-sensing outputs with underwater measurements and transmit compact response-oriented semantic alerts.

For oil-spill response, semantic messages should include not only a spill label but also boundary uncertainty, suspected source, drift direction, sensor modality, weather or sea-state context, and confidence. This is important because SAR dark regions may be confused with look-alike phenomena, such as low-wind areas, natural films, rain cells, or current-related surface patterns. Hence, SC should avoid overconfident binary alerts and instead support multi-sensor validation, uncertainty-aware response planning, and selective requests for additional data from AUVs, buoys, or surface vessels.

\subsection{Disaster Response and Adaptive Routing}

Disaster response in marine and coastal environments requires timely delivery of high-priority information under uncertain and degraded network conditions. Tsunami monitoring and early warning systems rely on seismic observations, ocean-bottom pressure sensors, coastal gauges, numerical models, and offshore monitoring networks \cite{Mori2022TsunamiReview}. Recent work has investigated Global Navigation Satellite System (GNSS)-based approaches for tsunami waveform forecasting \cite{Rim2022GNSSTsunami}, while island-based GNSS interferometric reflectometry (GNSS-IR) networks have been proposed as complementary tsunami detection and warning infrastructure \cite{Li2024GNSSIRTsunami}. These systems illustrate the importance of combining heterogeneous sensing modalities with rapid and reliable communication.

SC can contribute to such systems by prioritizing semantic messages that directly affect emergency decisions. A pressure anomaly, for example, may be encoded as an urgent semantic packet containing anomaly type, location, timestamp, confidence, and required response level. Compared with forwarding all raw measurements, this semantic representation is better suited for constrained underwater links when rapid decision support is required. However, raw observations remain important for model assimilation and verification, so SC should support layered reporting: urgent semantic alerts first, followed by additional data when bandwidth and time permit.

Adaptive routing is also critical during disasters because underwater links may be intermittent, damaged, or congested. In SC-enabled routing, packets should be scheduled according to semantic priority and deadline rather than only packet size or link quality. AoI is relevant in this context because stale hazard information may lose operational value even if it is eventually delivered \cite{Yates2021AoISurvey}. Semantic AoI extends this idea by considering whether the received content is still useful for the decision being made. For example, a delayed background salinity summary may remain useful for environmental analysis, while a delayed tsunami alert or AUV collision warning may not.

Federated and distributed learning can also support disaster response when local models must adapt without uploading raw data from every node. Semantic-oriented FL has been studied for hybrid ground--aqua computing systems \cite{Picano2024SemanticFL}, suggesting a path toward collaborative model updates across underwater, surface, and terrestrial components. Nevertheless, during disasters, communication resources are scarce, so model updates should not compete with urgent semantic alerts unless they directly improve response performance. A practical emergency design should therefore separate control-plane learning updates from high-priority semantic alerts and reserve communication resources for safety-critical traffic.

\subsection{Autonomous Underwater Vehicles}

AUVs are central to IoUT because they can collect data, act as mobile sensors, relay information, inspect infrastructure, and coordinate missions in regions where fixed infrastructure is sparse. AUVs have been widely used in marine geoscience, including seafloor mapping, environmental monitoring, and deep-sea observation \cite{Wynn2014AUVGeoscience}. AUV navigation and localization remain challenging because Global Positioning System (GPS) signals are unavailable underwater, acoustic localization is delayed and sparse, and inertial navigation accumulates error over time \cite{Paull2014AUVNavigation,Zhang2023AUVNavigation}. Subsea AUV technology reviews further identify communication, positioning, endurance, and maneuverability as key design challenges for long-term underwater deployment \cite{Zhou2023SubseaAUV}.

\begin{figure}[t]
\centering
\includegraphics[width=1.0\columnwidth]{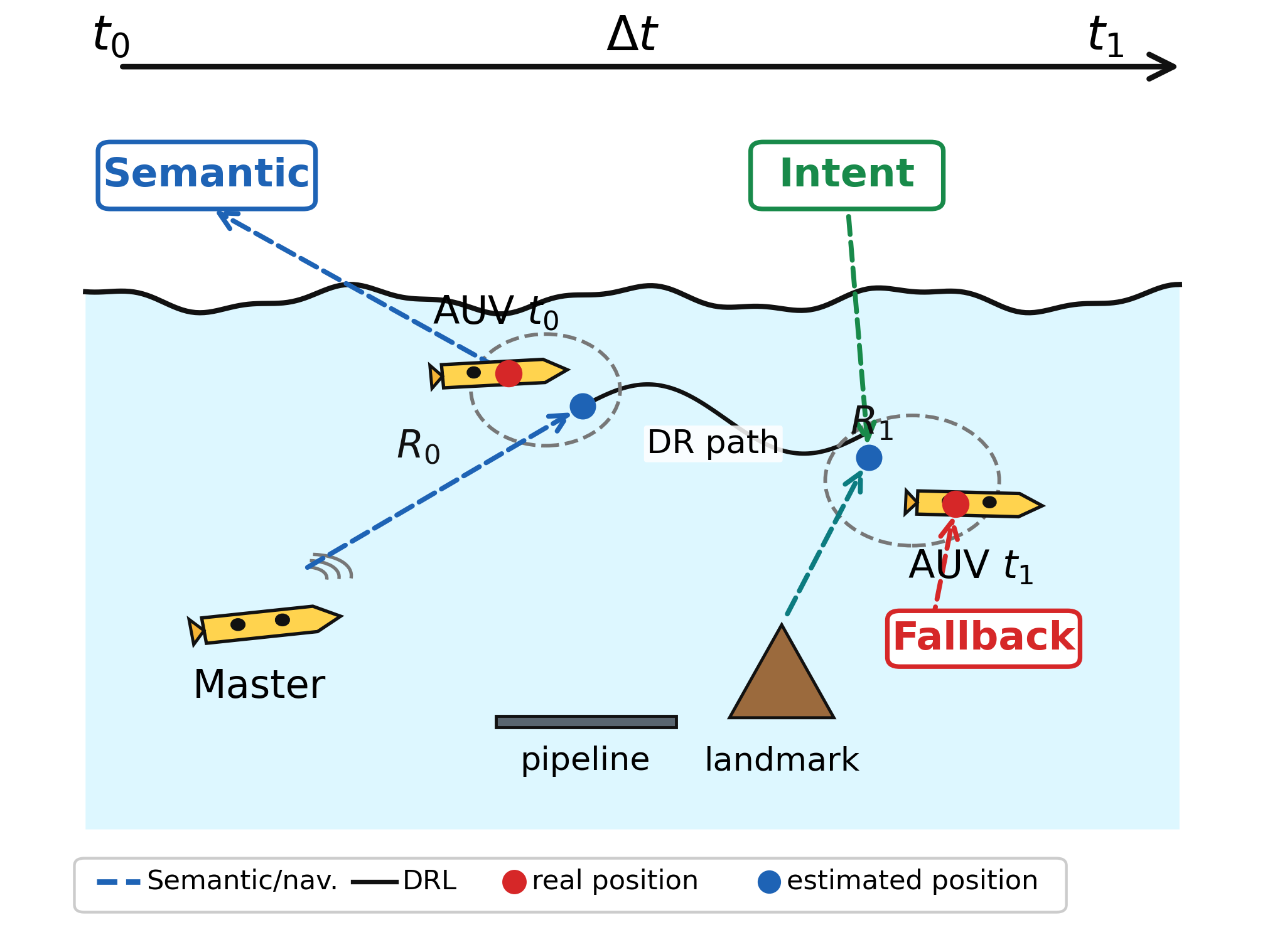}
\caption{SC-based AUV cooperation and localization in IoUT. Semantic messages support landmark-aware dead-reckoning correction, intent exchange, and fallback raw-data requests when uncertainty is high.}
\label{fig:auv_sc_localization}
\end{figure}

SC can support AUV operation in three main ways. First, AUVs can transmit semantic summaries of observations instead of full raw data. These summaries may include detected objects, semantic landmarks, inspection results, or map updates. Second, AUVs can exchange semantic intent messages, such as ``coverage completed,'' ``obstacle detected,'' or ``request high-resolution scan,'' to improve coordination among multiple vehicles. Third, AUVs can use semantic maps to support navigation and mission planning when acoustic localization is unreliable. In each case, SC reduces unnecessary communication while preserving the information needed for control, coordination, or decision-making. Fig.~\ref{fig:auv_sc_localization} illustrates how SC can support AUV cooperation by exchanging semantic summaries, intent messages, and landmark-based updates to improve localization and mission coordination under underwater communication constraints.

The communication benefit of SC is especially important for cooperative AUV missions. A state-of-the-art review of AUV technologies highlights the growing role of autonomy, sensing, underwater communication, and AI-enabled decision-making in future underwater systems \cite{Ma2025AUVKeyTech}. A broader survey of underwater vehicles for smart-ocean systems classifies AUVs, ROVs, unmanned surface vehicles, hybrid underwater vehicles, and supporting technologies, showing that future marine operations will likely involve heterogeneous robotic teams \cite{Xu2024UnderwaterVehicles}. In such teams, semantic map updates and task-level messages are often more useful than raw data streams because they reduce bandwidth usage and support faster coordination.

However, AUV-based SC should be designed conservatively. AUVs operate under strict energy budgets and may face strong currents, communication dropouts, navigation drift, and limited onboard computation. If a semantic model misclassifies an obstacle or suppresses a relevant observation, the mission may fail. Therefore, SC for AUVs should include confidence-aware decision-making, fallback communication modes, raw-data buffering for later validation, and task-dependent thresholds for requesting additional information. These safeguards are essential for using semantic information in safety-critical underwater autonomy rather than treating SC only as a bandwidth-saving mechanism.
\section{Future Research Directions}
\label{sec05}

\begin{figure*}[t]
\centering
\includegraphics[width=1.0\textwidth]{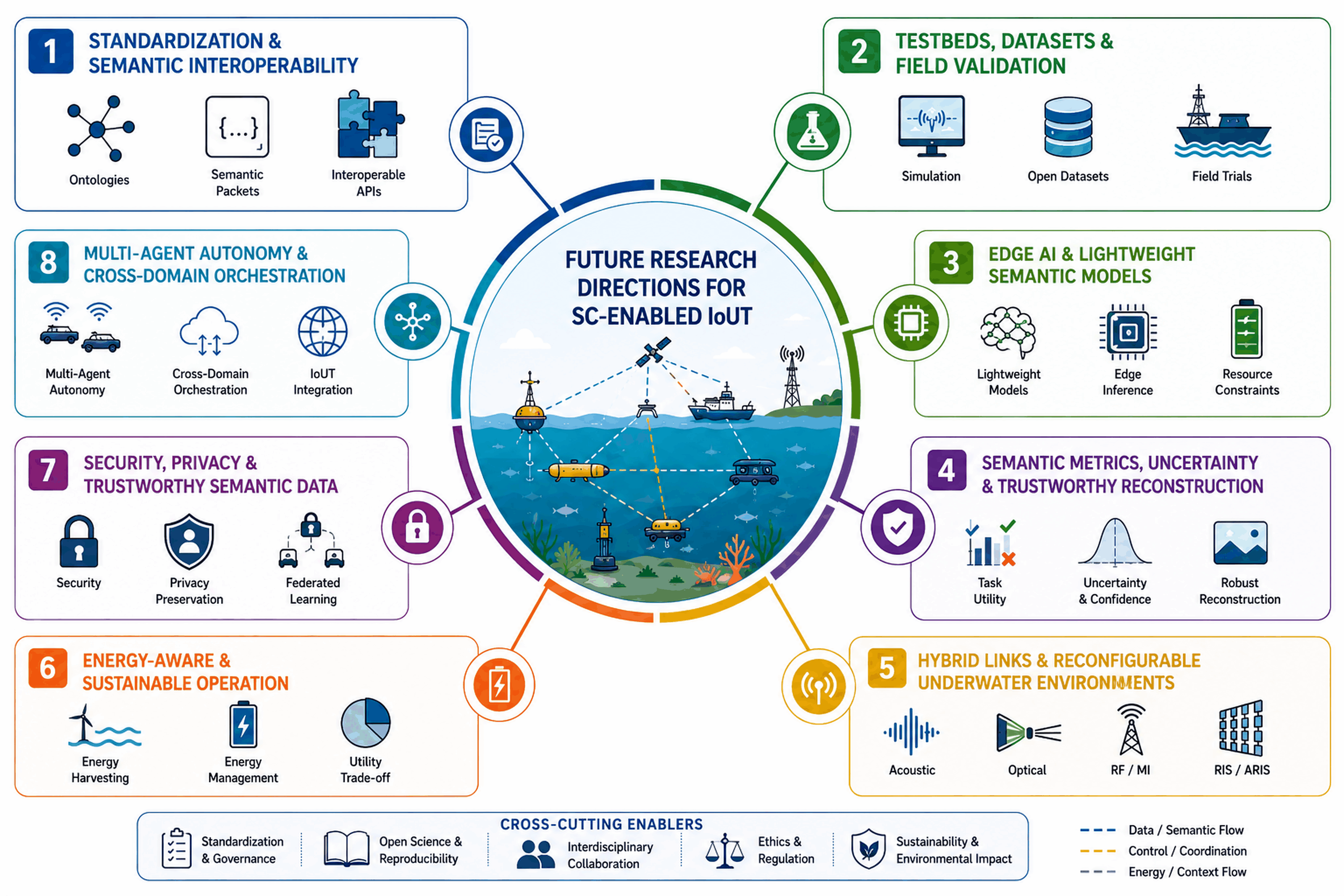}
\caption{Future research directions for SC-enabled IoUT systems.}
\label{fig:future_research_sc_iout}
\end{figure*}

Although SC offers a promising direction for improving bandwidth efficiency, energy sustainability, and task-oriented information delivery in IoUT systems, its practical deployment remains in its early stages. Existing studies are often evaluated under simplified channel models, limited datasets, idealized hardware assumptions, and application-specific scenarios. Therefore, future research must move from conceptual SC frameworks toward standardized, testable, resource-aware, secure, and trustworthy underwater systems. This section discusses key research directions required to advance SC-enabled IoUT networks from theoretical designs to practical marine deployments. Fig.~\ref{fig:future_research_sc_iout} provides a roadmap of the major research directions needed for practical, trustworthy, and sustainable SC-enabled IoUT systems. Table~\ref{tab:future_research_directions_sc_iout} summarizes these directions by highlighting current gaps, required research tasks, evaluation criteria, and representative references.

\begin{table*}[ht!]
\centering
\footnotesize
\caption{Future research directions for SC-enabled IoUT systems: gaps, research tasks, evaluation criteria, and representative references.}
\label{tab:future_research_directions_sc_iout}
\renewcommand{\arraystretch}{1.22}
\setlength{\tabcolsep}{3pt}
\begin{tabular}{p{3.0cm} p{4.0cm} p{4.5cm} p{3.25cm} p{2.0cm}}
\toprule
\textbf{Research Direction} 
& \textbf{Current Gap / Limitation} 
& \textbf{Key Research Tasks} 
& \textbf{Evaluation Criteria} 
& \textbf{Representative References} \\
\midrule

Standardized semantic representations and protocol interfaces
& IoUT systems still rely on heterogeneous metadata formats, device descriptions, sensing vocabularies, and application-specific semantic labels
& Develop lightweight semantic packet formats, underwater-domain ontologies, provenance-aware metadata, context models, and interoperable application programming interfaces for semantic data exchange
& Semantic consistency, interoperability, metadata overhead, parsing complexity, cross-domain compatibility
& \cite{OGCSensorThings2016,ETSISAREF2025,ETSINGSILD2021,W3CWoT2023,Lebo2013PROVO,Wilkinson2016FAIR} \\
\midrule

Reproducible testbeds, datasets, and field validation
& Many SC-enabled IoUT studies rely on simplified simulations, idealized channels, limited datasets, or hardware-independent assumptions
& Build open underwater semantic datasets, extend underwater simulators with semantic-layer modules, define repeatable acoustic/optical benchmarks, and validate SC models through field trials
& Reproducibility, channel realism, dataset diversity, field-test scalability, semantic outage, hardware-aware reporting
& \cite{Masiero2012DESERT,Guerra2009WOSS,Porter2011BELLHOP,vanWalree2017Watermark,Xie2009AquaSim,Martin2015AquaSimNG,Martins2014SUNRISE} \\
\midrule

Feasibility-aware edge AI and lightweight semantic models
& Large AI models may improve semantic extraction but often exceed underwater-node limits in memory, energy, and inference latency
& Design compact encoders, apply pruning, quantization, neural architecture search, TinyML, and hierarchical model placement across sensors, AUVs, buoys, and cloud servers
& Model size, memory footprint, inference latency, inference energy, task accuracy, communication savings
& \cite{Sze2017EfficientDNN,Howard2017MobileNets,Sandler2018MobileNetV2,Tan2019EfficientNet,Cai2019ProxylessNAS,Lin2020MCUNet,Warden2019TinyML} \\
\midrule

Semantic metrics, uncertainty, and trustworthy reconstruction
& Conventional metrics such as BER, throughput, and visual quality do not fully capture semantic correctness, task utility, or reconstruction risk
& Develop task-aware semantic metrics, semantic outage models, uncertainty-calibrated semantic packets, robustness benchmarks, and guidelines for generative reconstruction
& Semantic fidelity, task utility, uncertainty calibration, hallucination risk, robustness under domain shift, semantic reliability
& \cite{Guo2017Calibration,Gal2016Dropout,Lakshminarayanan2017DeepEnsembles,Ovadia2019Uncertainty,Hendrycks2019Robustness,Gawlikowski2023UncertaintySurvey,Ji2023HallucinationSurvey,Zhong2024TheorySurvey} \\
\midrule

Hybrid links, cross-media connectivity, and reconfigurable environments
& Acoustic, optical, RF, MI, and cross-media links differ in range, latency, bandwidth, energy cost, and environmental sensitivity; existing designs rarely optimize for semantic utility
& Develop semantic-aware link selection, acoustic--optical switching, cross-media semantic gateways, RIS-assisted links, and semantic utility-driven channel adaptation
& Semantic utility, link availability, latency, energy per semantic message, coverage, robustness under turbidity and mobility
& \cite{Hanson2008UWOC,Jamali2018UWOCFading,Tonolini2018CrossBoundary,Sun2022ARIS,Wang2023ARIS,Luo2024UARIS,Luo2025MLARIS} \\
\midrule

Energy-aware and sustainable semantic operation
& Semantic extraction may reduce transmission load but can increase local inference, decoding, and model-update energy
& Develop energy-causal semantic scheduling, adaptive semantic granularity, energy-harvesting-aware SC, ultra-low-power semantic sensing, and backscatter-assisted semantic messaging
& Node lifetime, energy neutrality, semantic bits/Joule, inference energy, harvested-energy stability, mission completion rate
& \cite{Khan2022EnergyHarvesting,Alamu2023EnergyHarvesting,Nordfjord2025OceanEnergy,Faria2022UnderwaterEnergyHarvesting,Jang2019UnderwaterBackscatter,Ghaffarivardavagh2020UBL,Afzal2022BatteryFreeImaging,Bereketli2022BackscatterCoverage} \\
\midrule

Security, privacy, and trustworthy semantic data
& Semantic packets may reveal mission intent, asset status, vehicle location, or environmental-event meaning even when raw data are not transmitted
& Design semantic authentication, provenance-aware access control, secure aggregation, privacy-preserving learning, and defenses against semantic manipulation and model poisoning
& Semantic integrity, privacy leakage, attack detection rate, trust score, robustness to poisoning, provenance correctness
& \cite{Dwork2006DifferentialPrivacy,Abadi2016DeepLearningDP,Shokri2017MembershipInference,Nasr2019PrivacyAnalysis,Zhu2019DeepLeakage,Bagdasaryan2020BackdoorFL,Blanchard2017Byzantine,Mothukuri2021FLSecuritySurvey} \\
\midrule

Digital twins, multi-agent autonomy, and cross-domain orchestration
& IoUT systems require coordination among underwater nodes, AUVs, surface gateways, cloud platforms, and maritime stakeholders, but semantic state synchronization remains immature
& Build semantic digital twins, integrate semantic maps into AUV mission planning, support multi-agent coordination, and connect underwater semantic packets to digital ocean infrastructures
& Twin fidelity, synchronization latency, semantic-map consistency, coordination efficiency, mission success rate, cross-domain interoperability
& \cite{Tzachor2023DigitalTwinOcean,Wang2025DTUnderwaterIoT,Barbie2021UnderwaterDT,Kaklis2023MaritimeDT,OGCILIAD2023,EuropeanCommissionDTO2024} \\
\bottomrule
\end{tabular}
\end{table*}

\subsection{Standardized Semantic Representations and Protocol Interfaces}

A major barrier to practical SC-enabled IoUT is the absence of standardized semantic representations for underwater data. Existing IoUT devices may describe temperature, salinity, sonar objects, AUV trajectories, ecological events, or infrastructure defects using different naming conventions, metadata structures, and data formats. This fragmentation complicates interoperability among underwater sensors, AUVs, surface gateways, cloud platforms, and domain-specific applications. For sensor interoperability, the Open Geospatial Consortium (OGC) SensorThings application programming interface (API) provides a geospatially enabled framework for connecting IoT devices, observations, and applications \cite{OGCSensorThings2016}. However, IoUT-specific SC requires extending such frameworks to include semantic priority, task relevance, confidence, uncertainty, provenance, and underwater channel context. The FAIR principles also emphasize that scientific data should be findable, accessible, interoperable, and reusable, which is directly relevant to semantic data sharing across marine stakeholders \cite{Wilkinson2016FAIR}.

Future SC-enabled IoUT protocols should support a common semantic packet structure. A possible representation is
\begin{equation}
    p_s = \{s,g,\ell,t,q,u,\mathcal{C},\mathcal{P}\},
    \label{eq:future_semantic_packet}
\end{equation}
where $s$ denotes the semantic payload, $g$ is the mission goal, $\ell$ is location, $t$ is timestamp, $q$ is confidence, $u$ is urgency, $\mathcal{C}$ represents environmental or channel context, and $\mathcal{P}$ stores provenance/security metadata. The Smart Applications REFerence (SAREF) ontology provides a useful starting point for semantic interoperability in IoT applications \cite{ETSISAREF2025}, while Next Generation Service Interfaces--Linked Data (NGSI-LD) offers a context-information model and API for publishing, querying, and subscribing to context data across heterogeneous systems \cite{ETSINGSILD2021}. For IoUT, these ideas should be adapted to represent underwater-specific concepts such as acoustic-link quality, turbidity level, AUV mission phase, depth-dependent channel state, semantic freshness, and sensor confidence.

Another important direction is the development of shared vocabularies for marine events and underwater tasks. The Web of Things (WoT) Thing Description can help describe device capabilities, interaction affordances, and service endpoints \cite{W3CWoT2023}, while the PROV Ontology (PROV-O) provides a standard mechanism for representing provenance across systems \cite{Lebo2013PROVO}. Such provenance is critical in SC because a compact semantic alert without sensor origin, processing history, confidence, and acquisition context may be difficult to trust. For ecology and biodiversity-related IoUT applications, community standards such as Darwin Core provide useful vocabulary elements for species observations, occurrence records, and sampling events \cite{Wieczorek2012DarwinCore}. JavaScript Object Notation for Linked Data (JSON-LD) can support lightweight linked-data serialization of semantic messages \cite{Sporny2020JSONLD}, while Resource Description Framework (RDF) 1.1 \cite{Beckett2014RDF11} and Web Ontology Language (OWL) 2 \cite{Hitzler2012OWL2} can support machine-interpretable semantic relations. A practical research challenge is to develop underwater-compatible profiles of these standards so that interoperability can be achieved without excessive packet overhead.

\subsection{Reproducible Testbeds, Datasets, and Field Validation}

A second major research direction is realistic validation. Many SC studies rely on simulations, idealized channel assumptions, or terrestrial datasets, which limits confidence in their suitability for underwater deployment. The DESERT Underwater framework was designed to support simulation, emulation, and realization of underwater network protocols \cite{Masiero2012DESERT}, while the World Ocean Simulation System (WOSS) allows underwater network simulations to incorporate realistic acoustic propagation modeling \cite{Guerra2009WOSS}. Future SC studies should use such tools not only for physical-layer simulation, but also for semantic-layer evaluation, including task accuracy, semantic outage, semantic AoI, confidence calibration, semantic priority handling, and compute--communication trade-offs.

Channel modeling must also be made more reproducible. BELLHOP is widely used for underwater acoustic ray tracing and propagation analysis \cite{Porter2011BELLHOP}, and the Watermark benchmark provides realistic and repeatable acoustic channels for comparing modulation schemes \cite{vanWalree2017Watermark}. These tools could be extended to benchmark semantic transmission under controlled but realistic conditions. For example, one benchmark could evaluate how semantic descriptors degrade under Doppler spread, packet loss, multipath fading, and long propagation delay, while another could test whether receiver-side task decisions remain correct when semantic packets are partially corrupted. Such benchmarks would help avoid misleading comparisons among SC methods evaluated under incompatible channel models.

Simulation platforms should also support AI-native protocols. Aqua-Sim was developed as an NS-2-based underwater network simulator \cite{Xie2009AquaSim}, and Aqua-Sim Next Generation transitioned to an NS-3-based underwater network simulator \cite{Martin2015AquaSimNG}. Future platforms should integrate neural encoders, semantic routers, FL modules, uncertainty estimators, and edge-inference energy models. Field testbeds remain equally important: the SUNRISE Porto testbed demonstrated the value of federated underwater experimentation for future Internet technologies \cite{Martins2014SUNRISE}. Digital-twin-assisted underwater field deployments, such as the Baltic Sea ocean-observation network reported by Barbie \textit{et al.} \cite{Barbie2021UnderwaterDT}, also suggest a path toward reproducible evaluation pipelines that combine real measurements, simulation, emulation, and virtual replay.

Future experimental reporting should include dataset characteristics, channel model, water conditions, deployment depth, node mobility, sensor type, semantic representation type, model size, inference latency, energy consumption, hardware platform, and task-specific metrics. Without such reporting, performance claims cannot be fairly compared across SC-enabled IoUT studies. Reproducibility should also include open code, parameter settings, and data-processing pipelines whenever possible, because small changes in underwater image enhancement, sonar preprocessing, or semantic labeling can substantially affect reported task accuracy.

\subsection{Feasibility-Aware Edge AI and Lightweight Semantic Models}
SC-enabled IoUT will not be practical unless semantic encoders can operate under underwater hardware constraints. Efficient DL processing has been studied extensively for embedded systems \cite{Sze2017EfficientDNN}, but underwater nodes face additional limitations due to expensive communication, difficult battery replacement, and restricted physical access. Future research should therefore move away from assuming that large transformers, LLMs, or diffusion models can run on every underwater node. Instead, semantic processing should be hierarchically distributed across sensor nodes, AUVs, buoys, edge gateways, shore stations, and cloud servers.

A feasibility-aware SC model should satisfy
\begin{equation}
    M_{\theta} + M_{\mathrm{act}} \leq M_{\mathrm{hw}}, \qquad
    E_{\mathrm{enc}} + b_s e_{\mathrm{tx}}(d,m) < b_x e_{\mathrm{tx}}(d,m),
    \label{eq:future_feasibility}
\end{equation}
where $M_{\theta}$ is model memory, $M_{\mathrm{act}}$ is activation memory, $M_{\mathrm{hw}}$ is available hardware memory, $E_{\mathrm{enc}}$ is semantic-encoding energy, $b_s$ is semantic payload size, $b_x$ is raw payload size, and $e_{\mathrm{tx}}(d,m)$ is per-bit transmission energy for distance $d$ and modality $m$. This condition emphasizes that SC is useful only when computation cost is justified by communication savings and task improvement. For low-power underwater nodes, this balance is often more important than raw model accuracy.

Lightweight model families are therefore important. MobileNets introduced depthwise separable convolutions for efficient vision models \cite{Howard2017MobileNets}, and MobileNetV2 improved mobile inference through inverted residual blocks \cite{Sandler2018MobileNetV2}. EfficientNet uses compound scaling to balance model depth, width, and resolution \cite{Tan2019EfficientNet}, while ProxylessNAS searches neural architectures directly on target hardware \cite{Cai2019ProxylessNAS}. TinyML research further demonstrates that ML inference can be deployed on ultra-low-power microcontrollers when models and runtimes are carefully co-designed \cite{Warden2019TinyML}. For extremely constrained underwater devices, binary or low-precision models such as XNOR-Net \cite{Rastegari2016XNORNet} and BinaryConnect \cite{Courbariaux2015BinaryConnect} may reduce memory and arithmetic cost, although their semantic accuracy must be validated under underwater degradation. MCUNet further demonstrates that neural architecture search and inference-runtime co-design can enable DL on microcontrollers \cite{Lin2020MCUNet}. Future IoUT research should adapt these approaches to underwater sensing tasks and explicitly report the trade-off among model size, semantic accuracy, inference latency, power consumption, and node lifetime.

Another important issue is model placement. Sensor-level models may perform event-triggered sensing and lightweight semantic filtering; AUV-level models may perform multimodal fusion and mission-aware prioritization; buoy-level models may aggregate data from multiple underwater nodes; and cloud-level models may perform expensive reasoning or generative visualization. This hierarchical design avoids unrealistic assumptions about underwater edge hardware while still allowing sophisticated semantic processing where resources are available. Future work should therefore treat model placement, semantic compression ratio, and link modality as a joint optimization problem rather than independent design choices.

\subsection{Semantic Metrics, Uncertainty, and Trustworthy Reconstruction}
Another open challenge is the absence of universally accepted metrics for semantic correctness in IoUT. Physical-layer metrics such as BER, throughput, and PDR remain useful, but they do not measure whether a semantic alert, object label, map update, or mission decision is correct. Future work should develop task-dependent semantic metrics that jointly account for meaning preservation, decision utility, uncertainty, freshness, and safety risk. Such metrics should be tied to the application: a semantic error in habitat classification may be tolerable for long-term monitoring, while a semantic error in AUV collision avoidance or pipeline-defect detection may be unacceptable.

For safety-critical applications, uncertainty calibration is essential. Neural-network calibration studies show that high prediction confidence does not always imply correctness \cite{Guo2017Calibration}. Bayesian approximation through dropout provides one approach for uncertainty estimation in DL \cite{Gal2016Dropout}, while deep ensembles offer a scalable method for predictive uncertainty estimation \cite{Lakshminarayanan2017DeepEnsembles}. Large-scale evaluations also show that model uncertainty can degrade under distribution shift \cite{Ovadia2019Uncertainty}. Since underwater data are affected by turbidity, lighting, acoustic noise, sensor drift, biofouling, and environmental variability, SC systems should report uncertainty together with semantic messages.

A practical semantic confidence condition can be expressed as
\begin{equation}
    \mathrm{Transmit}(s)=
    \begin{cases}
    1, & q(s) \geq q_{\min}(g) \ \mathrm{or}\ u(s)\geq u_{\min}(g),\\
    0, & \mathrm{otherwise},
    \end{cases}
    \label{eq:semantic_confidence_transmission}
\end{equation}
where $q(s)$ is the confidence of semantic content $s$, $u(s)$ is urgency, and $q_{\min}(g)$ and $u_{\min}(g)$ are task-dependent thresholds. This formulation helps distinguish routine ecological summaries from urgent disaster, inspection, or navigation messages. It also implies that SC should support uncertainty-aware prioritization, not only data compression.

Robustness evaluation should include corruption and domain-shift tests. Benchmarking work on common corruptions shows that neural models can be sensitive to perturbations that are not reflected by clean-test accuracy \cite{Hendrycks2019Robustness}. A broader review of uncertainty in DL emphasizes the need to distinguish aleatoric and epistemic uncertainty \cite{Gawlikowski2023UncertaintySurvey}, and computer-vision work has shown that modeling both uncertainty types can be useful for perception tasks \cite{Kendall2017Uncertainties}. For generative SC, hallucination risk is especially important: surveys of natural-language generation show that generated outputs may be fluent yet unsupported by the source \cite{Ji2023HallucinationSurvey}. Similarly, a generated underwater image may appear plausible yet be incorrect for inspection or scientific analysis. Therefore, generative reconstruction should be evaluated as task-oriented visualization rather than exact recovery. Future semantic information theory should also clarify semantic rate, semantic distortion, semantic outage, and semantic capacity under task constraints; recent theoretical surveys identify these as unresolved issues in SC \cite{Zhong2024TheorySurvey}.

\subsection{Hybrid Links, Cross-Media Connectivity, and Reconfigurable Underwater Environments}

Future SC-enabled IoUT systems will likely rely on hybrid acoustic, optical, RF, MI, and cross-media links. Optical links can support high data rates over short distances, as shown in early high-bandwidth underwater optical communication experiments \cite{Hanson2008UWOC}. However, underwater optical channels are sensitive to absorption, scattering, air bubbles, temperature gradients, and salinity variations, and experimental fading studies highlight the need for realistic channel-aware designs \cite{Jamali2018UWOCFading}. Acoustic links remain useful for long-range connectivity, while optical links may be used for short-range semantic updates, AUV docking, high-priority visual descriptors, or local data offloading. RF and MI links may be suitable for specialized near-surface or short-range scenarios, but their semantic utility must be evaluated together with range, power, and hardware constraints.

Cross-media communication is another important future direction. Tonolini and Adib demonstrated that wireless communication across the water--air boundary requires special treatment because radio and acoustic waves behave differently at the interface \cite{Tonolini2018CrossBoundary}. In SC-enabled IoUT, surface buoys and gateway nodes should therefore perform not only protocol conversion but also semantic translation between underwater networks and terrestrial/cloud systems. This translation may involve changing payload granularity, confidence representation, semantic priority, or task context. For example, an underwater node may transmit a compact alert to a buoy, while the buoy may enrich that message with geospatial metadata, satellite context, or historical sensor records before forwarding it to a cloud platform.

RISs and acoustic intelligent surfaces may also reshape underwater communication. Sun \textit{et al.} proposed acoustic RISs for high-data-rate and long-range underwater communications \cite{Sun2022ARIS}, while Wang \textit{et al.} studied the design of acoustic RISs for underwater communication links \cite{Wang2023ARIS}. More recent underwater acoustic RIS (UA-RIS) studies have begun to shift from principles toward experimental prototypes \cite{Luo2024UARIS}, and multilayer acoustic RIS designs have been proposed for beam steering and passive modulation \cite{Luo2025MLARIS}. For SC, the open question is not only whether RISs improve SNR, but whether they improve semantic utility per unit energy. A semantic-aware link-selection objective can be written as
\begin{equation}
    m^{\star}=\arg\max_{m\in\mathcal{M}}
    \left[
    U_g^{(m)}-\lambda_E E_m-\lambda_T T_m-\lambda_D d_s^{(m)}
    \right],
    \label{eq:future_hybrid_link_selection}
\end{equation}
where $U_g^{(m)}$ is task utility under modality $m$, $E_m$ is energy cost, $T_m$ is latency, and $d_s^{(m)}$ is semantic distortion. Future hybrid-link research should optimize this semantic utility rather than only physical throughput, received power, or channel rate.

\subsection{Energy-Aware and Sustainable Semantic Operation}

Energy remains one of the most important constraints for SC-enabled IoUT. Energy harvesting in underwater acoustic sensor networks has been surveyed as a way to extend network lifetime and support long-term deployments \cite{Khan2022EnergyHarvesting}. Energy harvesting techniques for underwater wireless communication networks include wave, current, thermal-gradient, microbial, piezoelectric, and electromagnetic mechanisms \cite{Alamu2023EnergyHarvesting}. More recent work on powering underwater robotic sensor networks emphasizes that the suitability of ocean energy harvesting and underwater wireless power transfer depends strongly on device type, depth, mobility, mission duration, and environmental conditions \cite{Nordfjord2025OceanEnergy}. Therefore, SC should be designed together with energy harvesting, duty cycling, wake-up communication, and mission-aware sensing.

Future SC systems should use energy-aware semantic scheduling. If a node has residual energy $e_i(t)$ and harvests energy $h_i(t)$, a semantic transmission policy should satisfy
\begin{equation}
    \sum_{t=1}^{T}
    \left(E_{\mathrm{sense}}(t)+E_{\mathrm{enc}}(t)+E_{\mathrm{tx}}(t)\right)
    \leq
    e_i(0)+\sum_{t=1}^{T} h_i(t).
    \label{eq:energy_causality}
\end{equation}
This constraint prevents semantic inference from consuming energy that should be reserved for sensing, localization, emergency communication, or recovery procedures. In practice, semantic granularity may need to adapt to available energy: a node with high energy may transmit richer semantic descriptions, while a low-energy node may transmit only critical event labels.

Underwater energy harvesting for submersible sensors has been studied to extend operational time \cite{Faria2022UnderwaterEnergyHarvesting}, but energy harvesting alone does not eliminate the need for energy-aware SC. Battery-free and near-zero-power systems offer another direction. Piezo-acoustic backscatter networking enables underwater communication by reflecting acoustic signals rather than generating new active transmissions \cite{Jang2019UnderwaterBackscatter}. Underwater backscatter localization extends this idea toward battery-free localization \cite{Ghaffarivardavagh2020UBL}, while battery-free underwater imaging demonstrates that acoustic energy harvesting and backscatter can support low-power visual sensing \cite{Afzal2022BatteryFreeImaging}. Coverage and source deployment for underwater acoustic backscatter networks have also been studied \cite{Bereketli2022BackscatterCoverage}. Future SC should investigate how semantic extraction can be simplified enough to operate with such ultra-low-power devices, for example by transmitting event-level labels, threshold-crossing alerts, acoustic signatures, or compressed binary semantic states.

\subsection{Security, Privacy, and Trustworthy Semantic Data}

SC introduces new security and privacy challenges because a compact semantic message may reveal sensitive meaning even when raw data are not transmitted. A semantic packet indicating ``pipeline defect,'' ``AUV location,'' or ``restricted-area activity'' may be more sensitive than a noisy raw signal. Therefore, future SC-enabled IoUT security should protect the meaning, provenance, confidence, and decision implications of information, not only the bitstream. This requires threat models that explicitly consider semantic leakage, semantic manipulation, confidence spoofing, and provenance forgery.

Differential privacy provides a principled framework for limiting leakage of individual data contributions \cite{Dwork2006DifferentialPrivacy}. DL with differential privacy further shows how privacy-preserving training can be implemented through gradient clipping and noise addition \cite{Abadi2016DeepLearningDP}. However, ML models can still leak sensitive information through membership inference attacks \cite{Shokri2017MembershipInference}, white-box privacy attacks \cite{Nasr2019PrivacyAnalysis}, or gradient leakage during distributed learning \cite{Zhu2019DeepLeakage}. This is relevant for IoUT because federated or distributed semantic models may be trained across AUVs, buoys, seabed nodes, and shore stations. Future systems should therefore combine lightweight privacy-preserving learning with semantic access control and provenance-aware auditing.

Trustworthy aggregation is another challenge. FL can be vulnerable to backdoor or model-replacement attacks, as shown by Bagdasaryan \textit{et al.} \cite{Bagdasaryan2020BackdoorFL}. Byzantine-resilient learning methods such as Krum address malicious or faulty workers in distributed learning \cite{Blanchard2017Byzantine}. In SC-enabled IoUT, semantic attacks may be even more subtle: an attacker may not corrupt many bits but may change the meaning of a semantic label, confidence score, or provenance field. Adversarial examples show that small perturbations can mislead neural models \cite{Goodfellow2015Adversarial}, and FL security surveys emphasize that poisoning, inference, and availability attacks must be handled together \cite{Mothukuri2021FLSecuritySurvey}. Threat analyses of FL further show that privacy and robustness cannot be treated independently \cite{Truong2021FLThreats}. Future semantic security should therefore define threat models for semantic manipulation, semantic eavesdropping, confidence spoofing, provenance forgery, and model-update poisoning, and should evaluate defense mechanisms under realistic underwater communication constraints.

\subsection{Digital Twins, Multi-Agent Autonomy, and Cross-Domain Orchestration}

Digital twins can provide a bridge between physical IoUT deployments and semantic decision-making. Ocean digital twins have been proposed as tools for sustainable marine management, planning, and forecasting \cite{Tzachor2023DigitalTwinOcean}. In SC-enabled IoUT, a digital twin can store semantic maps, link-quality histories, AUV trajectories, sensor confidence, environmental context, mission plans, and uncertainty estimates. Such a twin can then simulate alternative communication, routing, and sensing policies before applying them in the physical network. This capability is particularly important for underwater systems because field testing is expensive and failures may be difficult to recover.

System-level digital twin modeling for underwater wireless IoT networks has recently been explored with FL and multi-server cooperation \cite{Wang2025DTUnderwaterIoT}. Field-oriented digital-twin prototypes for underwater observation systems also show how real deployments can be connected with virtual models for testing and coordination \cite{Barbie2021UnderwaterDT}. For broader maritime systems, AI-enabled digital twins have been proposed to support sensing, control, and decision-making across the vessel life cycle \cite{Kaklis2023MaritimeDT}. These ideas are relevant to SC because semantic representations can serve as the interface between physical observations and digital-twin reasoning. For example, a digital twin does not always require raw sonar frames; it may require semantic map updates, confidence-aware object labels, estimated link states, and event-level summaries.

Cross-domain orchestration will also require open marine data infrastructures. The ILIAD Digital Twin of the Ocean initiative emphasizes semantic standards, OGC APIs, and interoperable data models for real-time marine insights \cite{OGCILIAD2023}. The European Digital Twin Ocean similarly aims to integrate data from satellites, sensors, models, and citizen observations to support ocean monitoring and prediction \cite{EuropeanCommissionDTO2024}. For SC-enabled IoUT, the future challenge is to connect underwater semantic packets to large-scale digital ocean infrastructures while preserving uncertainty, provenance, security, and application-specific meaning. This requires coordination among underwater network designers, robotics researchers, ocean scientists, standardization bodies, and domain users.

Conclusively, the next generation of SC-enabled IoUT should be standardized, testable, energy-aware, secure, and interoperable. Progress will require not only better AI models, but also realistic field validation, semantic metrics, lightweight edge inference, hybrid-link orchestration, trustworthy semantic data management, and cross-domain digital-twin integration.

\section{Limitations}
\label{sec06}
This survey provides a feasibility-aware synthesis of SC for IoUT systems, but several limitations should be acknowledged. 
\begin{itemize}
    \item First, SC for underwater networks is still an emerging research area, and many existing studies are conceptual, simulation-based, or limited to early prototypes. As a result, reported gains across different works cannot be directly compared without considering differences in datasets, channel models, communication modalities, hardware assumptions, semantic representations, and evaluation metrics. Therefore, the comparative discussions in this paper focus on design principles, trade-offs, and feasibility considerations rather than ranking individual methods purely by reported numerical performance.
    \item Second, this paper does not present a complete experimental benchmark or a new underwater prototype. The illustrative simulation results included in the paper are intended to clarify metric behavior and design trade-offs, such as task utility, semantic reliability, and energy-normalized semantic efficiency, rather than to claim measured field performance.
    \item Third, semantic utility is inherently task-dependent: a representation that is sufficient for ecological monitoring may be inadequate for AUV navigation or pipeline inspection. Hence, no single semantic metric can universally evaluate all SC-enabled IoUT applications.
    \item Finally, the rapid development of AI models, underwater sensing platforms, and semantic-networking standards means that SC-enabled IoUT remains a moving target. 
\end{itemize}
These limitations reinforce the need for standardized datasets, reproducible field testbeds, hardware-aware evaluation, and application-specific semantic metrics.

\section{Conclusion}
\label{sec07}
This paper reviewed SC for IoUT, emphasizing its role in enabling meaning-driven underwater communication under limited bandwidth, long propagation delay, time-varying channels, intermittent connectivity, and strict energy constraints. Unlike conventional bit-oriented communication, SC prioritizes task-relevant information, including semantic descriptors, anomaly alerts, confidence scores, and decision-oriented messages. The survey covered SC fundamentals, layered architectures, semantic channel modeling, evaluation metrics, learning-driven techniques, and representative applications, including environmental monitoring, marine ecology, subsea inspection, disaster response, and autonomous underwater vehicle coordination. The main insight is that SC can reduce unnecessary transmissions and improve task-oriented communication, but its benefits depend on the balance among communication savings, computation cost, semantic reliability, uncertainty, and application risk. SC should therefore not be treated as a universal replacement for raw-data transmission, since semantic compression may discard useful information and generative reconstruction may produce plausible but incorrect outputs. Future progress requires standardized semantic models, realistic testbeds, lightweight edge deployment, hybrid underwater links, uncertainty-aware reconstruction, semantic security, digital twins, and cross-domain interoperability. Addressing these challenges is essential for developing reliable, efficient, and sustainable SC-enabled IoUT systems.

\section*{Acknowledgment}
The authors would like to thank the Associate Provost for Research at the United Arab Emirates University (UAEU), UAE, for their kind support.

\bibliographystyle{IEEEtran}
\bibliography{References_v2}

@misc{MRFUnderwaterMarineIoT2026,
  author       = {{Market Research Future}},
  title        = {{Underwater Marine IoT Market Research Report}},
  year         = {2026},
  howpublished = {\url{https://www.marketresearchfuture.com/reports/underwater-marine-iot-market-39429}},
  note         = {Accessed: 30 May 2026}
}

@book{ShannonWeaver1949,
  author    = {Claude E. Shannon and Warren Weaver},
  title     = {The Mathematical Theory of Communication},
  publisher = {University of Illinois Press},
  address   = {Urbana, IL, USA},
  year      = {1949}
}

@article{Domingo2012IoUT,
  author  = {Mari Carmen Domingo},
  title   = {An Overview of the Internet of Underwater Things},
  journal = {Journal of Network and Computer Applications},
  volume  = {35},
  number  = {6},
  pages   = {1879--1890},
  year    = {2012},
  doi     = {10.1016/j.jnca.2012.07.012}
}

@article{Kao2017IoUT,
  author  = {Chien-Chi Kao and Yi-Shan Lin and Geng-De Wu and Chun-Ju Huang},
  title   = {A Comprehensive Study on the Internet of Underwater Things: Applications, Challenges, and Channel Models},
  journal = {Sensors},
  volume  = {17},
  number  = {7},
  pages   = {1477},
  year    = {2017},
  doi     = {10.3390/s17071477}
}

@article{Akyildiz2005UASN,
  author  = {Ian F. Akyildiz and Dario Pompili and Tommaso Melodia},
  title   = {Underwater Acoustic Sensor Networks: Research Challenges},
  journal = {Ad Hoc Networks},
  volume  = {3},
  number  = {3},
  pages   = {257--279},
  year    = {2005},
  doi     = {10.1016/j.adhoc.2005.01.004}
}

@article{Heidemann2012UnderwaterSensors,
  author  = {John Heidemann and Milica Stojanovic and Michele Zorzi},
  title   = {Underwater Sensor Networks: Applications, Advances and Challenges},
  journal = {Philosophical Transactions of the Royal Society A: Mathematical, Physical and Engineering Sciences},
  volume  = {370},
  number  = {1958},
  pages   = {158--175},
  year    = {2012},
  doi     = {10.1098/rsta.2011.0214}
}

@article{Qiu2020SmartOcean,
  author  = {Tie Qiu and Zhao Zhao and Tong Zhang and Chen Chen and C. L. Philip Chen},
  title   = {Underwater Internet of Things in Smart Ocean: System Architecture and Open Issues},
  journal = {IEEE Transactions on Industrial Informatics},
  volume  = {16},
  number  = {7},
  pages   = {4297--4307},
  year    = {2020},
  doi     = {10.1109/TII.2019.2946618}
}

@article{Gussen2016UWC,
  author  = {Camila M. G. Gussen and Paulo S. R. Diniz and Marcello L. R. Campos and Wallace A. Martins and Felipe M. Costa and Jonathan N. Gois},
  title   = {A Survey of Underwater Wireless Communication Technologies},
  journal = {Journal of Communication and Information Systems},
  volume  = {31},
  number  = {1},
  pages   = {242--255},
  year    = {2016},
  doi     = {10.14209/jcis.2016.22}
}

@article{Stojanovic2009UAC,
  author  = {Milica Stojanovic and James Preisig},
  title   = {Underwater Acoustic Communication Channels: Propagation Models and Statistical Characterization},
  journal = {IEEE Communications Magazine},
  volume  = {47},
  number  = {1},
  pages   = {84--89},
  year    = {2009},
  doi     = {10.1109/MCOM.2009.4752682}
}

@article{Chitre2008UnderwaterAcoustic,
  author  = {Mandar Chitre and Shiraz Shahabudeen and Milica Stojanovic},
  title   = {Underwater Acoustic Communications and Networking: Recent Advances and Future Challenges},
  journal = {Marine Technology Society Journal},
  volume  = {42},
  number  = {1},
  pages   = {103--116},
  year    = {2008},
  doi     = {10.4031/002533208786861263}
}

@article{Li2025UnderwaterAcoustic,
  author  = {Zhengnan Li and Mandar Chitre and Milica Stojanovic},
  title   = {Underwater Acoustic Communications},
  journal = {Nature Reviews Electrical Engineering},
  volume  = {2},
  number  = {2},
  pages   = {83--95},
  year    = {2025},
  doi     = {10.1038/s44287-024-00122-w}
}

@article{Kaushal2016UOWC,
  author  = {Hemani Kaushal and Georges Kaddoum},
  title   = {Underwater Optical Wireless Communication},
  journal = {IEEE Access},
  volume  = {4},
  pages   = {1518--1547},
  year    = {2016},
  doi     = {10.1109/ACCESS.2016.2552538}
}

@article{Zeng2017UOWCSurvey,
  author  = {Zhengquan Zeng and Shuping Fu and Huabing Zhang and Ying Dong and Jun Cheng},
  title   = {A Survey of Underwater Optical Wireless Communications},
  journal = {IEEE Communications Surveys \& Tutorials},
  volume  = {19},
  number  = {1},
  pages   = {204--238},
  year    = {2017},
  doi     = {10.1109/COMST.2016.2618841}
}

@article{Saeed2019UOWC,
  author  = {Nasir Saeed and Abdulkadir Celik and Tareq Y. Al-Naffouri and Mohamed-Slim Alouini},
  title   = {Underwater Optical Wireless Communications, Networking, and Localization: A Survey},
  journal = {Ad Hoc Networks},
  volume  = {94},
  pages   = {101935},
  year    = {2019},
  doi     = {10.1016/j.adhoc.2019.101935}
}

@article{Akyildiz2015MI,
  author  = {Ian F. Akyildiz and Pu Wang and Zhi Sun},
  title   = {Realizing Underwater Communication Through Magnetic Induction},
  journal = {IEEE Communications Magazine},
  volume  = {53},
  number  = {11},
  pages   = {42--48},
  year    = {2015},
  doi     = {10.1109/MCOM.2015.7321970}
}

@article{Theocharidis2025UnderwaterComm,
  title={Underwater communication technologies: a review: T. Theocharidis, E. Kavallieratou},
  author={Theocharidis, Theocharis and Kavallieratou, Ergina},
  journal={Telecommunication systems},
  volume={88},
  number={2},
  pages={54},
  year={2025},
  publisher={Springer}
}

@article{Hasan2025EnergyMAC,
  author  = {Walid K. Hasan and Iftekhar Ahmad and Daryoush Habibi and Quoc Viet Phung and Mohammad Al-Fawa'reh and Kazi Yasin Islam and Ruba Zaheer and Haitham Khaled},
  title   = {A Survey on Energy Efficient Medium Access Control for Acoustic Wireless Communication Networks in Underwater Environments},
  journal = {Journal of Network and Computer Applications},
  volume  = {235},
  pages   = {104079},
  year    = {2025},
  doi     = {10.1016/j.jnca.2024.104079}
}

@article{Awan2019UWSNReview,
  author  = {Khalid Mahmood Awan and Peer Azmat Shah and Khalid Iqbal and Saira Andleeb Gillani and Waqas Ahmad and Yunyoung Nam},
  title   = {Underwater Wireless Sensor Networks: A Review of Recent Issues and Challenges},
  journal = {Wireless Communications and Mobile Computing},
  volume  = {2019},
  pages   = {6470359},
  year    = {2019},
  doi     = {10.1155/2019/6470359}
}

@article{Fattah2020UWSNSurvey,
  author  = {Salmah Fattah and Abdullah Gani and Ismail Ahmedy and Mohd Yamani Idna Idris and Ibrahim Abaker Targio Hashem},
  title   = {A Survey on Underwater Wireless Sensor Networks: Requirements, Taxonomy, Recent Advances, and Open Research Challenges},
  journal = {Sensors},
  volume  = {20},
  number  = {18},
  pages   = {5393},
  year    = {2020},
  doi     = {10.3390/s20185393}
}

@article{Jahanbakht2021IoUTBigData,
  author  = {Mohammad Jahanbakht and Wei Xiang and Lajos Hanzo and Mostafa Rahimi Azghadi},
  title   = {Internet of Underwater Things and Big Marine Data Analytics---A Comprehensive Survey},
  journal = {IEEE Communications Surveys \& Tutorials},
  volume  = {23},
  number  = {2},
  pages   = {904--956},
  year    = {2021},
  doi     = {10.1109/COMST.2021.3053118}
}

@article{Mohsan2022IoUTSurvey,
  author  = {Syed Agha Hassnain Mohsan and Alireza Mazinani and Nawaf Qasem Hamood Othman and Hussain Amjad},
  title   = {Towards the Internet of Underwater Things: A Comprehensive Survey},
  journal = {Earth Science Informatics},
  volume  = {15},
  number  = {2},
  pages   = {735--764},
  year    = {2022},
  doi     = {10.1007/s12145-021-00762-8}
}

@article{Mohsan2023IoUTReview,
  author  = {Syed Agha Hassnain Mohsan and Yanlong Li and Muhammad Sadiq and Junwei Liang and Muhammad Asghar Khan},
  title   = {Recent Advances, Future Trends, Applications and Challenges of Internet of Underwater Things ({IoUT}): A Comprehensive Review},
  journal = {Journal of Marine Science and Engineering},
  volume  = {11},
  number  = {1},
  pages   = {124},
  year    = {2023},
  doi     = {10.3390/jmse11010124}
}

@article{Bello2022IoUTCommunication,
  author  = {Oladayo Bello and Sherali Zeadally},
  title   = {Internet of Underwater Things Communication: Architecture, Technologies, Research Challenges and Future Opportunities},
  journal = {Ad Hoc Networks},
  volume  = {135},
  pages   = {102933},
  year    = {2022},
  doi     = {10.1016/j.adhoc.2022.102933}
}

@article{Raj2021UIoTSecurity,
  author  = {Delphin Raj Kesari Mary and Eunbi Ko and Seung-Geun Kim and Sun-Ho Yum and Soo-Young Shin and Soo-Hyun Park},
  title   = {A Systematic Review on Recent Trends, Challenges, Privacy and Security Issues of Underwater Internet of Things},
  journal = {Sensors},
  volume  = {21},
  number  = {24},
  pages   = {8262},
  year    = {2021},
  doi     = {10.3390/s21248262}
}

@article{Luo2021RoutingUWSN,
  title={A survey of routing protocols for underwater wireless sensor networks},
  author={Luo, Junhai and Chen, Yanping and Wu, Man and Yang, Yang},
  journal={IEEE Communications Surveys \& Tutorials},
  volume={23},
  number={1},
  pages={137--160},
  year={2021},
  publisher={IEEE}
}

@article{Khan2018RoutingUWSN,
  author  = {Anwar Khan and Ihsan Ali and Abdullah Ghani and Nawsher Khan and Mohammed Alsaqer and Atiq Ur Rahman and Hasan Mahmood},
  title   = {Routing Protocols for Underwater Wireless Sensor Networks: Taxonomy, Research Challenges, Routing Strategies and Future Directions},
  journal = {Sensors},
  volume  = {18},
  number  = {5},
  pages   = {1619},
  year    = {2018},
  doi     = {10.3390/s18051619}
}

@article{Strinati2021SemanticGoal,
  author  = {Emilio Calvanese Strinati and Sergio Barbarossa},
  title   = {{6G} Networks: Beyond Shannon Towards Semantic and Goal-Oriented Communications},
  journal = {Computer Networks},
  volume  = {190},
  pages   = {107930},
  year    = {2021},
  doi     = {10.1016/j.comnet.2021.107930}
}

@article{Xie2021DeepSC,
  author  = {Huiqiang Xie and Zhijin Qin and Geoffrey Ye Li and Biing-Hwang Juang},
  title   = {Deep Learning Enabled Semantic Communication Systems},
  journal = {IEEE Transactions on Signal Processing},
  volume  = {69},
  pages   = {2663--2675},
  year    = {2021},
  doi     = {10.1109/TSP.2021.3071210}
}

@article{Luo2022SemanticOverview,
  author  = {Xuewen Luo and Hsiao-Hwa Chen and Qing Guo},
  title   = {Semantic Communications: Overview, Open Issues, and Future Research Directions},
  journal = {IEEE Wireless Communications},
  volume  = {29},
  number  = {1},
  pages   = {210--219},
  year    = {2022},
  doi     = {10.1109/MWC.101.2100269}
}

@article{Gunduz2023BeyondBits,
  author  = {Deniz G{\"u}nd{\"u}z and Zhijin Qin and Inaki Estella Aguerri and Harpreet S. Dhillon and Zhaohui Yang and Aylin Yener and Kai-Kit Wong and Chan-Byoung Chae},
  title   = {Beyond Transmitting Bits: Context, Semantics, and Task-Oriented Communications},
  journal = {IEEE Journal on Selected Areas in Communications},
  volume  = {41},
  number  = {1},
  pages   = {5--41},
  year    = {2023},
  doi     = {10.1109/JSAC.2022.3223408}
}

@article{Yang2023SemanticFutureInternet,
  author  = {Wanting Yang and Hongyang Du and Zi Qin Liew and Wei Yang Bryan Lim and Zehui Xiong and Dusit Niyato and Xuefen Chi and Xuemin Shen and Chunyan Miao},
  title   = {Semantic Communications for Future Internet: Fundamentals, Applications, and Challenges},
  journal = {IEEE Communications Surveys \& Tutorials},
  volume  = {25},
  number  = {1},
  pages   = {213--250},
  year    = {2023},
  doi     = {10.1109/COMST.2022.3223224}
}

@article{Lu2024SemanticsEmpowered,
  author  = {Zhilin Lu and Rongpeng Li and Kun Lu and Xianfu Chen and Ekram Hossain and Zhifeng Zhao and Honggang Zhang},
  title   = {Semantics-Empowered Communications: A Tutorial-cum-Survey},
  journal = {IEEE Communications Surveys \& Tutorials},
  volume  = {26},
  number  = {1},
  pages   = {41--79},
  year    = {2024},
  doi     = {10.1109/COMST.2023.3333342}
}

@article{Liu2024SemanticSurvey,
  title={A survey of secure semantic communications},
  author={Meng, Rui and Gao, Song and Fan, Dayu and Gao, Haixiao and Wang, Yining and Xu, Xiaodong and Wang, Bizhu and Lv, Suyu and Zhang, Zhidi and Sun, Mengying and others},
  journal={Journal of Network and Computer Applications},
  volume={239},
  pages={104181},
  year={2025},
  publisher={Elsevier}
}

@misc{Getu2023SemanticGoalSurvey,
  author       = {Tilahun M. Getu and Georges Kaddoum and Mehdi Bennis},
  title        = {Tutorial-Cum-Survey on Semantic and Goal-Oriented Communication: Research Landscape, Challenges, and Future Directions},
  year         = {2023},
  eprint       = {2308.01913},
  archivePrefix = {arXiv},
  primaryClass = {cs.IT}
}

@article{Shao2024TheorySemantic,
  title={A theory of semantic communication},
  author={Shao, Yulin and Cao, Qi and G{\"u}nd{\"u}z, Deniz},
  journal={IEEE Transactions on Mobile Computing},
  volume={23},
  number={12},
  pages={12211--12228},
  year={2024},
  publisher={IEEE}
}

@inproceedings{Zhang2023UnderwaterSemComSurvey,
  author    = {Jiasen Zhang and Weikai Sun and Yuanke Zhao and Haiyang Du},
  title     = {Semantic Communication in Underwater Communication: Advantage, Problem and Solution---A Survey},
  booktitle = {Proceedings of the 2023 8th International Conference on Intelligent Computing and Signal Processing (ICSP)},
  pages     = {2120--2123},
  year      = {2023},
  doi       = {10.1109/ICSP58490.2023.10248874}
}

@inproceedings{Loureiro2024SAGE,
  author    = {Jo{\~a}o Pedro Loureiro and Afonso Mateus and Filipe B. Teixeira and Rui Campos},
  title     = {A Semantic-Oriented Approach for Underwater Wireless Communications using Generative {AI}},
  booktitle = {Proceedings of the 15th IFIP Wireless and Mobile Networking Conference (WMNC)},
  address   = {Venice, Italy},
  pages     = {70--74},
  year      = {2024},
  url       = {https://opendl.ifip-tc6.org/db/conf/wmnc2024/wmnc2024/1571063070.pdf}
}

@article{Xu2024UWOSC,
  author  = {Jie Xu and Zhitong Huang and Yi Gao and Wenmin Zhai and Hongcheng Qiu and Yuefeng Ji},
  title   = {Design and Experimental Demonstration of Underwater Wireless Optical Communication System Based on Semantic Communication Paradigm},
  journal = {Optics Express},
  volume  = {32},
  number  = {2},
  pages   = {2188--2201},
  year    = {2024},
  doi     = {10.1364/OE.507955}
}

@article{Zhang2025UASC,
  title={An underwater acoustic semantic communication approach to underwater image transmission: Y. Zhang et al.},
  author={Zhang, Ying and Li, Huanyu and Li, Bingyu and Li, Li and Zhang, Weibo and Wang, Hao and Ren, Peng},
  journal={Intelligent Marine Technology and Systems},
  volume={3},
  number={1},
  pages={5},
  year={2025},
  publisher={Springer}
}

@article{Xu2025ESA_UWOSC,
  title={Environment semantics aided underwater wireless optical semantic communication},
  author={Xu, Jie and Huang, Zhitong and Zhai, Wenmin and Qiu, Hongcheng and Gao, Yi and Ji, Yuefeng},
  journal={Journal of Lightwave Technology},
  volume={43},
  number={3},
  pages={1186--1202},
  year={2024},
  publisher={IEEE}
}

@article{Picano2024SemanticFL,
  author  = {Benedetta Picano and Romano Fantacci},
  title   = {A Semantic-Oriented Federated Learning for Hybrid Ground--Aqua Computing Systems},
  journal = {IEEE Internet of Things Journal},
  volume  = {11},
  number  = {6},
  pages   = {10095--10103},
  year    = {2024},
  doi     = {10.1109/JIOT.2023.3325289}
}

@article{Loureiro2025Resilience,
  title={On the Resilience of Underwater Semantic Wireless Communications},
  author={Loureiro, Jo{\~a}o Pedro and Delgado, Patr{\'\i}cia and Ribeiro, Tom{\'a}s Feliciano and Teixeira, Filipe B and Campos, Rui},
  booktitle={OCEANS 2025 Brest},
  pages={1--7},
  year={2025},
  organization={IEEE}
}

@article{Chen2024LLMUnderwater,
 author={Chen, Weilong and Xu, Wenxuan and Chen, Haoran and Zhang, Xinran and Qin, Zhijin and Zhang, Yanru and Han, Zhu},
  journal={IEEE Transactions on Mobile Computing}, 
  title={Semantic Communication Based on Large Language Model for Underwater Image Transmission}, 
  year={2026},
  volume={25},
  number={2},
  pages={2060-2075},
  publisher={IEEE}
  }

@article{Guo2024SemComNet,
  title={A survey on semantic communication networks: Architecture, security, and privacy},
  author={Guo, Shaolong and Wang, Yuntao and Zhang, Ning and Su, Zhou and Luan, Tom H and Tian, Zhiyi and Shen, Xuemin},
  journal={IEEE communications surveys \& tutorials},
  volume={27},
  number={5},
  pages={2860--2894},
  year={2024},
  publisher={IEEE}
}

@article{Won2025SemComSecurityPrivacy,
  author  = {Dongwook Won and Geeranuch Woraphonbenjakul and Ayalneh Bitew Wondmagegn and Anh Tien Tran and Donghyun Lee and Demeke Shumeye Lakew and Sungrae Cho},
  title   = {Resource Management, Security, and Privacy Issues in Semantic Communications: A Survey},
  journal = {IEEE Communications Surveys \& Tutorials},
  volume  = {27},
  number  = {3},
  pages   = {1758--1797},
  year    = {2025},
  doi     = {10.1109/COMST.2024.3471685}
}

@article{Meng2025SecureSemCom,
  author  = {Rui Meng and Song Gao and Dayu Fan and Haixiao Gao and Yining Wang and Xiaodong Xu and Bizhu Wang and Suyu Lv and Zhidi Zhang and Mengying Sun and Shujun Han and Chen Dong and Xiaofeng Tao and Ping Zhang},
  title   = {A Survey of Secure Semantic Communications},
  journal = {Journal of Network and Computer Applications},
  year    = {2025},
  pages   = {104181},
  doi     = {10.1016/j.jnca.2025.104181}
}

@article{Muhammad2025UWSNModernization,
  author  = {Aman Muhammad and Fuzhong Li and Zahid Ullah Khan and Faheem Khan and Javed Khan and Sajid Ullah Khan},
  title   = {Exploration of Contemporary Modernization in {UWSNs} in the Context of Localization Including Opportunities for Future Research in Machine Learning and Deep Learning},
  journal = {Scientific Reports},
  volume  = {15},
  pages   = {5672},
  year    = {2025},
  doi     = {10.1038/s41598-025-89916-y}
}

@inproceedings{Partan2006PracticalUnderwater,
  author    = {Jim Partan and Jim Kurose and Brian Neil Levine},
  title     = {A Survey of Practical Issues in Underwater Networks},
  booktitle = {Proceedings of the 1st ACM International Workshop on Underwater Networks},
  series    = {WUWNet '06},
  pages     = {17--24},
  year      = {2006},
  publisher = {ACM},
  doi       = {10.1145/1161039.1161045}
}

@article{Climent2014UnderwaterAcousticWSN,
  author  = {Salvador Climent and Antonio Sanchez and Juan Vicente Capella and Nirvana Meratnia and Juan Jose Serrano},
  title   = {Underwater Acoustic Wireless Sensor Networks: Advances and Future Trends in Physical, {MAC} and Routing Layers},
  journal = {Sensors},
  volume  = {14},
  number  = {1},
  pages   = {795--833},
  year    = {2014},
  doi     = {10.3390/s140100795}
}

@article{Felemban2015USNApplications,
  author  = {Emad Felemban and Faisal Karim Shaikh and Umair Mujtaba Qureshi and Adil A. Sheikh and Saad Bin Qaisar},
  title   = {Underwater Sensor Network Applications: A Comprehensive Survey},
  journal = {International Journal of Distributed Sensor Networks},
  volume  = {11},
  number  = {11},
  pages   = {896832},
  year    = {2015},
  doi     = {10.1155/2015/896832}
}

@article{Sendra2015AcousticModems,
  author  = {Sandra Sendra and Jaime Lloret and Jose Miguel Jimenez and Lorena Parra},
  title   = {Underwater Acoustic Modems},
  journal = {IEEE Sensors Journal},
  volume  = {16},
  number  = {11},
  pages   = {4063--4071},
  year    = {2016},
  doi     = {10.1109/JSEN.2015.2434890}
}

@article{Sherlock2022MiniModems,
  author  = {Benjamin Sherlock and Nils Morozs and Jeffrey Neasham and Paul Mitchell},
  title   = {Ultra-Low-Cost and Ultra-Low-Power, Miniature Acoustic Modems Using Multipath Tolerant Spread-Spectrum Techniques},
  journal = {Electronics},
  volume  = {11},
  number  = {9},
  pages   = {1446},
  year    = {2022},
  doi     = {10.3390/electronics11091446}
}

@inproceedings{Pompili2006RoutingUWSN,
  author    = {Dario Pompili and Tommaso Melodia and Ian F. Akyildiz},
  title     = {Routing Algorithms for Delay-Insensitive and Delay-Sensitive Applications in Underwater Sensor Networks},
  booktitle = {Proceedings of the 12th Annual International Conference on Mobile Computing and Networking},
  series    = {MobiCom '06},
  pages     = {298--309},
  year      = {2006},
  publisher = {ACM}
}

@inproceedings{Xie2006VBF,
  author    = {Peng Xie and Jun-Hong Cui and Li Lao},
  title     = {{VBF}: Vector-Based Forwarding Protocol for Underwater Sensor Networks},
  booktitle = {Networking 2006: Networking Technologies, Services, and Protocols; Performance of Computer and Communication Networks; Mobile and Wireless Communications Systems},
  series    = {Lecture Notes in Computer Science},
  volume    = {3976},
  pages     = {1216--1221},
  year      = {2006},
  publisher = {Springer},
  doi       = {10.1007/11753810_111}
}

@inproceedings{Yan2008DBR,
  author    = {Hai Yan and Zhijie Jerry Shi and Jun-Hong Cui},
  title     = {{DBR}: Depth-Based Routing for Underwater Sensor Networks},
  booktitle = {Networking 2008 Ad Hoc and Sensor Networks, Wireless Networks, Next Generation Internet},
  series    = {Lecture Notes in Computer Science},
  volume    = {4982},
  pages     = {72--86},
  year      = {2008},
  publisher = {Springer},
  doi       = {10.1007/978-3-540-79549-0_7}
}

@inproceedings{Bao2011TheorySemantic,
  author    = {Jie Bao and Prithwish Basu and Michael Dean and Craig Partridge and Ananthram Swami and William Leland and James A. Hendler},
  title     = {Towards a Theory of Semantic Communication},
  booktitle = {2011 IEEE Network Science Workshop},
  pages     = {110--117},
  year      = {2011},
  doi       = {10.1109/NSW.2011.6004632}
}

@inproceedings{Tishby1999InformationBottleneck,
  author    = {Naftali Tishby and Fernando C. Pereira and William Bialek},
  title     = {The Information Bottleneck Method},
  booktitle = {Proceedings of the 37th Annual Allerton Conference on Communication, Control, and Computing},
  pages     = {368--377},
  year      = {1999}
}

@inproceedings{Farsad2018TextJSCC,
  author    = {Nariman Farsad and Milind Rao and Andrea Goldsmith},
  title     = {Deep Learning for Joint Source-Channel Coding of Text},
  booktitle = {2018 IEEE International Conference on Acoustics, Speech and Signal Processing},
  pages     = {2326--2330},
  year      = {2018},
  doi       = {10.1109/ICASSP.2018.8461983}
}

@article{Bourtsoulatze2019DeepJSCC,
  author  = {Eirina Bourtsoulatze and David Burth Kurka and Deniz G{\"u}nd{\"u}z},
  title   = {Deep Joint Source-Channel Coding for Wireless Image Transmission},
  journal = {IEEE Transactions on Cognitive Communications and Networking},
  volume  = {5},
  number  = {3},
  pages   = {567--579},
  year    = {2019},
  doi     = {10.1109/TCCN.2019.2919300}
}

@article{Weng2021SpeechSC,
  author  = {Zhenzi Weng and Zhijin Qin},
  title   = {Semantic Communication Systems for Speech Transmission},
  journal = {IEEE Journal on Selected Areas in Communications},
  volume  = {39},
  number  = {8},
  pages   = {2434--2444},
  year    = {2021},
  doi     = {10.1109/JSAC.2021.3087240}
}

@article{ZhangSTBL2023TaskUnawareSC,
  author  = {Hongwei Zhang and Shuo Shao and Meixia Tao and Xiaoyan Bi and Khaled B. Letaief},
  title   = {Deep Learning-Enabled Semantic Communication Systems With Task-Unaware Transmitter and Dynamic Data},
  journal = {IEEE Journal on Selected Areas in Communications},
  volume  = {41},
  number  = {1},
  pages   = {170--185},
  year    = {2023},
  doi     = {10.1109/JSAC.2022.3221991}
}

@article{Shi2021SemanticAwareNetworking,
  author  = {Guangming Shi and Yong Xiao and Yingyu Li and Xuemei Xie},
  title   = {From Semantic Communication to Semantic-Aware Networking: Model, Architecture, and Open Problems},
  journal = {IEEE Communications Magazine},
  volume  = {59},
  number  = {8},
  pages   = {44--50},
  year    = {2021},
  doi     = {10.1109/MCOM.001.2001239}
}

@article{Lu2021Rethinking,
  title={Rethinking modern communication from semantic coding to semantic communication},
  author={Lu, Kun and Zhou, Qingyang and Li, Rongpeng and Zhao, Zhifeng and Chen, Xianfu and Wu, Jianjun and Zhang, Honggang},
  journal={IEEE Wireless Communications},
  volume={30},
  number={1},
  pages={158--164},
  year={2022},
  publisher={IEEE}
}

@article{Zhou2022AdaptiveHARQ,
  author  = {Qingyang Zhou and Rongpeng Li and Zhifeng Zhao and Yong Xiao and Honggang Zhang},
  title   = {Adaptive Bit Rate Control in Semantic Communication With Incremental Knowledge-Based {HARQ}},
  journal = {IEEE Open Journal of the Communications Society},
  volume  = {3},
  pages   = {1076--1089},
  year    = {2022},
  doi     = {10.1109/OJCOMS.2022.3189023}
}

@article{Xiao2023ImplicitSemantic,
  author  = {Yong Xiao and Zijian Sun and Guangming Shi and Dusit Niyato},
  title   = {Imitation Learning-Based Implicit Semantic-Aware Communication Networks: Multi-Layer Representation and Collaborative Reasoning},
  journal = {IEEE Journal on Selected Areas in Communications},
  volume  = {41},
  number  = {3},
  pages   = {639--658},
  year    = {2023},
  doi     = {10.1109/JSAC.2022.3229419}
}

@article{Uysal2022DataSignificance,
  author  = {Elif Uysal and Onur Kaya and Anthony Ephremides and James Gross and Marian Codreanu and Petar Popovski and Mohamad Assaad and Gianluigi Liva and Andrea Munari and Beatriz Soret and Touraj Soleymani and Karl Henrik Johansson},
  title   = {Semantic Communications in Networked Systems: A Data Significance Perspective},
  journal = {IEEE Network},
  volume  = {36},
  number  = {4},
  pages   = {233--240},
  year    = {2022},
  doi     = {10.1109/MNET.106.2100636}
}

@article{Wen2023TaskOriented,
  author  = {Yuanming Shi and Yong Zhou and Dingzhu Wen and Youlong Wu and Chunxiao Jiang and Khaled B. Letaief},
  title   = {Task-Oriented Communications for {6G}: Vision, Principles, and Technologies},
  journal = {IEEE Wireless Communications},
  volume  = {30},
  number  = {3},
  pages   = {78--85},
  year    = {2023},
  doi     = {10.1109/MWC.002.2200468}
}

@article{Satyanarayanan2017EdgeComputing,
  author  = {Mahadev Satyanarayanan},
  title   = {The Emergence of Edge Computing},
  journal = {Computer},
  volume  = {50},
  number  = {1},
  pages   = {30--39},
  year    = {2017},
  doi     = {10.1109/MC.2017.9}
}

@article{Hogan2021KnowledgeGraphs,
  author  = {Aidan Hogan and Eva Blomqvist and Michael Cochez and Claudia d'Amato and Gerard de Melo and Claudio Gutierrez and Sabrina Kirrane and Jos{\'e} Emilio Labra Gayo and Roberto Navigli and Sebastian Neumaier and Axel-Cyrille Ngonga Ngomo and Axel Polleres and Sabbir M. Rashid and Anisa Rula and Lukas Schmelzeisen and Juan Sequeda and Steffen Staab and Antoine Zimmermann},
  title   = {Knowledge Graphs},
  journal = {ACM Computing Surveys},
  volume  = {54},
  number  = {4},
  pages   = {71:1--71:37},
  year    = {2021},
  doi     = {10.1145/3447772}
}

@inproceedings{McMahan2017FedAvg,
  author    = {H. Brendan McMahan and Eider Moore and Daniel Ramage and Seth Hampson and Blaise Ag{\"u}era y Arcas},
  title     = {Communication-Efficient Learning of Deep Networks from Decentralized Data},
  booktitle = {Proceedings of the 20th International Conference on Artificial Intelligence and Statistics},
  series    = {Proceedings of Machine Learning Research},
  volume    = {54},
  pages     = {1273--1282},
  year      = {2017},
  publisher = {PMLR}
}

@article{Kairouz2021FL,
  author  = {Peter Kairouz and H. Brendan McMahan and Brendan Avent and Aur{\'e}lien Bellet and Mehdi Bennis and Arjun Nitin Bhagoji and Kallista Bonawitz and Zachary Charles and Graham Cormode and Rachel Cummings and Rafael G. L. D'Oliveira and Hubert Eichner and Salim El Rouayheb and David Evans and Josh Gardner and Zachary Garrett and Adri{\`a} Gasc{\'o}n and Badih Ghazi and Phillip B. Gibbons and Marco Gruteser and Za{\"i}d Harchaoui and Chaoyang He and Lie He and Zhouyuan Huo and Ben Hutchinson and Justin Hsu and Martin Jaggi and Tara Javidi and Gauri Joshi and Mikhail Khodak and Jakub Kone{\v{c}}n{\'y} and Aleksandra Korolova and Farinaz Koushanfar and Sanmi Koyejo and Tancr{\`e}de Lepoint and Yang Liu and Prateek Mittal and Mehryar Mohri and Richard Nock and Ayfer {\"O}zg{\"u}r and Rasmus Pagh and Mariana Raykova and Hang Qi and Daniel Ramage and Ramesh Raskar and Dawn Song and Weikang Song and Sebastian U. Stich and Ziteng Sun and Ananda Theertha Suresh and Florian Tram{\`e}r and Praneeth Vepakomma and Jianyu Wang and Li Xiong and Zheng Xu and Qiang Yang and Felix X. Yu and Han Yu and Sen Zhao},
  title   = {Advances and Open Problems in Federated Learning},
  journal = {Foundations and Trends in Machine Learning},
  volume  = {14},
  number  = {1--2},
  pages   = {1--210},
  year    = {2021},
  doi     = {10.1561/2200000083}
}

@inproceedings{Papineni2002BLEU,
  author    = {Kishore Papineni and Salim Roukos and Todd Ward and Wei-Jing Zhu},
  title     = {{BLEU}: A Method for Automatic Evaluation of Machine Translation},
  booktitle = {Proceedings of the 40th Annual Meeting of the Association for Computational Linguistics},
  pages     = {311--318},
  year      = {2002},
  doi       = {10.3115/1073083.1073135}
}

@article{Wang2004SSIM,
  author  = {Zhou Wang and Alan C. Bovik and Hamid R. Sheikh and Eero P. Simoncelli},
  title   = {Image Quality Assessment: From Error Visibility to Structural Similarity},
  journal = {IEEE Transactions on Image Processing},
  volume  = {13},
  number  = {4},
  pages   = {600--612},
  year    = {2004},
  doi     = {10.1109/TIP.2003.819861}
}

@inproceedings{Radford2021CLIP,
  author    = {Alec Radford and Jong Wook Kim and Chris Hallacy and Aditya Ramesh and Gabriel Goh and Sandhini Agarwal and Girish Sastry and Amanda Askell and Pamela Mishkin and Jack Clark and Gretchen Krueger and Ilya Sutskever},
  title     = {Learning Transferable Visual Models From Natural Language Supervision},
  booktitle = {Proceedings of the 38th International Conference on Machine Learning},
  series    = {Proceedings of Machine Learning Research},
  volume    = {139},
  pages     = {8748--8763},
  year      = {2021},
  publisher = {PMLR}
}

@article{Yates2021AoISurvey,
  author  = {Roy D. Yates and Yin Sun and D. Richard Brown III and Sanjit K. Kaul and Eytan Modiano and Sennur Ulukus},
  title   = {Age of Information: An Introduction and Survey},
  journal = {IEEE Journal on Selected Areas in Communications},
  volume  = {39},
  number  = {5},
  pages   = {1183--1210},
  year    = {2021},
  doi     = {10.1109/JSAC.2021.3065072}
}

@inproceedings{Ronneberger2015UNet,
  author    = {Olaf Ronneberger and Philipp Fischer and Thomas Brox},
  title     = {{U-Net}: Convolutional Networks for Biomedical Image Segmentation},
  booktitle = {Medical Image Computing and Computer-Assisted Intervention -- MICCAI 2015},
  series    = {Lecture Notes in Computer Science},
  volume    = {9351},
  pages     = {234--241},
  year      = {2015},
  publisher = {Springer},
  doi       = {10.1007/978-3-319-24574-4_28}
}

@inproceedings{Islam2020SUIM,
  author    = {Md Jahidul Islam and Chelsey Edge and Yuyang Xiao and Peigen Luo and Muntaqim Mehtaz and Christopher Morse and Sadman Sakib Enan and Junaed Sattar},
  title     = {Semantic Segmentation of Underwater Imagery: Dataset and Benchmark},
  booktitle = {Proceedings of the IEEE/RSJ International Conference on Intelligent Robots and Systems (IROS)},
  pages     = {1769--1776},
  year      = {2020},
  doi       = {10.1109/IROS45743.2020.9340821}
}

@article{Islam2020FUnIEGAN,
  author  = {Md Jahidul Islam and Youya Xia and Junaed Sattar},
  title   = {Fast Underwater Image Enhancement for Improved Visual Perception},
  journal = {IEEE Robotics and Automation Letters},
  volume  = {5},
  number  = {2},
  pages   = {3227--3234},
  year    = {2020},
  doi     = {10.1109/LRA.2020.2974710}
}

@inproceedings{Vaswani2017Attention,
  author    = {Ashish Vaswani and Noam Shazeer and Niki Parmar and Jakob Uszkoreit and Llion Jones and Aidan N. Gomez and {\L}ukasz Kaiser and Illia Polosukhin},
  title     = {Attention Is All You Need},
  booktitle = {Advances in Neural Information Processing Systems},
  volume    = {30},
  pages     = {5998--6008},
  year      = {2017}
}

@inproceedings{Dosovitskiy2021ViT,
  author    = {Alexey Dosovitskiy and Lucas Beyer and Alexander Kolesnikov and Dirk Weissenborn and Xiaohua Zhai and Thomas Unterthiner and Mostafa Dehghani and Matthias Minderer and Georg Heigold and Sylvain Gelly and Jakob Uszkoreit and Neil Houlsby},
  title     = {An Image is Worth 16x16 Words: Transformers for Image Recognition at Scale},
  booktitle = {International Conference on Learning Representations},
  year      = {2021},
  url       = {https://openreview.net/forum?id=YicbFdNTTy}
}

@inproceedings{Liu2021Swin,
  author    = {Ze Liu and Yutong Lin and Yue Cao and Han Hu and Yixuan Wei and Zheng Zhang and Stephen Lin and Baining Guo},
  title     = {Swin Transformer: Hierarchical Vision Transformer using Shifted Windows},
  booktitle = {Proceedings of the IEEE/CVF International Conference on Computer Vision},
  pages     = {10012--10022},
  year      = {2021},
  doi       = {10.1109/ICCV48922.2021.00986}
}

@inproceedings{Li2022BLIP,
  author    = {Junnan Li and Dongxu Li and Caiming Xiong and Steven Hoi},
  title     = {{BLIP}: Bootstrapping Language-Image Pre-training for Unified Vision-Language Understanding and Generation},
  booktitle = {Proceedings of the 39th International Conference on Machine Learning},
  series    = {Proceedings of Machine Learning Research},
  volume    = {162},
  pages     = {12888--12900},
  year      = {2022},
  publisher = {PMLR}
}

@inproceedings{Rombach2022LDM,
  author    = {Robin Rombach and Andreas Blattmann and Dominik Lorenz and Patrick Esser and Bj{\"o}rn Ommer},
  title     = {High-Resolution Image Synthesis with Latent Diffusion Models},
  booktitle = {Proceedings of the IEEE/CVF Conference on Computer Vision and Pattern Recognition},
  pages     = {10684--10695},
  year      = {2022},
  doi       = {10.1109/CVPR52688.2022.01042}
}

@inproceedings{Zhang2023ControlNet,
  author    = {Lvmin Zhang and Anyi Rao and Maneesh Agrawala},
  title     = {Adding Conditional Control to Text-to-Image Diffusion Models},
  booktitle = {Proceedings of the IEEE/CVF International Conference on Computer Vision},
  pages     = {3836--3847},
  year      = {2023}
}

@article{Tian2025AquaScope,
  title={AquaScope: Reliable Underwater Image Transmission on Mobile Devices},
  author={Tian, Beitong and Zhao, Lingzhi and Chen, Bo and Wu, Mingyuan and Zheng, Haozhen and Vasisht, Deepak and Yan, Francis Y and Nahrstedt, Klara},
  journal={arXiv preprint arXiv:2502.10891},
  year={2025}
}

@article{Tian2025AquaVLM,
  title={AquaVLM: Improving Underwater Situation Awareness with Mobile Vision Language Models},
  author={Tian, Beitong and Zhao, Lingzhi and Chen, Bo and Zheng, Haozhen and Yang, Jingcheng and Wu, Mingyuan and Vasisht, Deepak and Nahrstedt, Klara},
  journal={arXiv preprint arXiv:2510.21722},
  year={2025}
}

@inproceedings{Hinton2015Distillation,
  author    = {Geoffrey Hinton and Oriol Vinyals and Jeff Dean},
  title     = {Distilling the Knowledge in a Neural Network},
  booktitle = {NIPS Deep Learning and Representation Learning Workshop},
  year      = {2015},
  url       = {https://arxiv.org/abs/1503.02531}
}

@inproceedings{Han2016DeepCompression,
  author    = {Song Han and Huizi Mao and William J. Dally},
  title     = {Deep Compression: Compressing Deep Neural Networks with Pruning, Trained Quantization and Huffman Coding},
  booktitle = {International Conference on Learning Representations},
  year      = {2016},
  url       = {https://arxiv.org/abs/1510.00149}
}

@article{khalil2020toward,
  title={Toward the internet of underwater things: Recent developments and future challenges},
  author={Khalil, Ruhul Amin and Saeed, Nasir and Babar, Mohammad Inayatullah and Jan, Tariqullah},
  journal={IEEE Consumer Electronics Magazine},
  volume={10},
  number={6},
  pages={32--37},
  year={2020},
  publisher={IEEE}
}

@inproceedings{Jacob2018Quantization,
  author    = {Benoit Jacob and Skirmantas Kligys and Bo Chen and Menglong Zhu and Matthew Tang and Andrew Howard and Hartwig Adam and Dmitry Kalenichenko},
  title     = {Quantization and Training of Neural Networks for Efficient Integer-Arithmetic-Only Inference},
  booktitle = {Proceedings of the IEEE/CVF Conference on Computer Vision and Pattern Recognition},
  pages     = {2704--2713},
  year      = {2018},
  doi       = {10.1109/CVPR.2018.00286}
}

@inproceedings{Hu2021LoRA,
  author    = {Edward J. Hu and Yelong Shen and Phillip Wallis and Zeyuan Allen-Zhu and Yuanzhi Li and Shean Wang and Lu Wang and Weizhu Chen},
  title     = {{LoRA}: Low-Rank Adaptation of Large Language Models},
  booktitle = {International Conference on Learning Representations},
  year      = {2022},
  url       = {https://openreview.net/forum?id=nZeVKeeFYf9}
}

@article{Gu2023Mamba,
  title={Mamba: Linear-time sequence modeling with selective state spaces},
  author={Gu, Albert and Dao, Tri},
  journal={arXiv preprint arXiv:2312.00752},
  year={2023}
}

@inproceedings{Zhu2024VisionMamba,
  author    = {Lianghui Zhu and Bencheng Liao and Qian Zhang and Xinlong Wang and Wenyu Liu and Xinggang Wang},
  title     = {Vision Mamba: Efficient Visual Representation Learning with Bidirectional State Space Model},
  booktitle = {Proceedings of the 41st International Conference on Machine Learning},
  series    = {Proceedings of Machine Learning Research},
  volume    = {235},
  pages     = {62429--62442},
  year      = {2024},
  publisher = {PMLR}
}

@inproceedings{Yu2025MambaOut,
  author    = {Weihao Yu and Xinchao Wang},
  title     = {{MambaOut}: Do We Really Need Mamba for Vision?},
  booktitle = {Proceedings of the IEEE/CVF Conference on Computer Vision and Pattern Recognition},
  year      = {2025}
}

@misc{W3CSSN2017,
  author       = {{W3C} and {OGC}},
  title        = {Semantic Sensor Network Ontology},
  year         = {2017},
  howpublished = {\url{https://www.w3.org/TR/vocab-ssn/}},
  note         = {W3C Recommendation}
}

@article{vanHarmelen2021NeuroSymbolic,
  author  = {Frank van Harmelen and Artur d'Avila Garcez and Pascal Hitzler and Lars Kai Hansen},
  title   = {Neuro-symbolic Artificial Intelligence},
  journal = {AI Communications},
  volume  = {34},
  number  = {3},
  pages   = {197--209},
  year    = {2021},
  doi     = {10.3233/AIC-210084}
}

@article{Yu2023NeuralSymbolic,
  author  = {Dongran Yu and Bo Yang and Dayou Liu and Hui Wang and Shirui Pan},
  title   = {A Survey on Neural-Symbolic Learning Systems},
  journal = {Neural Networks},
  volume  = {166},
  pages   = {105--126},
  year    = {2023},
  doi     = {10.1016/j.neunet.2023.06.028}
}

@article{khalil2021bayesian,
  title={Bayesian multidimensional scaling for location awareness in hybrid-Internet of Underwater Things},
  author={Khalil, Ruhul Amin and Saeed, Nasir and Babar, Mohammad Inayatullah and Jan, Tariqullah and Din, Sadia},
  journal={IEEE/CAA Journal of Automatica Sinica},
  volume={9},
  number={3},
  pages={496--509},
  year={2021},
  publisher={IEEE}
}

@inproceedings{Schlichtkrull2018RGCN,
  author    = {Michael Schlichtkrull and Thomas N. Kipf and Peter Bloem and Rianne van den Berg and Ivan Titov and Max Welling},
  title     = {Modeling Relational Data with Graph Convolutional Networks},
  booktitle = {The Semantic Web},
  series    = {Lecture Notes in Computer Science},
  volume    = {10843},
  pages     = {593--607},
  year      = {2018},
  publisher = {Springer},
  doi       = {10.1007/978-3-319-93417-4_38}
}

@inproceedings{Li2020FedProx,
  author    = {Tian Li and Anit Kumar Sahu and Manzil Zaheer and Maziar Sanjabi and Ameet Talwalkar and Virginia Smith},
  title     = {Federated Optimization in Heterogeneous Networks},
  booktitle = {Proceedings of Machine Learning and Systems},
  volume    = {2},
  pages     = {429--450},
  year      = {2020}
}

@inproceedings{Karimireddy2020SCAFFOLD,
  author    = {Sai Praneeth Karimireddy and Satyen Kale and Mehryar Mohri and Sashank Reddi and Sebastian Stich and Ananda Theertha Suresh},
  title     = {{SCAFFOLD}: Stochastic Controlled Averaging for Federated Learning},
  booktitle = {Proceedings of the 37th International Conference on Machine Learning},
  series    = {Proceedings of Machine Learning Research},
  volume    = {119},
  pages     = {5132--5143},
  year      = {2020},
  publisher = {PMLR}
}

@inproceedings{Reddi2021FedOpt,
  author    = {Sashank J. Reddi and Zachary Charles and Manzil Zaheer and Zachary Garrett and Keith Rush and Jakub Kone{\v{c}}n{\'y} and Sanjiv Kumar and H. Brendan McMahan},
  title     = {Adaptive Federated Optimization},
  booktitle = {International Conference on Learning Representations},
  year      = {2021},
  url       = {https://openreview.net/forum?id=LkFG3lB13U5}
}

@inproceedings{Bonawitz2017SecureAggregation,
  author    = {Keith Bonawitz and Vladimir Ivanov and Ben Kreuter and Antonio Marcedone and H. Brendan McMahan and Sarvar Patel and Daniel Ramage and Aaron Segal and Karn Seth},
  title     = {Practical Secure Aggregation for Privacy-Preserving Machine Learning},
  booktitle = {Proceedings of the ACM SIGSAC Conference on Computer and Communications Security},
  pages     = {1175--1191},
  year      = {2017},
  publisher = {ACM},
  doi       = {10.1145/3133956.3133982}
}

@article{Li2020WaterNet,
  author  = {Chongyi Li and Chunle Guo and Wenqi Ren and Runmin Cong and Junhui Hou and Sam Kwong and Dacheng Tao},
  title   = {An Underwater Image Enhancement Benchmark Dataset and Beyond},
  journal = {IEEE Transactions on Image Processing},
  volume  = {29},
  pages   = {4376--4389},
  year    = {2020},
  doi     = {10.1109/TIP.2019.2955241}
}

@article{Menna2018Underwater3D,
  author  = {Fabio Menna and Panagiotis Agrafiotis and Andreas Georgopoulos},
  title   = {State of the Art and Applications in Archaeological Underwater 3D Recording and Mapping},
  journal = {Journal of Cultural Heritage},
  volume  = {33},
  pages   = {231--248},
  year    = {2018},
  doi     = {10.1016/j.culher.2018.02.017}
}

@article{JohnsonRoberson2017UnderwaterMapping,
  author  = {Matthew Johnson-Roberson and Mitch Bryson and Ariell Friedman and Oscar Pizarro and Giancarlo Troni and Peter Ozog and Jon C. Henderson},
  title   = {High-Resolution Underwater Robotic Vision-Based Mapping and Three-Dimensional Reconstruction for Archaeology},
  journal = {Journal of Field Robotics},
  volume  = {34},
  number  = {4},
  pages   = {625--643},
  year    = {2017},
  doi     = {10.1002/rob.21658}
}

@article{Zhang2021PipelineLeak,
  author  = {Hongwei Zhang and Shitong Zhang and Yanhui Wang and Yuhong Liu and Yanan Yang and Tian Zhou and Hongyu Bian},
  title   = {Subsea Pipeline Leak Inspection by Autonomous Underwater Vehicle},
  journal = {Applied Ocean Research},
  volume  = {107},
  pages   = {102321},
  year    = {2021},
  doi     = {10.1016/j.apor.2020.102321}
}

@article{Bobkov2023PipelineTracking,
  author  = {Valery Bobkov and Antonina Shupikova and Alexander Inzartsev},
  title   = {Recognition and Tracking of an Underwater Pipeline from Stereo Images during AUV-Based Inspection},
  journal = {Journal of Marine Science and Engineering},
  volume  = {11},
  number  = {10},
  pages   = {2002},
  year    = {2023},
  doi     = {10.3390/jmse11102002}
}

@article{Li2024UATSSIC,
  author  = {Jiaxu Li and Guangyao Han and Shuai Chang and Xiaomei Fu},
  title   = {A Semantic Segmentation-Based Underwater Acoustic Image Transmission Framework for Cooperative SLAM},
  journal = {Defence Technology},
  volume  = {33},
  pages   = {339--351},
  year    = {2024},
  doi     = {10.1016/j.dt.2023.05.012}
}

@article{Saleh2022FishSurvey,
  author  = {Alzayat Saleh and Marcus Sheaves and Mostafa Rahimi Azghadi},
  title   = {Computer Vision and Deep Learning for Fish Classification in Underwater Habitats: A Survey},
  journal = {Fish and Fisheries},
  volume  = {23},
  number  = {4},
  pages   = {977--999},
  year    = {2022},
  doi     = {10.1111/faf.12666}
}

@article{Marrable2022FishLabeling,
  author  = {Daniel Marrable and Kathryn Barker and Sawitchaya Tippaya and Mathew Wyatt and Scott Bainbridge and Marcus Stowar and Jason Larke},
  title   = {Accelerating Species Recognition and Labelling of Fish From Underwater Video With Machine-Assisted Deep Learning},
  journal = {Frontiers in Marine Science},
  volume  = {9},
  pages   = {944582},
  year    = {2022},
  doi     = {10.3389/fmars.2022.944582}
}

@article{Wang2023VDMOSurvey,
  author  = {Ning Wang and Tingkai Chen and Shaoman Liu and Rongfeng Wang and Hamid Reza Karimi and Yejin Lin},
  title   = {Deep Learning-Based Visual Detection of Marine Organisms: A Survey},
  journal = {Neurocomputing},
  volume  = {532},
  pages   = {1--32},
  year    = {2023},
  doi     = {10.1016/j.neucom.2023.02.018}
}

@article{Shaban2021OilSpillSAR,
  author  = {Mohamed Shaban and Reem Salim and Hadil Abu Khalifeh and Adel Khelifi and Ahmed Shalaby and Shady El-Mashad and Ali Mahmoud and Mohammed Ghazal and Ayman El-Baz},
  title   = {A Deep-Learning Framework for the Detection of Oil Spills from SAR Data},
  journal = {Sensors},
  volume  = {21},
  number  = {7},
  pages   = {2351},
  year    = {2021},
  doi     = {10.3390/s21072351}
}

@article{Dehghani2023OilSpillHybrid,
  author  = {Saeed Dehghani-Dehcheshmeh and Mohammad Akhoondzadeh and Saeid Homayouni},
  title   = {Oil Spills Detection from SAR Earth Observations Based on a Hybrid CNN Transformer Networks},
  journal = {Marine Pollution Bulletin},
  volume  = {190},
  pages   = {114834},
  year    = {2023},
  doi     = {10.1016/j.marpolbul.2023.114834}
}

@article{Trujillo2024OilSpillTwoStep,
  author  = {Rubicel Trujillo-Acatitla and Jos{\'e} Tuxpan-Vargas and Cesar{\'e} Ovando-V{\'a}zquez and Erandi Monterrubio-Mart{\'i}nez},
  title   = {Marine Oil Spill Detection and Segmentation in SAR Data with Two Steps Deep Learning Framework},
  journal = {Marine Pollution Bulletin},
  volume  = {204},
  pages   = {116549},
  year    = {2024},
  doi     = {10.1016/j.marpolbul.2024.116549}
}

@article{Mori2022TsunamiReview,
  author  = {Nobuhito Mori and Kenji Satake and Daniel Cox and Katsuichiro Goda and Patricio A. Catalan and Tung-Cheng Ho and Fumihiko Imamura and Tori Tomiczek and Patrick Lynett and Takuya Miyashita and Abdul Muhari and Vasily Titov and Rick Wilson},
  title   = {Giant Tsunami Monitoring, Early Warning and Hazard Assessment},
  journal = {Nature Reviews Earth \& Environment},
  volume  = {3},
  pages   = {557--572},
  year    = {2022},
  doi     = {10.1038/s43017-022-00327-3}
}

@article{Rim2022GNSSTsunami,
  author  = {Donsub Rim and Robert Baraldi and Christopher M. Liu and Randall J. LeVeque and Kenjiro Terada},
  title   = {Tsunami Early Warning From Global Navigation Satellite System Data Using Convolutional Neural Networks},
  journal = {Geophysical Research Letters},
  volume  = {49},
  number  = {20},
  pages   = {e2022GL099511},
  year    = {2022},
  doi     = {10.1029/2022GL099511}
}

@article{Li2024GNSSIRTsunami,
  author  = {Linlin Li and Qiang Qiu and Mai Ye and Dongju Peng and Ya-Ju Hsu and Peitao Wang and Huabin Shi and Kristine M. Larson and Peizhen Zhang},
  title   = {Island-Based GNSS-IR Network for Tsunami Detecting and Warning},
  journal = {Coastal Engineering},
  volume  = {190},
  pages   = {104501},
  year    = {2024},
  doi     = {10.1016/j.coastaleng.2024.104501}
}

@article{Wynn2014AUVGeoscience,
  author  = {Russell B. Wynn and Veerle A. I. Huvenne and Timothy P. Le Bas and Bramley J. Murton and Douglas P. Connelly and Brian J. Bett and Henry A. Ruhl and Kirsty J. Morris and John Peakall and Daniel R. Parsons and Esther J. Sumner and Stephen E. Darby and Robert M. Dorrell and James E. Hunt},
  title   = {Autonomous Underwater Vehicles ({AUVs}): Their Past, Present and Future Contributions to the Advancement of Marine Geoscience},
  journal = {Marine Geology},
  volume  = {352},
  pages   = {451--468},
  year    = {2014},
  doi     = {10.1016/j.margeo.2014.03.012}
}

@article{Zhang2023AUVNavigation,
  author  = {Bingbing Zhang and Daxiong Ji and Shuo Liu and Xinke Zhu and Wen Xu},
  title   = {Autonomous Underwater Vehicle Navigation: A Review},
  journal = {Ocean Engineering},
  volume  = {273},
  pages   = {113861},
  year    = {2023},
  doi     = {10.1016/j.oceaneng.2023.113861}
}

@article{Zhou2023SubseaAUV,
  author  = {Jing Zhou and Yulin Si and Ying Chen},
  title   = {A Review of Subsea AUV Technology},
  journal = {Journal of Marine Science and Engineering},
  volume  = {11},
  number  = {6},
  pages   = {1119},
  year    = {2023},
  doi     = {10.3390/jmse11061119}
}

@article{Ma2025AUVKeyTech,
  title={The state of the art in key technologies for autonomous underwater vehicles: A review},
  author={Ma, Dong and Li, Ye and Ma, Teng and Pascoal, Ant{\'o}nio M},
  journal={Engineering},
  year={2025},
  publisher={Elsevier}
}

@article{Xu2024UnderwaterVehicles,
  title={State-of-the-art underwater vehicles and technologies enabling smart ocean: Survey and classifications},
  author={Xu, Jiajie and Irigoien, Xabier and Alouini, Mohamed-Slim},
  journal={arXiv preprint arXiv:2412.18667},
  year={2024}
}

@techreport{OGCSensorThings2016,
  author      = {{Open Geospatial Consortium}},
  title       = {{OGC SensorThings API Part 1: Sensing}},
  institution = {{Open Geospatial Consortium}},
  type        = {Implementation Standard},
  number      = {15-078r6},
  year        = {2016},
  url         = {https://docs.ogc.org/is/15-078r6/15-078r6.html}
}

@techreport{ETSISAREF2025,
  author      = {{ETSI}},
  title       = {{SmartM2M; Smart Applications; Reference Ontology and oneM2M Mapping}},
  institution = {{European Telecommunications Standards Institute}},
  number      = {{ETSI TS 103 264 V4.1.1}},
  year        = {2025},
  url         = {https://saref.etsi.org/}
}

@techreport{ETSINGSILD2021,
  author      = {{ETSI}},
  title       = {{Context Information Management (CIM); NGSI-LD API}},
  institution = {{European Telecommunications Standards Institute}},
  number      = {{ETSI GS CIM 009 V1.5.1}},
  year        = {2021},
  url         = {https://www.etsi.org/deliver/etsi_gs/CIM/001_099/009/01.05.01_60/gs_CIM009v010501p.pdf}
}

@techreport{W3CWoT2023,
  author      = {Sebastian Kaebisch and Michael McCool and Victor Charpenay and Michael Koster},
  title       = {{Web of Things (WoT) Thing Description 1.1}},
  institution = {{World Wide Web Consortium}},
  type        = {W3C Recommendation},
  year        = {2023},
  url         = {https://www.w3.org/TR/wot-thing-description11/}
}

@techreport{Lebo2013PROVO,
  author      = {Timothy Lebo and Satya Sahoo and Deborah McGuinness and Khalid Belhajjame and James Cheney and Daniel Corsar and Daniel Garijo and Stian Soiland-Reyes and Stephan Zednik and Jun Zhao},
  title       = {{PROV-O: The PROV Ontology}},
  institution = {{World Wide Web Consortium}},
  type        = {W3C Recommendation},
  year        = {2013},
  url         = {https://www.w3.org/TR/prov-o/}
}

@techreport{Sporny2020JSONLD,
  author      = {Manu Sporny and Dave Longley and Gregg Kellogg and Markus Lanthaler and Pierre-Antoine Champin and Niklas Lindstr{\"o}m},
  title       = {{JSON-LD 1.1: A JSON-based Serialization for Linked Data}},
  institution = {{World Wide Web Consortium}},
  type        = {W3C Recommendation},
  year        = {2020},
  url         = {https://www.w3.org/TR/json-ld11/}
}

@techreport{Beckett2014RDF11,
  author      = {Richard Cyganiak and David Wood and Markus Lanthaler},
  title       = {{RDF 1.1 Concepts and Abstract Syntax}},
  institution = {{World Wide Web Consortium}},
  type        = {W3C Recommendation},
  year        = {2014},
  url         = {https://www.w3.org/TR/rdf11-concepts/}
}

@techreport{Hitzler2012OWL2,
  author      = {Pascal Hitzler and Markus Kr{\"o}tzsch and Bijan Parsia and Peter F. Patel-Schneider and Sebastian Rudolph},
  title       = {{OWL 2 Web Ontology Language Primer}},
  institution = {{World Wide Web Consortium}},
  type        = {W3C Recommendation},
  year        = {2012},
  url         = {https://www.w3.org/TR/owl2-primer/}
}

@inproceedings{Masiero2012DESERT,
  author    = {Riccardo Masiero and Saiful Azad and Federico Favaro and Matteo Petrani and Giovanni Toso and Federico Guerra and Paolo Casari and Michele Zorzi},
  title     = {{DESERT Underwater: An NS-Miracle-based Framework to Design, Simulate, Emulate and Realize Test-beds for Underwater Network Protocols}},
  booktitle = {Proceedings of OCEANS 2012 MTS/IEEE Yeosu},
  pages     = {1--10},
  year      = {2012},
  doi       = {10.1109/OCEANS-Yeosu.2012.6263524}
}

@inproceedings{Guerra2009WOSS,
  author    = {Federico Guerra and Paolo Casari and Michele Zorzi},
  title     = {{World Ocean Simulation System (WOSS): A Simulation Tool for Underwater Networks with Realistic Propagation Modeling}},
  booktitle = {Proceedings of the Fourth ACM International Workshop on Underwater Networks},
  year      = {2009}
}

@misc{Porter2011BELLHOP,
  author       = {Michael B. Porter},
  title        = {{The BELLHOP Manual and User's Guide: Preliminary Draft}},
  year         = {2011},
  howpublished = {Heat, Light, and Sound Research, Inc.},
  url          = {https://oalib-acoustics.org/AcousticsToolbox/}
}

@article{vanWalree2017Watermark,
  author  = {Paul A. van Walree and Fran{\c{c}}ois-Xavier Socheleau and Roald Otnes and Tron Jenserud},
  title   = {The Watermark Benchmark for Underwater Acoustic Modulation Schemes},
  journal = {IEEE Journal of Oceanic Engineering},
  volume  = {42},
  number  = {4},
  pages   = {1007--1018},
  year    = {2017},
  doi     = {10.1109/JOE.2017.2699078}
}

@inproceedings{Xie2009AquaSim,
  author    = {Peng Xie and Zhong Zhou and Zheng Peng and Hai Yan and Tiansi Hu and Jun-Hong Cui and Zhijie Shi and Yunsi Fei and Shengli Zhou},
  title     = {{Aqua-Sim: An NS-2 Based Simulator for Underwater Sensor Networks}},
  booktitle = {Proceedings of OCEANS 2009 MTS/IEEE Biloxi},
  pages     = {1--7},
  year      = {2009}
}

@inproceedings{Martin2015AquaSimNG,
  author    = {Robert Martin and Yujie Zhu and Lina Pu and Fangming Dou and Zheng Peng and Jun-Hong Cui and Sanguthevar Rajasekaran},
  title     = {{Aqua-Sim Next Generation: An NS-3 Based Simulator for Underwater Sensor Networks}},
  booktitle = {Proceedings of the 10th International Conference on Underwater Networks \& Systems},
  year      = {2015}
}

@inproceedings{Martins2014SUNRISE,
  author    = {Ricardo Martins and Jo{\~a}o Borges de Sousa and Renato Caldas and Chiara Petrioli and John Potter},
  title     = {{SUNRISE Project: Porto University Testbed}},
  booktitle = {Proceedings of the Underwater Communications and Networking Conference},
  pages     = {1--5},
  year      = {2014},
  doi       = {10.1109/UComms.2014.7017143}
}

@article{Barbie2021UnderwaterDT,
  title={Developing an underwater network of ocean observation systems with digital twin prototypes—a field report from the baltic sea},
  author={Barbie, Alexander and Pech, Niklas and Hasselbring, Wilhelm and Fl{\"o}gel, Sascha and Wenzh{\"o}fer, Frank and Walter, Michael and Shchekinova, Elena and Busse, Marc and T{\"u}rk, Matthias and Hofbauer, Michael and others},
  journal={IEEE Internet Computing},
  volume={26},
  number={3},
  pages={33--42},
  year={2021},
  publisher={IEEE}
}

@article{Sze2017EfficientDNN,
  author  = {Vivienne Sze and Yu-Hsin Chen and Tien-Ju Yang and Joel S. Emer},
  title   = {Efficient Processing of Deep Neural Networks: A Tutorial and Survey},
  journal = {Proceedings of the IEEE},
  volume  = {105},
  number  = {12},
  pages   = {2295--2329},
  year    = {2017},
  doi     = {10.1109/JPROC.2017.2761740}
}

@misc{Howard2017MobileNets,
  author        = {Andrew G. Howard and Menglong Zhu and Bo Chen and Dmitry Kalenichenko and Weijun Wang and Tobias Weyand and Marco Andreetto and Hartwig Adam},
  title         = {{MobileNets}: Efficient Convolutional Neural Networks for Mobile Vision Applications},
  year          = {2017},
  eprint        = {1704.04861},
  archivePrefix = {arXiv},
  primaryClass  = {cs.CV}
}

@inproceedings{Sandler2018MobileNetV2,
  author    = {Mark Sandler and Andrew Howard and Menglong Zhu and Andrey Zhmoginov and Liang-Chieh Chen},
  title     = {{MobileNetV2}: Inverted Residuals and Linear Bottlenecks},
  booktitle = {Proceedings of the IEEE Conference on Computer Vision and Pattern Recognition},
  pages     = {4510--4520},
  year      = {2018},
  doi       = {10.1109/CVPR.2018.00474}
}

@inproceedings{Tan2019EfficientNet,
  author    = {Mingxing Tan and Quoc V. Le},
  title     = {{EfficientNet}: Rethinking Model Scaling for Convolutional Neural Networks},
  booktitle = {Proceedings of the 36th International Conference on Machine Learning},
  series    = {Proceedings of Machine Learning Research},
  volume    = {97},
  pages     = {6105--6114},
  year      = {2019}
}

@inproceedings{Cai2019ProxylessNAS,
  author    = {Han Cai and Ligeng Zhu and Song Han},
  title     = {{ProxylessNAS}: Direct Neural Architecture Search on Target Task and Hardware},
  booktitle = {International Conference on Learning Representations},
  year      = {2019},
  url       = {https://openreview.net/forum?id=HylVB3AqYm}
}

@inproceedings{Rastegari2016XNORNet,
  author    = {Mohammad Rastegari and Vicente Ordonez and Joseph Redmon and Ali Farhadi},
  title     = {{XNOR-Net}: ImageNet Classification Using Binary Convolutional Neural Networks},
  booktitle = {European Conference on Computer Vision},
  pages     = {525--542},
  year      = {2016},
  publisher = {Springer},
  doi       = {10.1007/978-3-319-46493-0_32}
}

@inproceedings{Courbariaux2015BinaryConnect,
  author    = {Matthieu Courbariaux and Yoshua Bengio and Jean-Pierre David},
  title     = {{BinaryConnect}: Training Deep Neural Networks with Binary Weights During Propagations},
  booktitle = {Advances in Neural Information Processing Systems},
  volume    = {28},
  year      = {2015}
}

@inproceedings{Lin2020MCUNet,
  author    = {Ji Lin and Wei-Ming Chen and Yujun Lin and Chuang Gan and Song Han},
  title     = {{MCUNet}: Tiny Deep Learning on IoT Devices},
  booktitle = {Advances in Neural Information Processing Systems},
  volume    = {33},
  pages     = {11711--11722},
  year      = {2020}
}

@inproceedings{Guo2017Calibration,
  author    = {Chuan Guo and Geoff Pleiss and Yu Sun and Kilian Q. Weinberger},
  title     = {On Calibration of Modern Neural Networks},
  booktitle = {Proceedings of the 34th International Conference on Machine Learning},
  series    = {Proceedings of Machine Learning Research},
  volume    = {70},
  pages     = {1321--1330},
  year      = {2017}
}

@inproceedings{Gal2016Dropout,
  author    = {Yarin Gal and Zoubin Ghahramani},
  title     = {Dropout as a Bayesian Approximation: Representing Model Uncertainty in Deep Learning},
  booktitle = {Proceedings of the 33rd International Conference on Machine Learning},
  series    = {Proceedings of Machine Learning Research},
  volume    = {48},
  pages     = {1050--1059},
  year      = {2016}
}

@inproceedings{Ovadia2019Uncertainty,
  author    = {Yaniv Ovadia and Emily Fertig and Jie Ren and Zachary Nado and D. Sculley and Sebastian Nowozin and Joshua Dillon and Balaji Lakshminarayanan and Jasper Snoek},
  title     = {Can You Trust Your Model's Uncertainty? Evaluating Predictive Uncertainty Under Dataset Shift},
  booktitle = {Advances in Neural Information Processing Systems},
  volume    = {32},
  year      = {2019}
}

@inproceedings{Hendrycks2019Robustness,
  author    = {Dan Hendrycks and Thomas Dietterich},
  title     = {Benchmarking Neural Network Robustness to Common Corruptions and Perturbations},
  booktitle = {International Conference on Learning Representations},
  year      = {2019},
  url       = {https://openreview.net/forum?id=HJz6tiCqYm}
}

@article{Gawlikowski2023UncertaintySurvey,
  author  = {Jakub Gawlikowski and Cedrique Rovile Njieutcheu Tassi and Mohsin Ali and Jungseok Lee and Matthias Humt and Jianxiang Feng and Anna Kruspe and Rudolph Triebel and Peter Jung and Ribana Roscher and Muhammad Shahzad and Wen Yang and Richard Bamler and Xiao Xiang Zhu},
  title   = {A Survey of Uncertainty in Deep Neural Networks},
  journal = {Artificial Intelligence Review},
  volume  = {56},
  pages   = {1513--1589},
  year    = {2023},
  doi     = {10.1007/s10462-023-10562-9}
}

@inproceedings{Kendall2017Uncertainties,
  author    = {Alex Kendall and Yarin Gal},
  title     = {What Uncertainties Do We Need in Bayesian Deep Learning for Computer Vision?},
  booktitle = {Advances in Neural Information Processing Systems},
  volume    = {30},
  year      = {2017}
}

@article{Ji2023HallucinationSurvey,
  author  = {Ziwei Ji and Nayeon Lee and Rita Frieske and Tiezheng Yu and Dan Su and Yan Xu and Etsuko Ishii and Ye Jin Bang and Andrea Madotto and Pascale Fung},
  title   = {Survey of Hallucination in Natural Language Generation},
  journal = {ACM Computing Surveys},
  volume  = {55},
  number  = {12},
  pages   = {1--38},
  year    = {2023},
  doi     = {10.1145/3571730}
}

@article{Zhong2024TheorySurvey,
  author  = {Yu Zhong and Feng Liu and Yuxuan Liu and Hengshuai Yao and Tao Han},
  title   = {Semantic Communication: A Survey of Its Theoretical Development},
  journal = {Entropy},
  volume  = {26},
  number  = {2},
  pages   = {102},
  year    = {2024},
  doi     = {10.3390/e26020102}
}

@article{Hanson2008UWOC,
  author  = {Frank Hanson and Stojan Radic},
  title   = {High Bandwidth Underwater Optical Communication},
  journal = {Applied Optics},
  volume  = {47},
  number  = {2},
  pages   = {277--283},
  year    = {2008},
  doi     = {10.1364/AO.47.000277}
}

@article{Jamali2018UWOCFading,
  author  = {Mohammad Vahid Jamali and Ali Mirani and Alireza Parsay and Bahman Abolhassani and Pooya Nabavi and Ata Chizari and Pirazh Khorramshahi and Sajjad AbdollahRamezani and Jawad A. Salehi},
  title   = {Statistical Studies of Fading in Underwater Wireless Optical Channels in the Presence of Air Bubble, Temperature, and Salinity Random Variations},
  journal = {IEEE Transactions on Communications},
  volume  = {66},
  number  = {10},
  pages   = {4706--4723},
  year    = {2018}
}

@inproceedings{Tonolini2018CrossBoundary,
  author    = {Francesco Tonolini and Fadel Adib},
  title     = {Networking Across Boundaries: Enabling Wireless Communication Through the Water-Air Interface},
  booktitle = {Proceedings of the ACM SIGCOMM 2018 Conference},
  pages     = {117--131},
  year      = {2018},
  publisher = {ACM}
}

@article{Sun2022ARIS,
  author  = {Zhi Sun and Huan Guo and Ian F. Akyildiz},
  title   = {High-Data-Rate Long-Range Underwater Communications via Acoustic Reconfigurable Intelligent Surfaces},
  journal = {IEEE Communications Magazine},
  volume  = {60},
  number  = {10},
  pages   = {96--102},
  year    = {2022}
}

@article{Wang2023ARIS,
  author  = {Hao Wang and Zhi Sun and Huan Guo and Pu Wang and Ian F. Akyildiz},
  title   = {Designing Acoustic Reconfigurable Intelligent Surface for Underwater Communications},
  journal = {IEEE Transactions on Wireless Communications},
  volume  = {22},
  number  = {12},
  pages   = {8934--8948},
  year    = {2023}
}

@article{Luo2024UARIS,
  title={Underwater acoustic reconfigurable intelligent surfaces: From principle to practice},
  author={Luo, Yu and Pu, Lina and Diao, Junming and Liu, Chun-Hung and Song, Aijun},
  journal={IEEE Communications Standards Magazine},
  year={2025},
  publisher={IEEE}
}

@article{Luo2025MLARIS,
  title={Multilayered Intelligent Reflecting Surface for Long-Range Underwater Acoustic Communication},
  author={Luo, Yu and Pu, Lina and Song, Aijun},
  journal={arXiv preprint arXiv:2501.18355},
  year={2025}
}

@article{Khan2022EnergyHarvesting,
  author  = {Anwar Khan and Muhammad Imran and Abdullah Alharbi and Ehab Mahmoud Mohamed and Mostafa M. Fouda},
  title   = {Energy Harvesting in Underwater Acoustic Wireless Sensor Networks: Design, Taxonomy, Applications, Challenges and Future Directions},
  journal = {IEEE Access},
  volume  = {10},
  pages   = {134606--134622},
  year    = {2022},
  doi     = {10.1109/ACCESS.2022.3230600}
}

@article{Alamu2023EnergyHarvesting,
  author  = {Olumide Alamu and Thomas O. Olwal and Adnan M. Abu-Mahfouz},
  title   = {Energy Harvesting Techniques for Sustainable Underwater Wireless Communication Networks: A Review},
  journal = {e-Prime -- Advances in Electrical Engineering, Electronics and Energy},
  volume  = {5},
  pages   = {100265},
  year    = {2023},
  doi     = {10.1016/j.prime.2023.100265}
}

@article{Nordfjord2025OceanEnergy,
  author  = {Sverrir Jan Nordfjord and Saemundur E. Thorsteinsson and Kristinn Andersen},
  title   = {Powering Underwater Robotics Sensor Networks Through Ocean Energy Harvesting and Wireless Power Transfer Methods: Systematic Review},
  journal = {Journal of Marine Science and Engineering},
  volume  = {13},
  number  = {9},
  pages   = {1728},
  year    = {2025},
  doi     = {10.3390/jmse13091728}
}

@article{Faria2022UnderwaterEnergyHarvesting,
  author  = {C. L. Faria and M. S. Martins and T. Matos and R. Lima and J. M. Miranda and L. M. Gon{\c{c}}alves},
  title   = {Underwater Energy Harvesting to Extend Operation Time of Submersible Sensors},
  journal = {Sensors},
  volume  = {22},
  number  = {4},
  pages   = {1341},
  year    = {2022},
  doi     = {10.3390/s22041341}
}

@inproceedings{Jang2019UnderwaterBackscatter,
  author    = {Junsu Jang and Fadel Adib},
  title     = {Underwater Backscatter Networking},
  booktitle = {Proceedings of the ACM Special Interest Group on Data Communication},
  pages     = {187--199},
  year      = {2019},
  publisher = {ACM},
  doi       = {10.1145/3341302.3342091}
}

@inproceedings{Ghaffarivardavagh2020UBL,
  author    = {Reza Ghaffarivardavagh and Sayed Saad Afzal and Osvy Rodriguez and Fadel Adib},
  title     = {Underwater Backscatter Localization: Toward a Battery-Free Underwater GPS},
  booktitle = {Proceedings of the 19th ACM Workshop on Hot Topics in Networks},
  pages     = {125--131},
  year      = {2020},
  publisher = {ACM},
  doi       = {10.1145/3422604.3425950}
}

@article{Afzal2022BatteryFreeImaging,
  author  = {Sayed Saad Afzal and Waleed Akbar and Osvy Rodriguez and Mario Doumet and Unsoo Ha and Reza Ghaffarivardavagh and Fadel Adib},
  title   = {Battery-Free Wireless Imaging of Underwater Environments},
  journal = {Nature Communications},
  volume  = {13},
  pages   = {5546},
  year    = {2022},
  doi     = {10.1038/s41467-022-33223-x}
}

@article{Bereketli2022BackscatterCoverage,
  author  = {Alper Bereketli},
  title   = {Interference-Free Source Deployment for Coverage in Underwater Acoustic Backscatter Networks},
  journal = {Peer-to-Peer Networking and Applications},
  volume  = {15},
  pages   = {1577--1594},
  year    = {2022},
  doi     = {10.1007/s12083-022-01312-9}
}

@inproceedings{Dwork2006DifferentialPrivacy,
  author    = {Cynthia Dwork},
  title     = {Differential Privacy},
  booktitle = {Automata, Languages and Programming},
  series    = {Lecture Notes in Computer Science},
  volume    = {4052},
  pages     = {1--12},
  year      = {2006},
  publisher = {Springer},
  doi       = {10.1007/11787006_1}
}

@inproceedings{Shokri2017MembershipInference,
  author    = {Reza Shokri and Marco Stronati and Congzheng Song and Vitaly Shmatikov},
  title     = {Membership Inference Attacks Against Machine Learning Models},
  booktitle = {2017 IEEE Symposium on Security and Privacy},
  pages     = {3--18},
  year      = {2017},
  doi       = {10.1109/SP.2017.41}
}

@inproceedings{Zhu2019DeepLeakage,
  author    = {Ligeng Zhu and Zhijian Liu and Song Han},
  title     = {Deep Leakage from Gradients},
  booktitle = {Advances in Neural Information Processing Systems},
  volume    = {32},
  year      = {2019}
}

@inproceedings{Bagdasaryan2020BackdoorFL,
  author    = {Eugene Bagdasaryan and Andreas Veit and Yiqing Hua and Deborah Estrin and Vitaly Shmatikov},
  title     = {How To Backdoor Federated Learning},
  booktitle = {Proceedings of the 23rd International Conference on Artificial Intelligence and Statistics},
  series    = {Proceedings of Machine Learning Research},
  volume    = {108},
  pages     = {2938--2948},
  year      = {2020}
}

@inproceedings{Blanchard2017Byzantine,
  author    = {Peva Blanchard and El Mahdi El Mhamdi and Rachid Guerraoui and Julien Stainer},
  title     = {Machine Learning with Adversaries: Byzantine Tolerant Gradient Descent},
  booktitle = {Advances in Neural Information Processing Systems},
  volume    = {30},
  year      = {2017}
}

@inproceedings{Goodfellow2015Adversarial,
  author    = {Ian J. Goodfellow and Jonathon Shlens and Christian Szegedy},
  title     = {Explaining and Harnessing Adversarial Examples},
  booktitle = {International Conference on Learning Representations},
  year      = {2015},
  url       = {https://arxiv.org/abs/1412.6572}
}

@article{Mothukuri2021FLSecuritySurvey,
  author  = {Viraaji Mothukuri and Reza M. Parizi and Seyedamin Pouriyeh and Ali Dehghantanha and Kim-Kwang Raymond Choo},
  title   = {A Survey on Security and Privacy of Federated Learning},
  journal = {Future Generation Computer Systems},
  volume  = {115},
  pages   = {619--640},
  year    = {2021},
  doi     = {10.1016/j.future.2020.10.007}
}

@article{Truong2021FLThreats,
  author  = {Nguyen Truong and Kai Sun and Siyao Wang and Florian Guitton and YiKe Guo},
  title   = {Privacy Preservation in Federated Learning: An Insightful Survey from the GDPR Perspective},
  journal = {Computers \& Security},
  volume  = {110},
  pages   = {102402},
  year    = {2021},
  doi     = {10.1016/j.cose.2021.102402}
}

@article{Tzachor2023DigitalTwinOcean,
  author  = {Asaf Tzachor and Ofir Hendel and Catherine E. Richards},
  title   = {Digital Twins: A Stepping Stone to Achieve Ocean Sustainability?},
  journal = {npj Ocean Sustainability},
  volume  = {2},
  pages   = {16},
  year    = {2023},
  doi     = {10.1038/s44183-023-00023-9}
}

@article{Wang2025DTUnderwaterIoT,
  author  = {Lei Wang and Lei Yan and Xinbin Li and Song Han},
  title   = {System-Level Digital Twin Modeling for Underwater Wireless IoT Networks},
  journal = {Journal of Marine Science and Engineering},
  volume  = {13},
  number  = {1},
  pages   = {32},
  year    = {2025},
  doi     = {10.3390/jmse13010032}
}

@article{Kaklis2023MaritimeDT,
  author  = {Dimitrios Kaklis and Iraklis Varlamis and George Giannakopoulos and Takis J. Varelas and Constantine D. Spyropoulos},
  title   = {Enabling Digital Twins in the Maritime Sector Through the Lens of AI and Industry 4.0},
  journal = {Journal of Industrial Information Integration},
  volume  = {34},
  pages   = {100178},
  year    = {2023},
  doi     = {10.1016/j.jii.2023.100178}
}

@misc{OGCILIAD2023,
  author       = {{Open Geospatial Consortium}},
  title        = {{ILIAD Digital Twin of the Ocean}},
  year         = {2023},
  howpublished = {\url{https://www.ogc.org/initiatives/iliad/}},
  note         = {Accessed: 31 May 2026}
}

@misc{EuropeanCommissionDTO2024,
  author       = {{European Commission}},
  title        = {{European Digital Twin Ocean}},
  year         = {2024},
  howpublished = {\url{https://research-and-innovation.ec.europa.eu/funding/funding-opportunities/funding-programmes-and-open-calls/horizon-europe/eu-missions-horizon-europe/restore-our-ocean-and-waters/european-digital-twin-ocean_en}},
  note         = {Accessed: 06 Jun 2026}
}

@article{Sozer2000UnderwaterAcousticNetworks,
  author  = {Ethem M. Sozer and Milica Stojanovic and John G. Proakis},
  title   = {Underwater Acoustic Networks},
  journal = {IEEE Journal of Oceanic Engineering},
  volume  = {25},
  number  = {1},
  pages   = {72--83},
  year    = {2000},
  doi     = {10.1109/48.820738}
}

@article{Stojanovic2007CapacityDistance,
  author  = {Milica Stojanovic},
  title   = {On the Relationship Between Capacity and Distance in an Underwater Acoustic Communication Channel},
  journal = {ACM SIGMOBILE Mobile Computing and Communications Review},
  volume  = {11},
  number  = {4},
  pages   = {34--43},
  year    = {2007},
  doi     = {10.1145/1347364.1347373}
}

@inproceedings{Fall2003DTN,
  author    = {Kevin Fall},
  title     = {A Delay-Tolerant Network Architecture for Challenged Internets},
  booktitle = {Proceedings of the ACM SIGCOMM 2003 Conference on Applications, Technologies, Architectures, and Protocols for Computer Communication},
  pages     = {27--34},
  year      = {2003},
  publisher = {ACM},
  doi       = {10.1145/863955.863960}
}

@article{Dunbabin2012RobotsEnvMonitoring,
  author  = {Matthew Dunbabin and Lino Marques},
  title   = {Robots for Environmental Monitoring: Significant Advancements and Applications},
  journal = {IEEE Robotics \& Automation Magazine},
  volume  = {19},
  number  = {1},
  pages   = {24--39},
  year    = {2012},
  doi     = {10.1109/MRA.2011.2181683}
}

@article{Ditria2020FishAbundance,
  author  = {Ellen M. Ditria and Sebastian Lopez-Marcano and Michael K. Sievers and Eric L. Jinks and Christopher J. Brown and Rod M. Connolly},
  title   = {Automating the Analysis of Fish Abundance Using Object Detection: Optimising Animal Ecology With Deep Learning},
  journal = {Frontiers in Marine Science},
  volume  = {7},
  pages   = {429},
  year    = {2020},
  doi     = {10.3389/fmars.2020.00429}
}

@article{Solberg2012OilSpillRemoteSensing,
  author  = {Anne H. Schistad Solberg},
  title   = {Remote Sensing of Ocean Oil-Spill Pollution},
  journal = {Proceedings of the IEEE},
  volume  = {100},
  number  = {10},
  pages   = {2931--2945},
  year    = {2012},
  doi     = {10.1109/JPROC.2012.2196250}
}

@article{Paull2014AUVNavigation,
  author  = {Liam Paull and Sajad Saeedi and Mae Seto and Howard Li},
  title   = {{AUV} Navigation and Localization: A Review},
  journal = {IEEE Journal of Oceanic Engineering},
  volume  = {39},
  number  = {1},
  pages   = {131--149},
  year    = {2014},
  doi     = {10.1109/JOE.2013.2278891}
}

@inproceedings{Kirillov2023SAM,
  author    = {Alexander Kirillov and Eric Mintun and Nikhila Ravi and Hanzi Mao and Chloe Rolland and Laura Gustafson and Tete Xiao and Spencer Whitehead and Alexander C. Berg and Wan-Yen Lo and Piotr Doll{\'a}r and Ross Girshick},
  title     = {Segment Anything},
  booktitle = {Proceedings of the IEEE/CVF International Conference on Computer Vision},
  pages     = {4015--4026},
  year      = {2023}
}

@inproceedings{Li2023BLIP2,
  author    = {Junnan Li and Dongxu Li and Silvio Savarese and Steven C. H. Hoi},
  title     = {{BLIP-2}: Bootstrapping Language-Image Pre-training with Frozen Image Encoders and Large Language Models},
  booktitle = {Proceedings of the 40th International Conference on Machine Learning},
  series    = {Proceedings of Machine Learning Research},
  volume    = {202},
  pages     = {19730--19742},
  year      = {2023},
  publisher = {PMLR}
}

@inproceedings{Ho2020DDPM,
  author    = {Jonathan Ho and Ajay Jain and Pieter Abbeel},
  title     = {Denoising Diffusion Probabilistic Models},
  booktitle = {Advances in Neural Information Processing Systems},
  volume    = {33},
  pages     = {6840--6851},
  year      = {2020}
}

@inproceedings{Frantar2023GPTQ,
  author    = {Elias Frantar and Saleh Ashkboos and Torsten Hoefler and Dan Alistarh},
  title     = {{GPTQ}: Accurate Post-Training Quantization for Generative Pre-trained Transformers},
  booktitle = {International Conference on Learning Representations},
  year      = {2023},
  url       = {https://openreview.net/forum?id=tcbBPnfwxS}
}

@inproceedings{Dettmers2023QLoRA,
  author    = {Tim Dettmers and Artidoro Pagnoni and Ari Holtzman and Luke Zettlemoyer},
  title     = {{QLoRA}: Efficient Finetuning of Quantized LLMs},
  booktitle = {Advances in Neural Information Processing Systems},
  volume    = {36},
  year      = {2023}
}

@article{Ganin2016DANN,
  author  = {Yaroslav Ganin and Evgeniya Ustinova and Hana Ajakan and Pascal Germain and Hugo Larochelle and Fran{\c{c}}ois Laviolette and Mario Marchand and Victor Lempitsky},
  title   = {Domain-Adversarial Training of Neural Networks},
  journal = {Journal of Machine Learning Research},
  volume  = {17},
  number  = {59},
  pages   = {1--35},
  year    = {2016}
}

@inproceedings{Li2021FedBN,
  author    = {Xiaoxiao Li and Meirui Jiang and Xiaofei Zhang and Michael Kamp and Qi Dou},
  title     = {{FedBN}: Federated Learning on Non-IID Features via Local Batch Normalization},
  booktitle = {International Conference on Learning Representations},
  year      = {2021},
  url       = {https://openreview.net/forum?id=6YEQUn0QICG}
}

@inproceedings{Lin2018DeepGradientCompression,
  author    = {Yujun Lin and Song Han and Huizi Mao and Yu Wang and William J. Dally},
  title     = {Deep Gradient Compression: Reducing the Communication Bandwidth for Distributed Training},
  booktitle = {International Conference on Learning Representations},
  year      = {2018},
  url       = {https://openreview.net/forum?id=SkhQHMW0W}
}

@inproceedings{Mehta2022MobileViT,
  author    = {Sachin Mehta and Mohammad Rastegari},
  title     = {{MobileViT}: Light-weight, General-purpose, and Mobile-friendly Vision Transformer},
  booktitle = {International Conference on Learning Representations},
  year      = {2022},
  url       = {https://openreview.net/forum?id=vh-0sUt8HlG}
}

@article{Wilkinson2016FAIR,
  author  = {Mark D. Wilkinson and Michel Dumontier and IJsbrand Jan Aalbersberg and Gabrielle Appleton and Myles Axton and Arie Baak and Niklas Blomberg and Jan-Willem Boiten and Luiz Bonino da Silva Santos and Philip E. Bourne and Jildau Bouwman and Anthony J. Brookes and Tim Clark and Merc{\`e} Crosas and Ingrid Dillo and Olivier Dumon and Scott Edmunds and Chris T. Evelo and Richard Finkers and Alejandra Gonzalez-Beltran and Alasdair J. G. Gray and Paul Groth and Carole Goble and Jeffrey S. Grethe and Jaap Heringa and Peter A. C. 't Hoen and Rob Hooft and Tobias Kuhn and Ruben Kok and Joost Kok and Scott J. Lusher and Maryann E. Martone and Albert Mons and Abel L. Packer and Bengt Persson and Philippe Rocca-Serra and Marco Roos and Rene van Schaik and Susanna-Assunta Sansone and Erik Schultes and Thierry Sengstag and Ted Slater and George Strawn and Morris A. Swertz and Mark Thompson and Johan van der Lei and Erik van Mulligen and Jan Velterop and Andra Waagmeester and Peter Wittenburg and Katherine Wolstencroft and Jun Zhao and Barend Mons},
  title   = {The {FAIR} Guiding Principles for Scientific Data Management and Stewardship},
  journal = {Scientific Data},
  volume  = {3},
  pages   = {160018},
  year    = {2016},
  doi     = {10.1038/sdata.2016.18}
}

@article{Wieczorek2012DarwinCore,
  author  = {John Wieczorek and David Bloom and Robert Guralnick and Stan Blum and Markus D{\"o}ring and Renato Giovanni and Tim Robertson and David Vieglais},
  title   = {Darwin Core: An Evolving Community-Developed Biodiversity Data Standard},
  journal = {PLoS ONE},
  volume  = {7},
  number  = {1},
  pages   = {e29715},
  year    = {2012},
  doi     = {10.1371/journal.pone.0029715}
}

@book{Warden2019TinyML,
  author    = {Pete Warden and Daniel Situnayake},
  title     = {TinyML: Machine Learning with TensorFlow Lite on Arduino and Ultra-Low-Power Microcontrollers},
  publisher = {O'Reilly Media},
  year      = {2019}
}

@inproceedings{Lakshminarayanan2017DeepEnsembles,
  author    = {Balaji Lakshminarayanan and Alexander Pritzel and Charles Blundell},
  title     = {Simple and Scalable Predictive Uncertainty Estimation Using Deep Ensembles},
  booktitle = {Advances in Neural Information Processing Systems},
  volume    = {30},
  year      = {2017}
}

@inproceedings{Abadi2016DeepLearningDP,
  author    = {Martin Abadi and Andy Chu and Ian Goodfellow and H. Brendan McMahan and Ilya Mironov and Kunal Talwar and Li Zhang},
  title     = {Deep Learning with Differential Privacy},
  booktitle = {Proceedings of the 2016 ACM SIGSAC Conference on Computer and Communications Security},
  pages     = {308--318},
  year      = {2016},
  publisher = {ACM},
  doi       = {10.1145/2976749.2978318}
}

@inproceedings{Nasr2019PrivacyAnalysis,
  author    = {Milad Nasr and Reza Shokri and Amir Houmansadr},
  title     = {Comprehensive Privacy Analysis of Deep Learning: Passive and Active White-box Inference Attacks Against Centralized and Federated Learning},
  booktitle = {2019 IEEE Symposium on Security and Privacy},
  pages     = {739--753},
  year      = {2019},
  doi       = {10.1109/SP.2019.00065}
}

\end{document}